\documentclass[a4paper,12pt]{article}
\usepackage{jheppub}

\usepackage[applemac]{inputenc}
\usepackage{amsmath}
\usepackage{slashed}
\usepackage{amscd,amstext,amsbsy,amsopn,amsthm,amsxtra,upref,amsfonts,amssymb,euscript,latexsym,graphicx,color}
\usepackage[dvipsnames]{xcolor}
\usepackage[all]{xy}
\usepackage[american]{babel}
\usepackage{multicol}%\usepackage{physics}
\usepackage{tikz}
\usepackage{tikz-3dplot}
\usepackage{mdframed}
\usepackage{todonotes}
\usetikzlibrary{patterns}
\usepackage{hyperref,pgfplots,mdframed,subcaption}
\usepackage[normalem]{ulem}

\usepackage{soul}
\newcommand\figref{Figure~\ref}
\usepackage{booktabs}
\usepackage{amsmath,bbm,array,cleveref,amsfonts,subcaption,graphicx,wrapfig,lscape,youngtab,bbold,float,multirow,longtable,rotating,epstopdf}
\tikzstyle{white}=[circle,draw,
inner sep=0pt,minimum size=2mm]
\tikzstyle{black}=[circle,fill=black,inner sep=0pt,minimum size=2mm]
\def\beqa{\begin{eqnarray}}
	\def\eeqa{\end{eqnarray}}

\newcommand\undermat[2]{%
	\makebox[0pt][l]{$\smash{\underbrace{\phantom{%
					\begin{matrix}#2\end{matrix}}}_{\text{\normalsize $#1$}}}$}#2}

\newcommand{\be}{\begin{equation}}
\newcommand{\ee}{\end{equation}}
\newcommand{\beq}{\begin{equation}}
\newcommand{\beql}[1]{\begin{equation}\label{#1}}
\newcommand{\eeq}{\end{equation}}
\newcommand{\ba}{\begin{array}}
\newcommand{\ea}{\end{array}}
\newcommand{\bea}{\begin{eqnarray}}
\newcommand{\beal}[1]{\begin{eqnarray}\label{#1}}
\newcommand{\eea}{\end{eqnarray}}
\newcommand{\ben}{\begin{enumerate}}
\newcommand{\een}{\end{enumerate}}
\newcommand{\bean}{\begin{eqnarray*}}
\newcommand{\eean}{\end{eqnarray*}}

\newcommand{\fref}[1]{Figure \ref{#1}}
\newcommand{\btab}[1]{\begin{tabular}{#1}}
\newcommand{\etab}{\end{tabular}}

\newcommand{\comment}[1]{}

\def\Re           {{\rm Re\hskip0.1em}}
\def\Im           {{\rm Im\hskip0.1em}}

\newcommand{\drawsquare}[2]{\hbox{%
		\rule{#2pt}{#1pt}\hskip-#2pt%  left vertical
		\rule{#1pt}{#2pt}\hskip-#1pt%  lower horizontal
		\rule[#1pt]{#1pt}{#2pt}}\rule[#1pt]{#2pt}{#2pt}\hskip-#2pt%  upper horizontal
	\rule{#2pt}{#1pt}}% right vertical

\newcommand{\fund}{~\raisebox{-.5pt}{\drawsquare{6.5}{0.4}}~}
\newcommand{\antifund}{~\overline{\raisebox{-.5pt}{\drawsquare{6.5}{0.4}}}~}

\newcommand{\symm}{~\raisebox{-.5pt}{\drawsquare{6.5}{0.4}}\hskip-0.4pt%
	\raisebox{-.5pt}{\drawsquare{6.5}{0.4}}~}%  symmetric second rank

\newcommand{\asymm}{~\raisebox{-3.5pt}{\drawsquare{6.5}{0.4}}\hskip-6.9pt%
	\raisebox{3pt}{\drawsquare{6.5}{0.4}}~}%  antisymmetric second rank

\newcommand{\antiasymm}{~\overline{\raisebox{-3.5pt}{\drawsquare{6.5}{0.4}}\hskip-6.9pt%
		\raisebox{3pt}{\drawsquare{6.5}{0.4}}}~}%  antisymmetric second rank

\newcommand{\antisymm}{~\overline{\raisebox{-.5pt}{\drawsquare{6.5}{0.4}}\hskip-0.4pt%
		\raisebox{-.5pt}{\drawsquare{6.5}{0.4}}}~}%  symmetric second rank
\newcommand{\AP}[1]{{\color{blue}#1}} % needs the package 'color'
\title{Dimers in a Bottle}
\author[a]{Eduardo Garc\'{\i}a-Valdecasas}
\author[b]{Shani Meynet}
\author[a]{Antoine Pasternak}
\author[c]{Valdo Tatitscheff}

\affiliation[a]{Physique Th\'eorique et Math\'ematique and International Solvay Institutes \\ Universit\'e Libre de Bruxelles; C.P. 231, 1050 Brussels, Belgium}

\affiliation[b]{SISSA and INFN, Via Bonomea 265; I 34136 Trieste, Italy}

\affiliation[c]{IRMA, UMR 7501, Universit\'e de Strasbourg et CNRS\\
	7 rue Ren\'e Descartes 67000 Strasbourg, France}

\emailAdd{eduardo.garcia.valdecasas@gmail.com, smeynet@sissa.it, antoine.pasternak@ulb.ac.be, valdotatitscheff@gmail.com}

\abstract{We revisit D3-branes at toric CY$_3$ singularities with orientifolds and their description in terms of dimer models. We classify orientifold actions on the dimer through smooth involutions of the torus. In particular, we describe new orientifold projections related to maps on the dimer without fixed points, leading to Klein bottles. These new orientifolds lead to novel $\mathcal{N}=1$ SCFT's that resemble, in many aspects, non-orientifolded theories. For instance, we recover the presence of fractional branes and some of them trigger a cascading RG-flow \`a la Klebanov-Strassler. The remaining involutions lead to non-supersymmetric setups, thus exhausting the possible orientifolds on dimers.}

\preprint{
}

\begin{document}
	
	\maketitle
	
	%==================================================%
	%==================================================%

	\newpage

		\section{Introduction}
D-branes at singularities extend the original $\mathcal{N}=4$ SYM AdS/CFT correspondence \cite{Maldacena:1997re,Witten:1998qj} to theories without conformal invariance and/or reduced supersymmetry \cite{Klebanov:1998hh,Klebanov:1999rd,Klebanov:2000nc,Klebanov:2000hb}. These setups have enriched our understanding of both QFT and String Theory by providing a geometric understanding of gauge dynamics and dualities while giving tools to tackle brane dynamics through field theory computations \cite{Feng:2000mi,Feng:2001xr,Beasley:2001zp,Feng:2001bn,Klebanov:2000hb,Berenstein:2005xa,Franco:2005zu,Bertolini:2005di,Ibanez:2007tu,Tenreiro:2017fon}. Furthermore, these setups allow for bottom-up constructions in string phenomenology where most features of the gauge theory depend only on the local features of the compactification \cite{Aldazabal:2000sa,Berenstein:2001nk,Verlinde:2005jr}. More recently, warped throats have been used in trying to uplift to de Sitter vacua \cite{Franco:2014hsa,Retolaza:2015sta} and in the search for SUSY breaking vacua in quantum gravity \cite{Franco:2005zu,Berenstein:2005xa,Argurio:2007qk,Buratti:2018onj,Argurio:2019eqb,Argurio:2020dkg,Argurio:2020npm}.
	
The correspondence is particularly sharp when one considers $4d$ $\mathcal{N} = 1 $ gauge theories arising in D3-branes probing singular, non-compact toric CY$_3$ varieties. The problem of finding the gauge theory given the CY$_3$ was solved by brane tilings, also dubbed dimer models \cite{Hanany:2005ve,Franco:2005sm,Franco:2005rj}, where all the informations of the gauge theory are encoded in a bipartite graph on a torus.

Besides propagating strings on singular backgrounds, one can consider a particular gauging of a $\mathbb{Z}_2$ isometry of space together with worldsheet parity, that is, an orientifold \cite{Pradisi:1988xd,Horava:1989vt,Dai:1989ua,Bianchi:1990yu,Bianchi:1990tb}. Its projection on the open string spectrum opens the D-brane/gauge correspondence to new possibilities. For instance, they extend the  available gauge and matter fields, which may break conformal invariance, and allow non-perturbative contributions to the superpotential through instantons \cite{Ibanez:2006da,Argurio:2007vqa,Blumenhagen:2009qh}. While direct construction in string theory is in practice only feasible for some orbifold theories, they may be constructed directly in the dimer model \cite{Franco:2007ii} by identifying gauge groups and fields according to a suitable involution of the graph and possibly assigning some signs to the fixed loci in the dimer, corresponding to the different choices in the orientifold projection. This makes possible for toric singularities to be orientifolded. In the same paper, orientifolds were classified in two groups, depending on the involution, those that leave four fixed points and those that leave a single or two fixed lines, in the dimer. Interestingly, these correspond to three of the five possible smooth involutions on the torus \cite{dugger2019involutions}, the remaining ones corresponding to a shift of the fundamental cell and a glide reflection, i.e. combining a shift and a reflection. 

The main purpose of this paper is to study the two last cases, that leave no fixed loci and assess whether they correspond to sensible orientifolds in string theory. We will argue that only the glide reflection leads to SUSY preserving orientifolds, while the shift is always breaking it. Moreover, we will show how the orientifold projection corresponding to a glide reflection has remarkable properties. Not only, the projected theory always has a conformal fixed point, but also admits, in some cases, a non-trivial RG-flow described by a cascade of Seiberg dualities, analogous to the one of the conifold \cite{Seiberg:1994pq,Klebanov:2000hb}.

The organization of the paper is as follows. In \Cref{Sec:Review} we review the basic tools of dimer models and orientifolds, putting them in the context of torus involutions to find the missing cases. In \Cref{Sec:Examples} we describe glide orientifolds starting from orbifolds and describing their general properties. The absence of fixed loci is tackled in \Cref{Sec:Oplane}, where we understand it to be dual to a pair of opposite sign orientifold planes and use T- and mirror duality to give a global picture. Finally, in \Cref{Sec:Toric} we study the action on the toric geometry through the Zig-Zag paths, allowing the study of fractional branes in the orientifolded theory. A proof of the non-existence of SUSY preserving shift orientifolds is also provided. Some string computations and a cascade analysis are left for the \Cref{Sec:Z4appendix,Sec:Appendix2}.

\section{Torus involutions and Orientifolds}\label{Sec:Review}
In this section we will give a brief review of Dimer models and their orientifolds. In particular, we emphasize the connection between orientifold projections and involutions of the torus on which the dimer model is defined.

\subsection{Review of Dimer models}

A large class of quiver gauge theories can be engineered in String Theory by placing a bunch of D3-branes at the tip of a toric CY$_3$ singularity. The special structure of these geometries allows us to describe the QFT using a combinatorial tool, a bipartite tiling of $\mathbb{T}^2$ called {\it dimer model}. The dictionary between the bipartite graph elements of the brane tiling and the corresponding gauge theory is presented in \Cref{Tab:Dict}. This bipartite graph is physically realized by double T-duality on the D3-brane setup as a 5-brane web (or {\it brane tiling}) on $\mathbb{T}^2$, where D5-branes are suspended between NS5-branes, or rather the NS5-brane wraps a holomorphic cycle in the $\mathbb{T}^2$ wrapped by the D5. We will use both expressions, {\it dimer model} and {\it brane tiling} interchangeably. 

\begin{figure}[h!]
	\centering
	\begin{subfigure}[t]{0.28\textwidth }
		\begin{center} 
			\includegraphics[width=\textwidth]{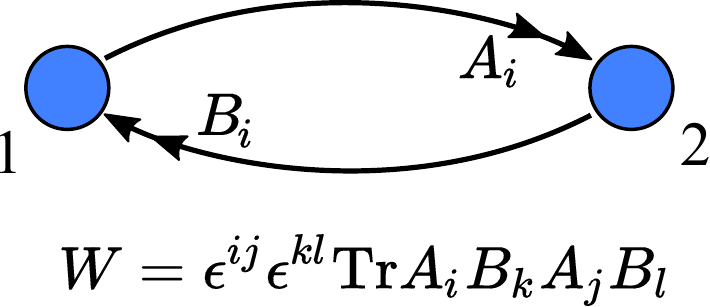}
			\caption{}
			\label{Fig:ConifoldQuiver}
		\end{center}
	\end{subfigure} \hspace{15mm}
	\begin{subfigure}[t]{0.18\textwidth } 
		\begin{center} 
			\includegraphics[width=\textwidth]{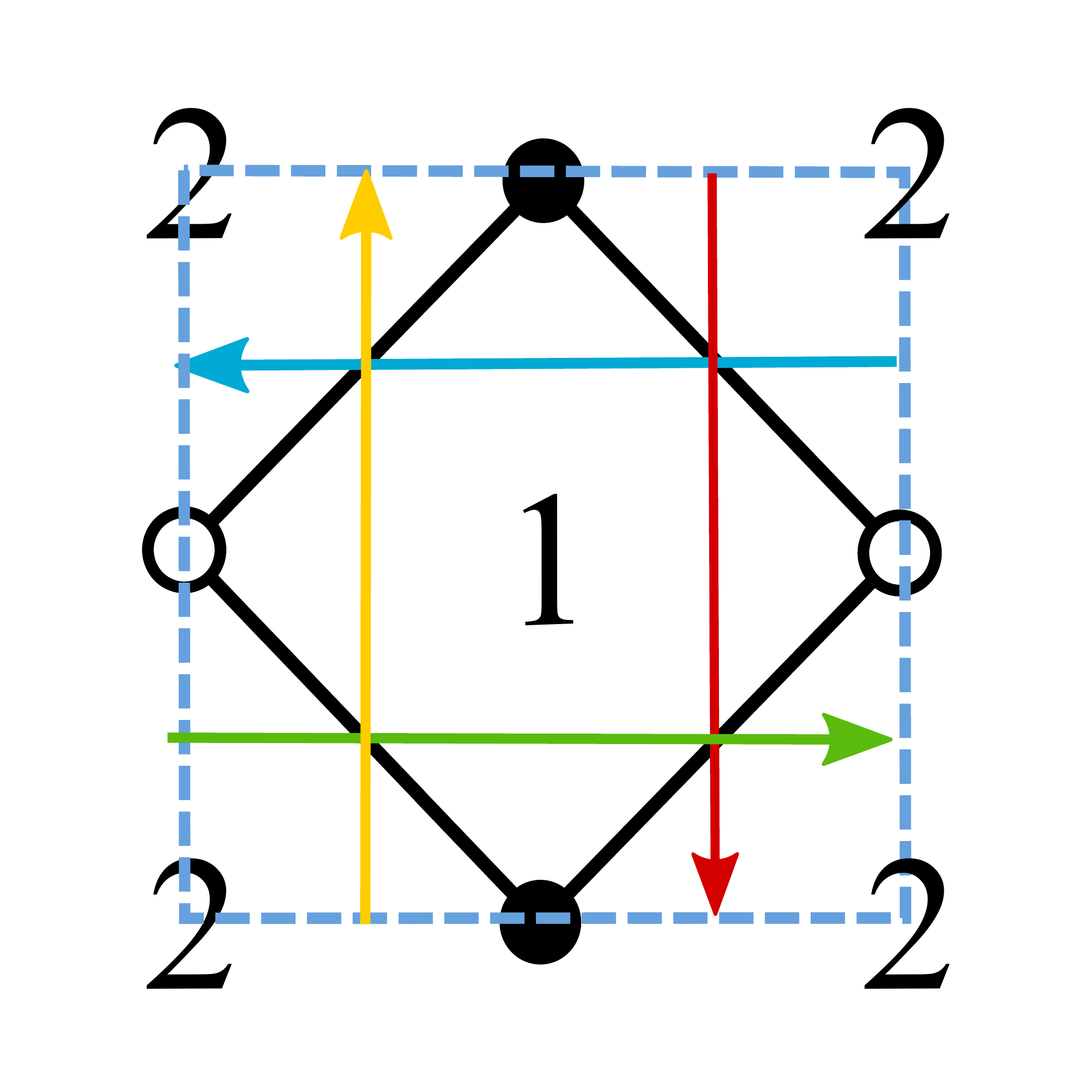}
			\caption{}
			\label{Fig:ConifoldZigZags}
		\end{center}
	\end{subfigure} \hspace{15mm}
	\begin{subfigure}[t]{0.18\textwidth } 
		\begin{center} 
			\includegraphics[width=\textwidth]{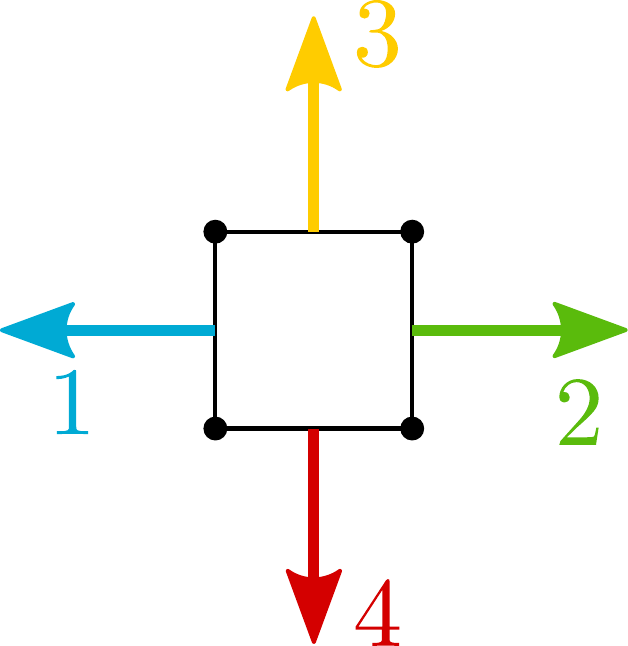}
			\caption{}
			\label{Fig:ToricDiagramConifold}
		\end{center}
	\end{subfigure}
	\caption{Conifold: (a) quiver diagram and superpotential, (b) dimer diagram with ZZPs, and (c) toric diagram with outward normals corresponding to ZZPs. }
	\label{Fig:Conifold} 
\end{figure}

We now present a brief summary of the basic notions of D3-branes at toric singularities and their dimer model description which the versed reader may safely skip. For thorough presentations, see e.g. \cite{Franco:2005rj,Franco:2005sm}. The prototypical example is given by the conifold singularity \cite{Klebanov:1998hh}. This theory has gauge group\footnote{We should stress that in this context gauge groups are in fact a product of several $U(N)$'s. Nevertheless, some $U(1)$ factors can be anomalous and decouple through a generalization of the Green-Schwarz mechanism, while the non-anomalous $U(1)$'s play the role of global symmetries \cite{Ibanez:1998qp}. In this paper, we thus restrict our attention to the non-abelian part of the gauge groups, in particular when computing local gauge anomalies.} $SU(N_1) \times SU(N_2)$, four bifundamental chiral fields $A_i = (\antifund_1,\fund_2)$ and $B_i = (\fund_1, \antifund_2) $, $i=1,2$, and superpotential $W = \epsilon^{ij}\epsilon^{kl} \text{Tr} A_i B_k A_j B_l$. The quiver, dimer and toric diagrams for the conifold are shown in \fref{Fig:Conifold}. 

%===============================================================================
\begin{table}
	\centering
	\begin{tabular}{|p{3.5cm}|p{9cm}|}
		\hline 
		\ \ \ {\bf Brane Tiling} &\ \ \ \ \ \ {\bf Quiver Gauge Theory} \\
		\hline \hline
		Face & $U(N_i)$ gauge factor \\ \hline
		Edge between faces $i$ and $j$ & Chiral superfield in the bifundamental representation of groups $i$ and $j$ (adjoint representation if $i = j$). The chirality, i.e.~orientation, of the bifundamental is such that it goes clockwise around black nodes and counter-clockwise around white nodes. \\ \hline
		$k$-valent node & Superpotential term made of $k$ chiral superfields. Its sign is $+/-$ for a white/black node, respectively. \\ \hline
	\end{tabular}
	\captionof{table}{Dictionary relating brane tilings to quiver gauge theories.\label{Tab:Dict}}
\end{table}	
%===============================================================================

%===============================================================================
%===============================================================================

Another ingredient that we will extensively use are the \textit{Zig-Zag paths} (ZZPs)\footnote{Zig-Zag paths can be computed as the oriented differences of \textit{external perfect matchings}.} \cite{Franco:2005rj,Hanany:2005ss,Feng:2005gw,Franco:2006gc}. The latter are oriented paths on the dimer that capture global properties of the geometry. They are constructed by following edges in the graph and turning maximally left (right) at white (black) nodes. The ZZPs of the conifold dimer are shown in \Cref{Fig:ConifoldZigZags}. These form non self-intersecting closed loops on the torus with non-trivial homology around the two fundamental cycles. These homology numbers can be associated to charges for two of the three $U(1)$ isometries of the toric CY 3-fold, the remaining one being associated to the $U(1)_R$ R-symmetry. They are also in correspondence with legs in the $(p,q)$ web diagram obtained as the dual graph to the toric diagram, their $(p,q)$ labels being exactly the homology charges of the ZZPs.

\subsection{Orientifold projections and Dimers}

String Theory admits sensible propagation on singular backgrounds. Most notable are orbifolds and orientifolds. These arise as suitable projections on the theory propagating in a smooth background. Such involutions are readily described in dimer models \cite{Franco:2007ii}.

Let us focus on orientifold projections in our setup of D3-branes in type IIB, defined by modding out by the action $\Omega R (-1)^{F_L}$, $\Omega$ being worldsheet parity, $R$ a geometric $\mathbb{Z}_2$ isometry of the CY$_3$ and $F_L$ the left-moving fermion number in spacetime. Extended objects are located at the fixed point of the $R$ action, the O-planes. They are non-dynamical objects with a tension and an RR charge as the ones of D-branes. The $\mathbb{Z}_2$ symmetry acts holomorphically on the internal coordinates, and as follows on the K\"ahler form $J$ and the holomorphic 3-form $\Omega_3$:
\begin{equation}
	J \rightarrow J \quad \text{ and }\quad \Omega_3 \rightarrow - \Omega_3 \, ,
\end{equation}
where the $-$ sign is necessary in order for the O-plane to preserve some common supercharges with the D3-branes. The resulting gauge theory is obtained by looking at the projected open string spectrum. The orientifold projection on Chan-Paton factors is essentially free. Denote by $\lambda$ the Chan-Paton matrix, the orientifold acts with a unitary matrix $\gamma_{\Omega}$:
\begin{equation}
	\Omega: \lambda \rightarrow \gamma_{\Omega} \lambda^T \gamma^{-1}_{\Omega} \, .
\end{equation}

\begin{figure}[h!]
	\centering
	\begin{subfigure}[t]{0.4\textwidth }
		\begin{center} 
			\includegraphics[width=\textwidth]{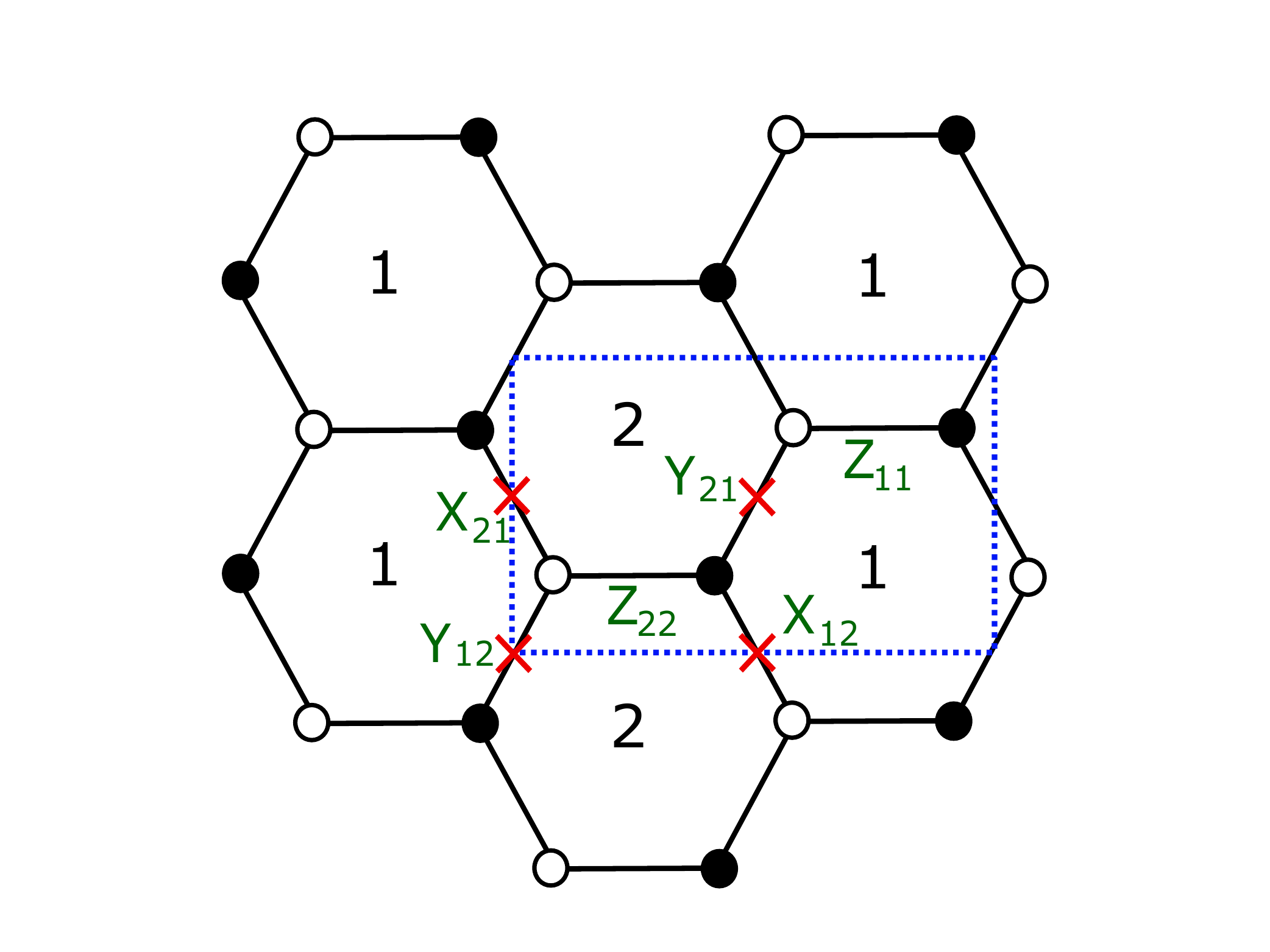}
			\caption{}
			\label{Fig:OrientifoldPoints}
		\end{center}
	\end{subfigure} \hspace{15mm}
	\begin{subfigure}[t]{0.4\textwidth } 
		\begin{center}
			\includegraphics[width=\textwidth]{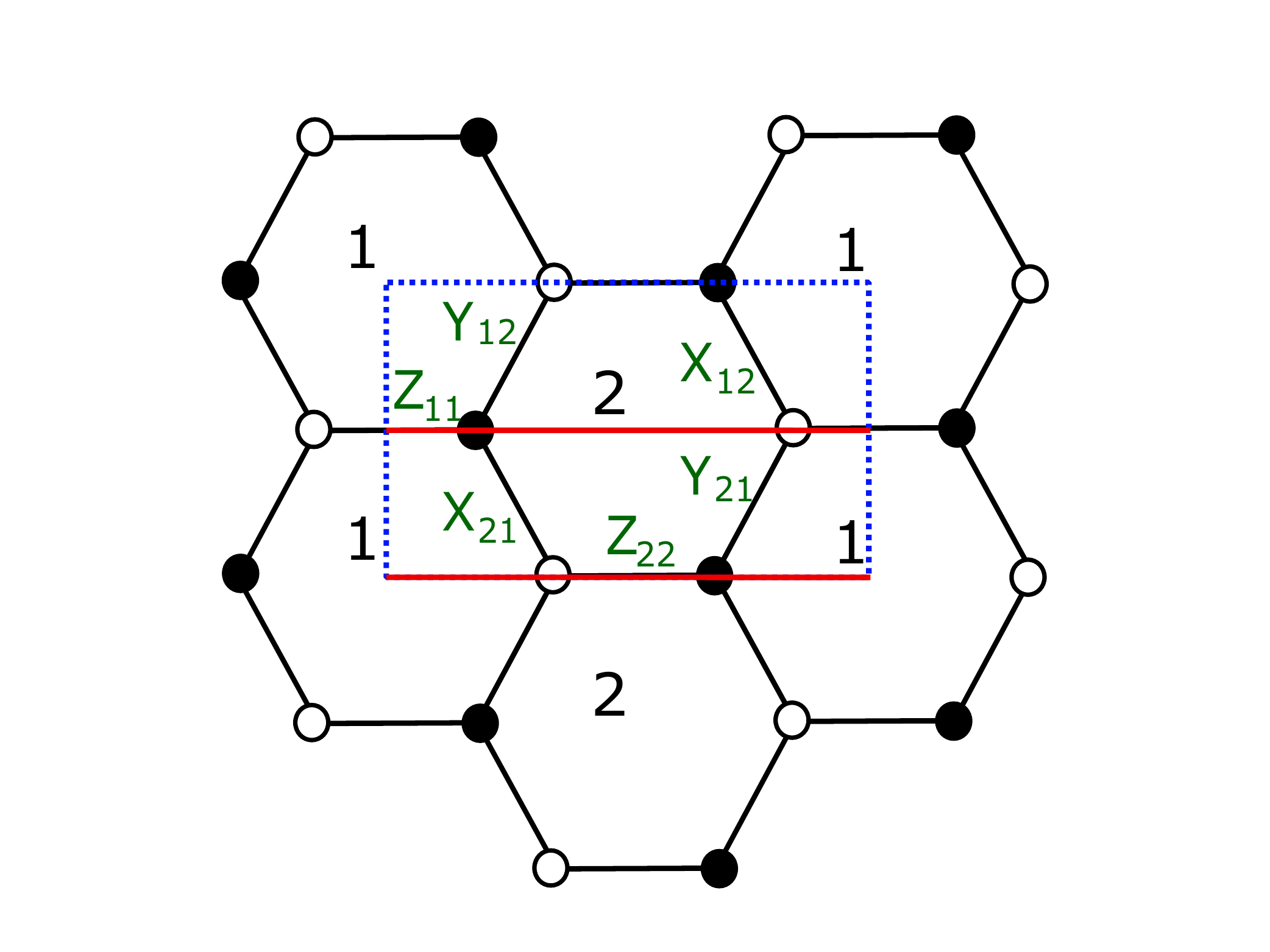}
			\caption{}
			\label{Fig:OrientifoldLine}
		\end{center}
	\end{subfigure} 
	\caption{(a) Orientifold of $\mathbb{C}^2/\mathbb{Z}_2$ with fixed points. (b) Orientifold of $\mathbb{C}^2/\mathbb{Z}_2$ with fixed lines.}\label{Fig:Orientifold} 
\end{figure}

Orientifold projections on D-branes at singularities and their description on dimers were studied in \cite{Franco:2007ii}. In this framework, the orientifold projection corresponds to a $\mathbb{Z}_2$ involution acting on the torus that identifies faces, edges and vertices in an appropriate way. The authors studied involutions with fixed loci (see \figref{Fig:Orientifold} for examples) resulting in a set of rules needed to construct the projected theory that we now summarize below.

%===============================================================================
%===============================================================================

\begin{enumerate}
	
	\item Self-identified faces project to $SO/USp$ groups, depending on the O-plane charge, $+$ or $-$ respectively. All other faces are identified with their image, merging to one $SU$ group.
	
	\item Every edge on top of a fixed locus becomes a symmetric or antisymmetric tensor (or their conjugate), depending on the O-plane charge, $+$ or $-$ respectively. The remaining edges are identified with their images, merging to bifundamental fields. More concretely, bifundamentals are identified as $(\antifund_i,\fund_j) \sim (\antifund_{j'},\fund_{i'})  \rightarrow  (\antifund_i,\fund_j) $, where $i',j'$ are the images of gauge groups $i,j$.
	
	\item The superpotential is found upon projection of the fields.
\end{enumerate}

Before moving on to the next section, we present in detail two examples of orientifold projections.

%===============================================================================
\paragraph{Fixed Points.} \label{Sec:ExampleFP}
%===============================================================================
In an orientifold of this type, there are four fixed points in a unit cell. In order to preserve SUSY, their signs must satisfy the so-called {\it sign rule}: their product must be $(-1)^{n_W/2}$ where $n_W$ is the number of superpotential terms.

In the example of \figref{Fig:OrientifoldPoints}, we chose the signs $(-+-+)$, starting with the fixed point at the origin of the unit cell and going clockwise. We have that face $1$ is identified with face $2$, meaning that the resulting theory will have only one gauge group $SU(N)$. The bifundamental fields are identified as follows
\begin{align}
	Y_{12} \sim Y_{12} \to \antiasymm  \,&\, ,\, X_{21} \sim X_{21} \to \symm \nonumber \\
	Y_{21} \sim Y_{21} \to \asymm\,&\, ,\,  X_{12} \sim X_{12} \to \antisymm\nonumber \\
	Z_{11} \sim Z_{22} & \to\, \text{Adj} \nonumber \, .
\end{align}
and the superpotential is given by
\begin{align}
	W = X_{12} Y_{21} Z_{11} -  X_{21} Y_{12} Z_{11} \, ,
\end{align}
where we implicitly take a trace over gauge indices.

To be sure that this projection preserves some supersymmetry, we need to check the action of the involution on $\Omega_3$. To do so, we compute the mesonic moduli space of our theory, which correspond to the singularity D3-branes are probing. Mesonic operators are given by
\begin{equation}
	\begin{array}{cccccc}
		x &=& X_{12}X_{21} \, ,& y&=& Y_{12}Y_{21} \\
		w_1 &=& Y_{12}X_{21} \, ,& w_2 &=& Y_{21}X_{12}  \\
		z_1 &=& Z_{11} \, ,& z_2 &=& Z_{22}  \, .
	\end{array}
\end{equation}

F-term equations impose $w_1=w_2=w$ and $z_1=z_2=z$, and the classical relation between the fields gives $xy=w_1w_2=w^2$. Thus, the mesonic moduli space is the symmetric product of $N$ copies of the $A_1$ singularity, $xy=w^2$, where $N$ is the number of probe D3-branes. The three form, $\Omega_3$, can be easily computed using the Poincar\'e residue formula:
\begin{align}
	\Omega_3=\text{Res} \frac{\text{d}x \wedge \text{d}y \wedge \text{d}w \wedge \text{d}z}{w^2-xy}=\frac{\text{d}x \wedge \text{d}y \wedge \text{d}z}{2w} \, .
\end{align}

Under the involution, the fields are mapped in the following way
\begin{align}
	x \to x \, ,& \quad y \to y  \, ,\nonumber \\
	w \to -w \, ,& \quad  z \to z \nonumber \, ,
\end{align}
where the sign taken by a meson is given by the product of the fixed point charges it crosses. The orientifold action on the holomorphic 3-form is thus odd, $\Omega_3 \to -\Omega_3$, meaning that the O-plane in compatible with the supersymmetry charges preserved by the D3-branes. It is easy to see that sign configuration not respecting the sign rule are not supersymmetric.

%===============================================================================
\paragraph{Fixed Lines.} 
%===============================================================================

In the example of \figref{Fig:OrientifoldLine}, we have two fixed lines, each one coming with a sign, $+$ or $-$, which is unconstrained. We chose to assign $-$ to the bottom line and $+$ to the other. The faces are self-identified, leading to a gauge group $USp(N_1) \times SO(N_2)$. The identification of fields gives
\begin{align}
	Y_{12} \sim X_{21} \to Q^1_{12} \,& \, ,  \quad X_{12} \sim Y_{21} \to Q^2_{12} \nonumber \\
	Z_{11} \sim Z_{11} \to \asymm \,& \, , \quad Z_{22} \sim Z_{22} \to \symm \, .
\end{align}
and the superpotential is given by
\begin{align}
	W= (Q_{12}^{2}Q_{12}^{2 T} - Q_{12}^{1}Q_{12}^{1 T})Z_{11} + (Q_{12}^{2 T}Q_{12}^{2} - Q_{12}^{1 T}Q_{12}^{1})Z_{22} \, .
\end{align}

The mesons are the same as in the previous example, since the geometry is the same, but the action of the orientifold is different and given by
\begin{align}
	x \leftrightarrow & \, y \, ,  \nonumber \\
	w \to -w \, , \quad &z \to -z \nonumber \, ,
\end{align}
where the fixed line exchanges two mesons and introduces a sign to the self-mapped mesons given by the product of the signs of the two fixed lines crossed. We can again see that the SUSY condition is respected
\begin{align}
	\Omega_3= \frac{\text{d}x \wedge \text{d}y \wedge \text{d}z}{2w} \, \to \frac{\text{d}y \wedge \text{d}x \wedge \text{d}z}{2w} \, = -\Omega_3 \, .
\end{align}
In particular, we see that the signs of the fixed lines play no role in the last relation.

\subsection{Torus involutions}\label{Torinv}

There are five inequivalent non-trivial smooth involutions \cite{dugger2019involutions}, i.e. involutive diffeomorphisms, on a torus\footnote{They are classified by the topology of their orbit set which is always one of the parabolic $2$-orbifolds listed in \cite{Thurston:gt3m}.}. Three of them have a fixed locus and the two others do not. To list all of them we consider a square torus, with complex structure\footnote{We are interested only in smooth involutions, the complex structure doesn't play any role in the analysis, thus we fixed it to a handy value. The use of complex coordinates will be useful for later observations.} $\tau=i$. We take $z$ as the complex coordinate on the torus, the periodicity condition is $z\sim z + m + n i$, with $m,n \in \mathbb{Z}$. The involutions are given by:

\begin{enumerate}
	\item Two fixed lines: $z\rightarrow \bar{z}$. The fixed loci are two parallel lines located at $\text{Im}(z)=0, 1/2$ along the real axis. Under this involution the torus is projected to an annulus. \label{Item1}
	
	\item Single fixed line: $z \rightarrow i \bar{z}$. The fixed line is $\text{Re}(z)=\text{Im}(z)$, corresponding to a diagonal line of the unit cell. The resulting surface is a Moebius strip.\label{Item2}
	
	\item Fixed points: $z\rightarrow -z$. In this case we have four fixed points, $z=0, 1/2, i/2$ and $(1+ i)/2$. The resulting topology is that of a sphere. \label{Item3}
	
	\item Glide reflection: $z \rightarrow \bar{z} + 1/2 $. There are no fixed loci. The resulting topology is that of a Klein bottle. \label{Item5}
	
	\item Shift: $z  \rightarrow z+ 1/2$. Again, the involution has no fixed loci. The torus is projected to another torus. \label{Item4}
	
\end{enumerate}

As already mentioned, \ref{Item1}, \ref{Item2} and \ref{Item3} are involutions with fixed loci correspond to orientifold operations already studied in the literature. In this paper, we will focus on \ref{Item5}, the glide reflection, studying the consistency of such projection and its properties. Regarding involution \ref{Item4}, we will show that the shift is not compatible with the required properties to preserve supersymmetry.

Let us conclude this section with few comments. First, involutions with fixed loci teach us that if the involution is holomorphic, $z \to f(z)$, nodes in the dimer are mapped to nodes of opposite color, while if it is antiholomorphic, $z \to f(\bar{z})$, nodes are mapped to nodes of the same color. This is a requirement from the orientifold mapping of chiral superfields. It gives us a hint for the unexplored involutions. Indeed, we expect \ref{Item4} to be consistent with an orientifold identification only if nodes are mapped to nodes of the opposite color, while \ref{Item5} would be consistent only if the mapping is between vertices of the same color. Second, we stress that the involution should be not only a symmetry for the torus, but also for the embedded dimer model. In particular, a generic fundamental cell for a dimer model has the shape of a parallelogram. The symmetry may be present in the abstract graph, but in order to be shown explicitly, consider the case of say \ref{Item2}, one has to deform the embedding in such a way that the resulting fundamental cell is now a rhombus, displaying a symmetry with respect to one of the diagonals. From this observation we conclude that in order to display a glide symmetry, the fundamental cell must be a rectangle. Third, a $\mathbb{Z}_2$ glide reflection with diagonal axis is described by the map $z \rightarrow i \bar z + (1+i)/2$ which has $\Re(z)=\Im(z)+1/2$ as fixed line, hence they are nothing else than reflections about a diagonal axis. In particular, they do not correspond to a class of smooth involutions not listed above.

Even if we can deform the embedding to make the involution explicit, it is possible that the model can be endowed with extra structures, capturing some physical properties. For example, isoradial embeddings described in \cite{Hanany:2005ss} encode the R-charges of the fields. In this paper, though, we are not interested in these particular cases.

\section{Glide Orientifolds}\label{Sec:Examples}

In this section we investigate glide reflection orientifolds. We start with orbifold examples, motivating our results in the dimer from the open string projection on the Chan-Paton indices. We also explicitly check that it preserves supersymmetry, in particular, it acts on the CY 3-form as $\Omega_3 \rightarrow -\Omega_3$. We extend our results to orbifolds of the conifold, considering the cascade in the presence of deformation fractional branes. Finally, we discuss anomalies, or rather their absence, and conformality in the presence of these orientifolds. 

\subsection{Orbifold $\mathbb{C}^2/ \mathbb{Z}_2$}\label{Sec:C2Z2}

We consider the recipe directly applied in the dimer and then check that it is indeed predicted by open-string computation.	

\paragraph{Projection on the dimer model.}
\begin{figure}[h!]
	\centering
	\begin{subfigure}[t]{0.4\textwidth }
		\begin{center} 
			\includegraphics[width=\textwidth]{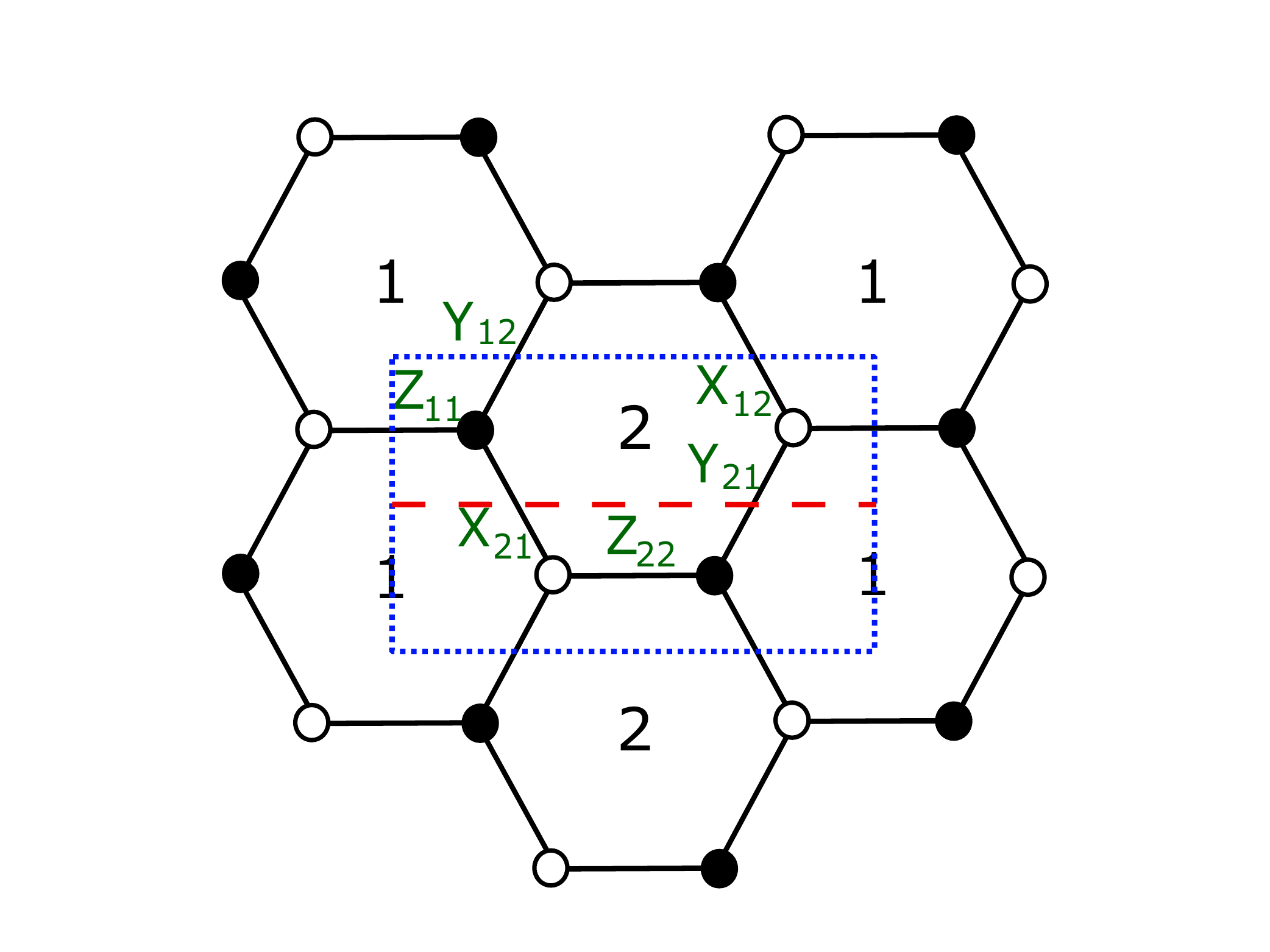}
			\caption{}
			\label{Fig:Z2linereflect}
		\end{center}
	\end{subfigure} \hspace{15mm}
	\begin{subfigure}[t]{0.4\textwidth }
		\begin{center} 
			\includegraphics[width=\textwidth]{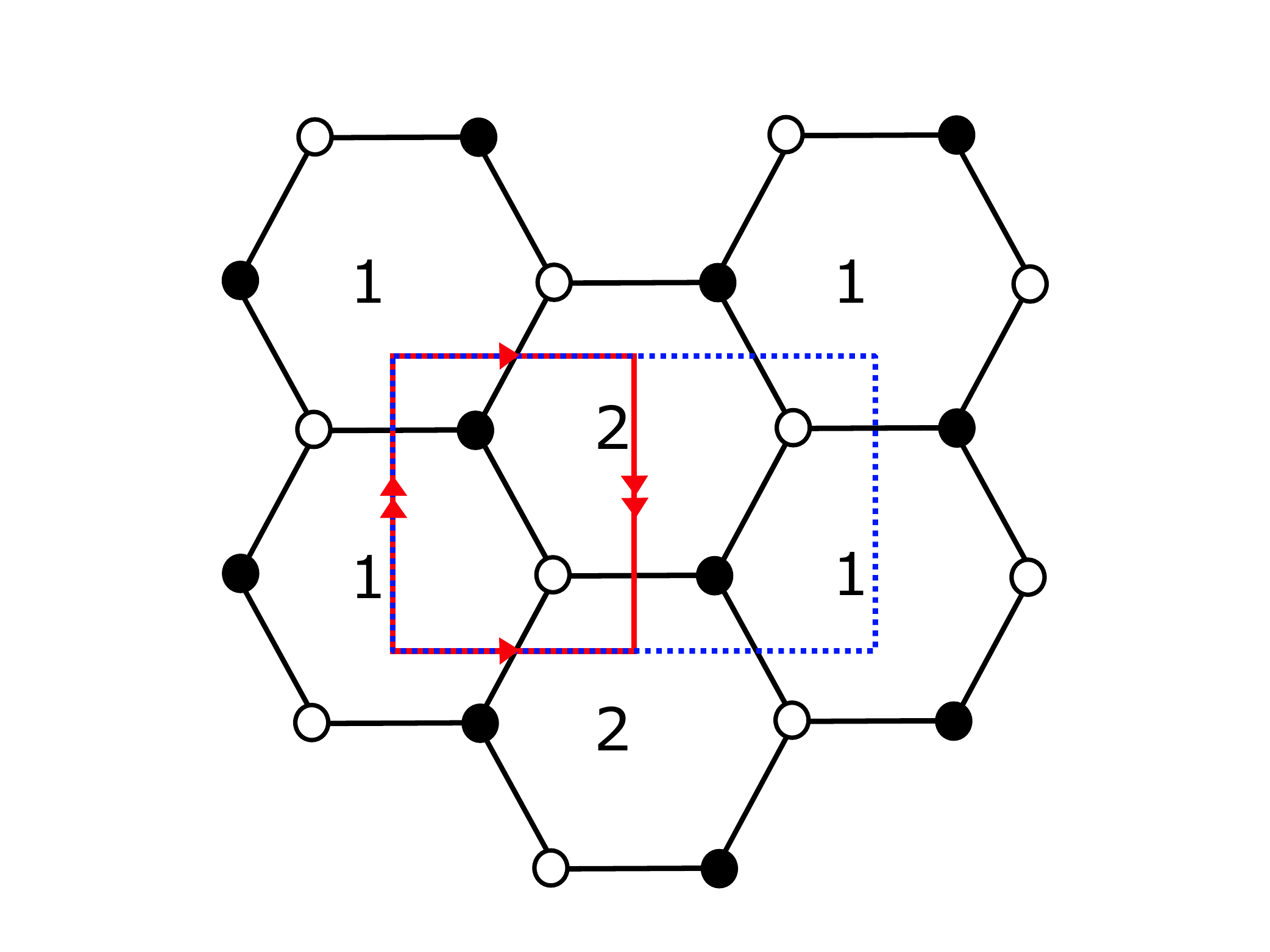}
			\caption{}
			\label{Fig:Z2KleinBottle}
		\end{center}
	\end{subfigure}
	\caption{(a) Dimer diagram for the orbifold $\mathbb{C}^2/\mathbb{Z}_2 \times \mathbb{C}$. The unit cell and the reflection axis are depicted in blue and red respectively. (b) The Klein bottle we obtain with the orientifold projection.}\label{Fig:Z2} 
\end{figure}
We present in \figref{Fig:Z2} the dimer for the orbifold $\mathbb{C}^2/ \mathbb{Z}_2$ where the glide reflection is a combined operation of a horizontal shift by one half of the length of the unit cell followed by a reflection with respect to the dashed red horizontal axis. Nodes are mapped to nodes of the same color, as we want from the analysis in \Cref{Torinv}. Note that this operation leaves no fixed loci in the unit cell. The projected theory is embedded in a Klein Bottle drawn on one half of the original unit cell, as illustrated in \figref{Fig:Z2KleinBottle}.

The edge $X_{12}$ is identified with $Y_{12}$, $X_{21}$ with $Y_{21}$ and $Z_{11}$ with $Z_{22}$. Following the rules summarized in \Cref{Sec:Review}, the resulting theory has gauge group $SU(N)_1$ with matter content given by two tensors\footnote{The two tensors are of the form $(\antifund_1,\antifund_1)$ and $(\fund_1,\fund_1)$.} and one adjoint field. Note that the tensor fields are not in an irreducible representation, so we split them in their symmetric and antisymmetric parts;
\begin{align}
	\mathcal{X}_{S,A} &=  \antisymm_1 , \antiasymm_1 \, \nonumber, \\
	\mathcal{Y}_{S,A} &=  \symm_1 , \asymm_1 \, , \\
	\mathcal{Z} &=  \mathrm{Adj}_1 \, .\nonumber
\end{align}
The superpotential is obtained by explicitly projecting the original one and keeping half of the terms,
\begin{align}
	W = \mathcal{X} \mathcal{Y} \mathcal{Z}^T -  \mathcal{Y}\mathcal{X} \mathcal{Z} =  \mathcal{X}_A \mathcal{Y}_S \mathcal{Z} - \mathcal{X}_S \mathcal{Y}_A \mathcal{Z} \, . \label{Eq:Z2sup2}
\end{align}

In a SUSY-preserving orientifold in type IIB, the holomorphic 3-form must map to minus itself. This is easy to check by noting that the orientifold action on the mesons is
\begin{align}
	x &\leftrightarrow y  \hspace{1cm} w \to w \hspace{1cm} z \to z  \, .
\end{align}
The action on the 3-form is then
\begin{align}
	\Omega_3=\frac{\text{d}x \wedge \text{d}y \wedge \text{d}z}{2w} \to \frac{\text{d}y \wedge \text{d}x \wedge \text{d}z}{2w} =-\Omega_3 \, .
\end{align}
It is also clear from the matter content and the first equality of \Cref{Eq:Z2sup2} that the gauge theory preserves $\mathcal{N}=2$ supersymmetry\footnote{The attentive reader might have noticed that this orientifolded theory is identical to the one obtained with fixed points in \Cref{Sec:ExampleFP}, although the involution acts differently on the coordinates. This is however an artifact of the orbifold $\mathbb{C}^2 / \mathbb{Z}_2$ since glide reflections will not provide tensors in general.}.

It is worth noting that the theory, unlike many examples of projections with fixed loci, is free from any local gauge anomaly, regardless of the gauge group rank. Although this example is rather trivial, we will see that this feature is general and related to tensor fields being absent or coming in pairs, symmetric and antisymmetric, cancelling each other's contribution to the anomaly cancellation conditions (ACC). We also note that the projected theory is actually conformal. Indeed, the $\beta$-function of the gauge group can be shown to be zero. The fact that these orientifolds naturally lead to SCFT's will be discussed in \Cref{Sec:KBproperties}.

\paragraph{Open string projection.}We now consider the orientifold projection on the Chan-Paton indices of the open string spectrum. For D-branes localized on the $\mathbb{C}^2/\mathbb{Z}_2\times \mathbb{C}$ singularity the open string spectrum is obtained by promoting the flat space one to $2N\times 2N$ matrices with a restricted set of non-zero entries:
\begin{equation}
	A_\mu = \left(\begin{array}{cc}
		A_{1\mu} & 0 \\
		0 & A_{2\mu}
	\end{array}\right) \, , \quad
	\Phi_1 = \left(\begin{array}{cc}
		0 & X_{12} \\
		X_{21} & 0
	\end{array}\right) \, , \quad \Phi_2 = \left(\begin{array}{cc}
		0 & Y_{12} \\
		Y_{21} & 0
	\end{array}\right)\, , \quad \Phi_3 = \left(\begin{array}{cc}
		Z_{11} & 0 \\
		0 & Z_{22}
	\end{array}\right) \, , 
\end{equation}
where the gauge group is $SU(N)_1 \times SU(N)_2$ and matther fields transform in the following representations, 
\begin{equation}
	X_{ij}, Y_{ij} = (\fund_i,\antifund_j), \quad Z_{ii} = \text{Adj}_i \, .
\end{equation}
Decomposing the $\mathbb{C}^3$ fields the orbifold superpotential becomes,
\begin{eqnarray}
	W &=& \left[\Phi_1 , \Phi_2\right] \Phi_3 \nonumber \\ 
	&=& X_{12} Y_{21} Z_{11} -  Y_{21} X_{12} Z_{22} + X_{21}  Y_{12} Z_{22} -  Y_{12} X_{21} Z_{11} \, ,
\end{eqnarray}
where an overall trace over gauge indices is understood.

A general orientifold projection on the $\mathbb{C}^3$ fields acts as,
\begin{eqnarray}
	A_\mu &=& - \gamma_\Omega A^T_\mu \gamma_\Omega^{-1} \, ,  \label{Eq:Z2orientVector}\\
	\Phi_i &=& R_{ij} \gamma_\Omega \Phi_j^T \gamma_\Omega^{-1} \, , \label{Eq:Z2orientMatter} 
\end{eqnarray}
where $\gamma_\Omega$ is a $2N\times 2N$ matrix acting on gauge group (Chan-Paton) indices and $R_{ij}$ acts on space indices $i, j$ running from 1 to 3. Different choices for these matrices lead to different orientifold projections. In order to reproduce the glide reflection orientifold, we specifically choose
\begin{equation}
	\gamma_\Omega = \left(\begin{array}{cc}
		0 & \mathbb{1}_N \\
		\mathbb{1}_N & 0
	\end{array}\right) \, , \quad \text{and} \quad
	R = \left(\begin{array}{ccc}
		0 \, & \, 1 \, & \, 0\\
		1 & 0 & 0 \\
		0 & 0 & 1
	\end{array}\right) \, ,
\end{equation}
so that $\Phi_1$ and $\Phi_2$ coordinates are exchanged by the orientifold. Equation \eqref{Eq:Z2orientVector} translates into
\begin{equation}
	A_{1\mu} = - A_{2\mu}^T \, ,
\end{equation}
which tells us that the two gauge groups are now identified as one $SU(N)_1$ in the orientifolded theory. Equation \eqref{Eq:Z2orientMatter} maps the superfields in the following way:
\begin{equation}
	\begin{array}{ccccc}
		X_{12} &=& Y_{12}^T  &\equiv&  \mathcal{X}_{A,S} \, , \\
		Y_{21} &=& X_{21}^T  &\equiv&  \mathcal{Y}_{A,S} \, , \\
		Z_{11} &=& Z_{22}^T  &\equiv&  \mathcal{Z}  \, .
	\end{array}
\end{equation}
We recognise the same field content of the theory obtained with the dimer technique. It is easy then to show that we recover the superpotential advertised in \Cref{Eq:Z2sup2} (up to an irrelevant numerical factor). We thus conclude that the glide reflection on the dimer reproduces the orientifold projection we just computed in string theory.

In the following, we discuss the dimer construction in more involved examples. It is clear that not all dimer models have the required symmetry, and in \Cref{Sec:Toric} we provide a necessary condition for a given toric CY$_3$ to admit a glide reflection directly from its toric diagram. %In the appropriate cases, we will assume that the worldsheet construction showed in the present section can always be generalized for the examined singularity and stick only to a analysis of its dimer model or toric data.

\subsection{More orbifold examples}\label{Sec:HigherOrbifolds}

\begin{figure}[h!]
	\centering
	\includegraphics[width=0.5\textwidth]{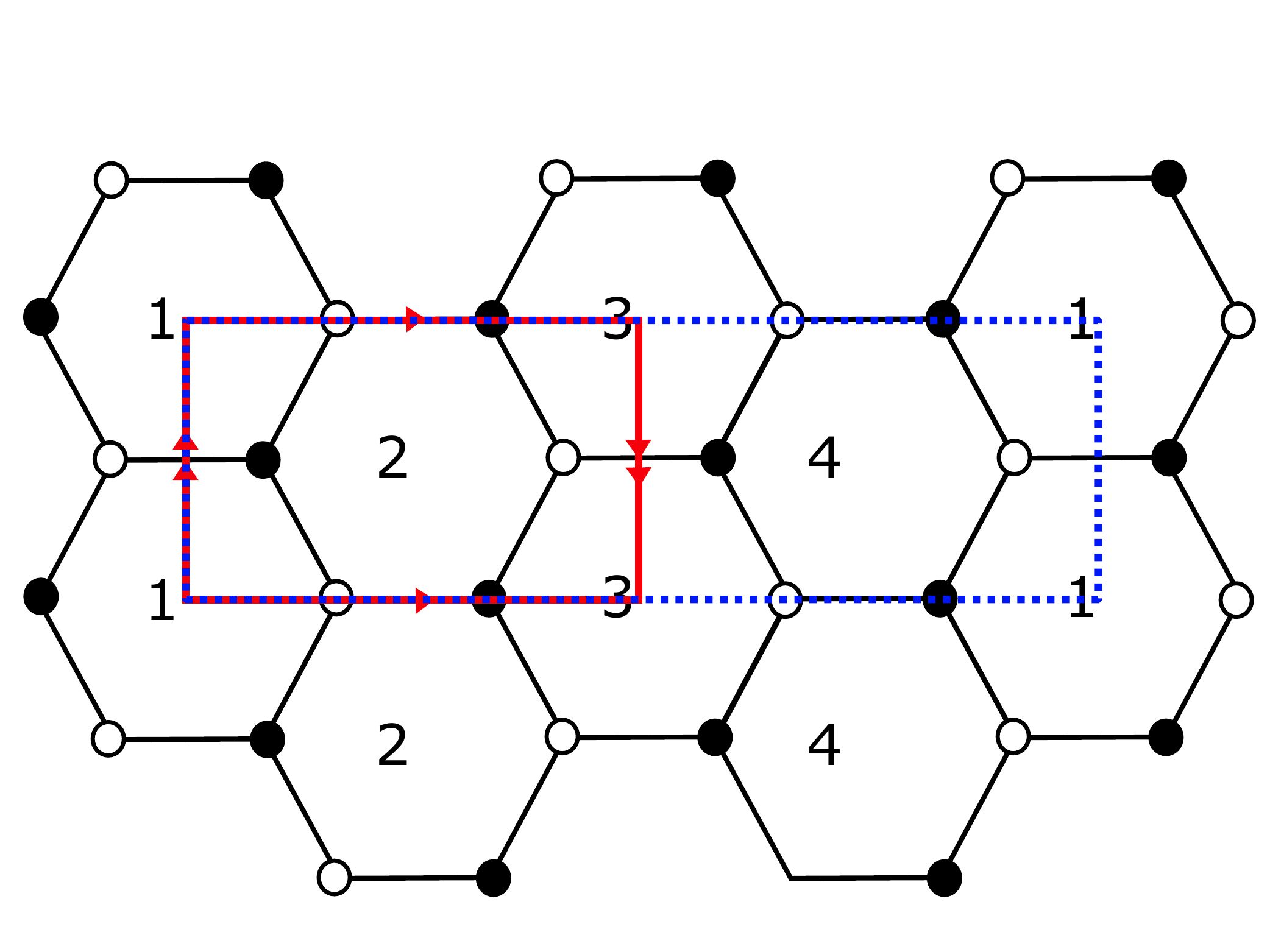}
	\caption{Dimer diagram for the orbifold $\mathbb{C}^2/ \mathbb{Z}_4$. The unit cell is depicted in blue and we show in red the Klein bottle obtained from the orientifold projection.}\label{Fig:Z4}
\end{figure}

The previous example has so much symmetry that it could be misleading. Let us start our journey to less symmetric theories by considering $\mathbb{C}^2 / \mathbb{Z}_4$, whose dimer model and relevant involution we present in \figref{Fig:Z4}.
From the four initial gauge groups, only two of them are kept after the projection, $SU(N_1)_1\times SU(N_2)_2$. The surviving fields are
\begin{equation}
	\begin{array}{ccc}
		\mathcal{X}_{12} =  (\antifund_1 , \fund_2) \, , \quad & \mathcal{X}_{21} = (\antifund_2 , \antifund_1) \, , \quad &\mathcal{Y}_{21} =  (\antifund_2 , \fund_1) \, , \\
		\mathcal{Y}_{12} = (\fund_1 , \fund_2) \, , \quad &\mathcal{Z}_{11} = \mathrm{Adj}_1 \, , \quad  &\mathcal{Z}_{22} = \mathrm{Adj}_2 \, .
	\end{array} \label{Eq:Z4Matter}
\end{equation}
and the resulting superpotential is found to be
\begin{equation}
	W =  \mathcal{X}_{12}\mathcal{Y}_{21}\mathcal{Z}_{11} -  \mathcal{Y}_{21}\mathcal{X}_{12}\mathcal{Z}_{22} + \mathcal{X}_{21}\mathcal{Y}_{12}\mathcal{Z}_{22} - \mathcal{Y}_{12}\mathcal{X}_{21}\mathcal{Z}_{11}^T \, . \label{Eq:Z4supo}
\end{equation}
The open string projection computation for this example can be found in \Cref{Sec:Z4appendix}. Note that despite its similarities with the orbifold $\mathbb{C}^2 / \mathbb{Z}_2$ (without orientifold), this model has a different matter content, which cannot be obtained from dimer models.

The mapping of the mesons is the same as for $\mathbb{C}^2 / \mathbb{Z}_2$ so that the holomorphic 3-form transforms as follows:
\begin{align}
	\Omega_3=\frac{\text{d}x \wedge \text{d}y \wedge \text{d}z}{4w^3} \to \frac{\text{d}y \wedge \text{d}x \wedge \text{d}z}{4w^3} =-\Omega_3 \, ,
\end{align}
and hence suggests that our projection is indeed supersymmetric and the resulting gauge theory preserves $\mathcal{N}=2$ supersymmetry. Note that the usual orientifold techniques in the dimer, fixed points and line(s), are not able to reproduce it.

Our observations make it clear that any orbifold $\mathbb{C}^2 / \mathbb{Z}_{2n} \times \mathbb{C}$ will admit a glide reflection, for any integer $n$. More general orbifolds, such as $\mathbb{C}^3 / \mathbb{Z}_n$ or $\mathbb{C}^3 / \mathbb{Z}_p \times \mathbb{Z}_q$, can also enjoy the glide reflections, see an example in \Cref{Fig:Z12KleinBottle}. In \Cref{Sec:Toric} we will discuss the general geometric condition a singularity should meet in order to admit such orientifold. 

\paragraph{$\mathcal{N}=2$ fractional branes.} Let us briefly comment on the fractional branes of the orientifolded theory \cite{Franco:2005zu}. The glide orientifold of $\mathbb{C}^2/ \mathbb{Z}_4$ is free of local gauge anomalies for any rank $N_1$ and $N_2$. Hence, it has a fractional brane. We find that it is an $\mathcal{N}=2$ fractional brane corresponding to a subset of the $\mathcal{N}=2$ fractional branes of the parent theory. In \Cref{Sec:FracZZP} we will discuss this fact in detail.

\subsection{Conifold-like singularities}
As we will explain in \Cref{Sec:Toric}, the conifold $\mathcal{C}$ itself does not admit a glide reflection, but conifold-like singularities like its orbifold $\mathcal{C}/ \mathbb{Z}_2$ or the zeroth Hirzebruch surface $F_0$ do. We now study those examples in turn.

\begin{figure}[h!]
	\centering
	\includegraphics[width=0.4\textwidth]{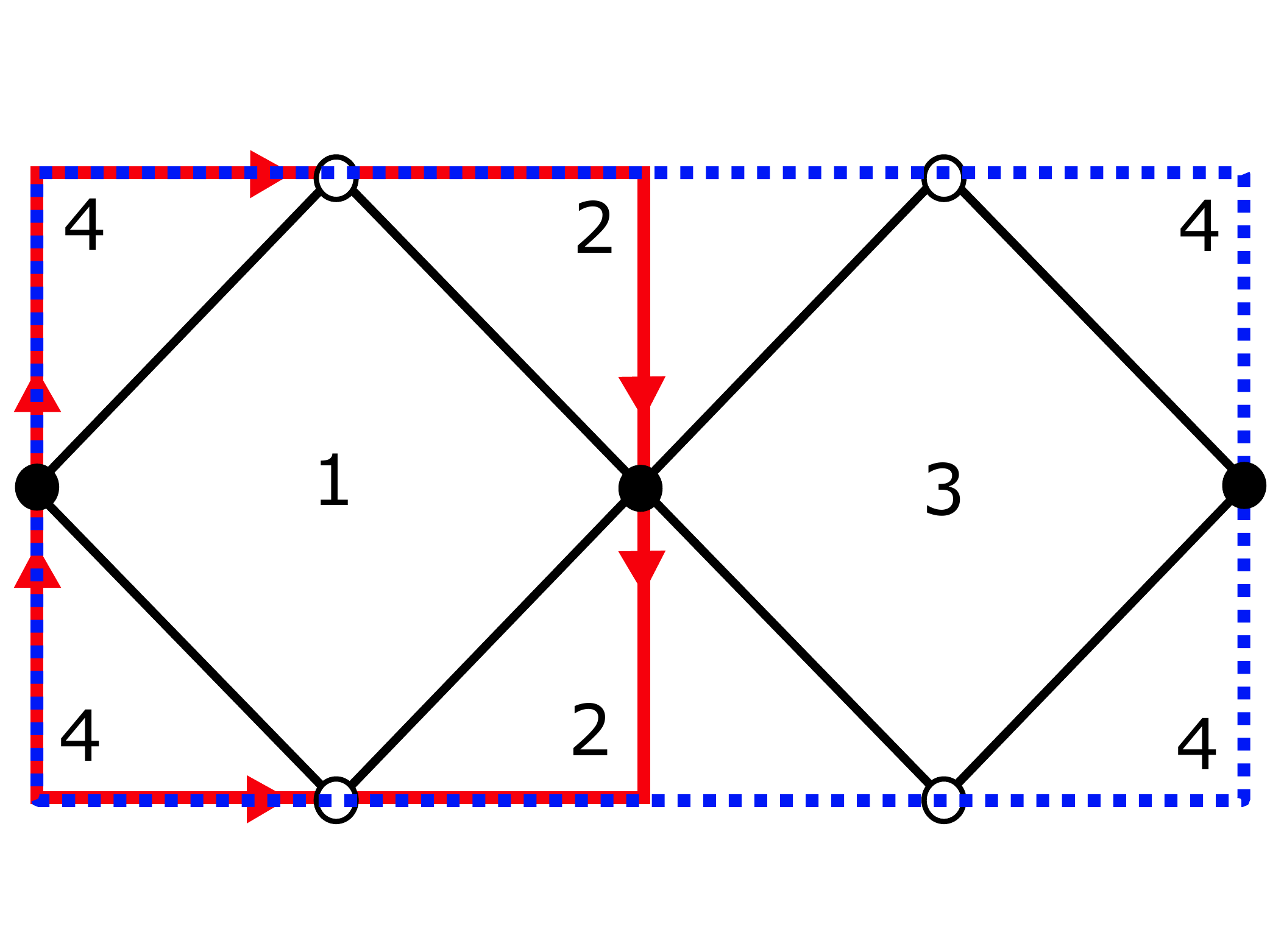}
	\caption{Dimer diagram for the orbifold of the conifold $\mathcal{C}/ \mathbb{Z}_2$. The unit cell is depicted in blue and we show in red the Klein bottle obtained from the orientifold projection.}\label{Fig:ConifoldZ2}
\end{figure}

\paragraph{Non-chiral orbifold of the conifold $\mathcal{C}/ \mathbb{Z}_2$.}The dimer model and the glide orientifold of $\mathcal{C}/ \mathbb{Z}_2$ are shown in \figref{Fig:ConifoldZ2}.
The resulting gauge theory has gauge group $SU(N_1) \times SU(N_2)$ with  matter content given by
\begin{equation}
	\begin{array}{ccccccc}
		A & = & (\antifund_1 , \antifund_2) \, ,  &\quad  & B & = & (\fund_1 , \fund_2)\, , \\
		C  & = & (\antifund_1 , \fund_2)\, ,&  \quad &	D  & = & (\fund_1 , \antifund_2)\, , 
	\end{array}
\end{equation}
Note in passing that the ACC do not impose any constraint on the ranks, so that $N_1$ and $N_2$ may be chosen independently. The superpotential reads
\begin{align}
	W= ABCD - BA C^T D^T \, . 
\end{align}
For details of computations using worldsheet techniques and a proof that the 3-form is odd under the orientifold action, see \Cref{Sec:ConifoldZ2Appendix}.

\begin{figure}
	\centering
	\includegraphics[width=0.4\textwidth]{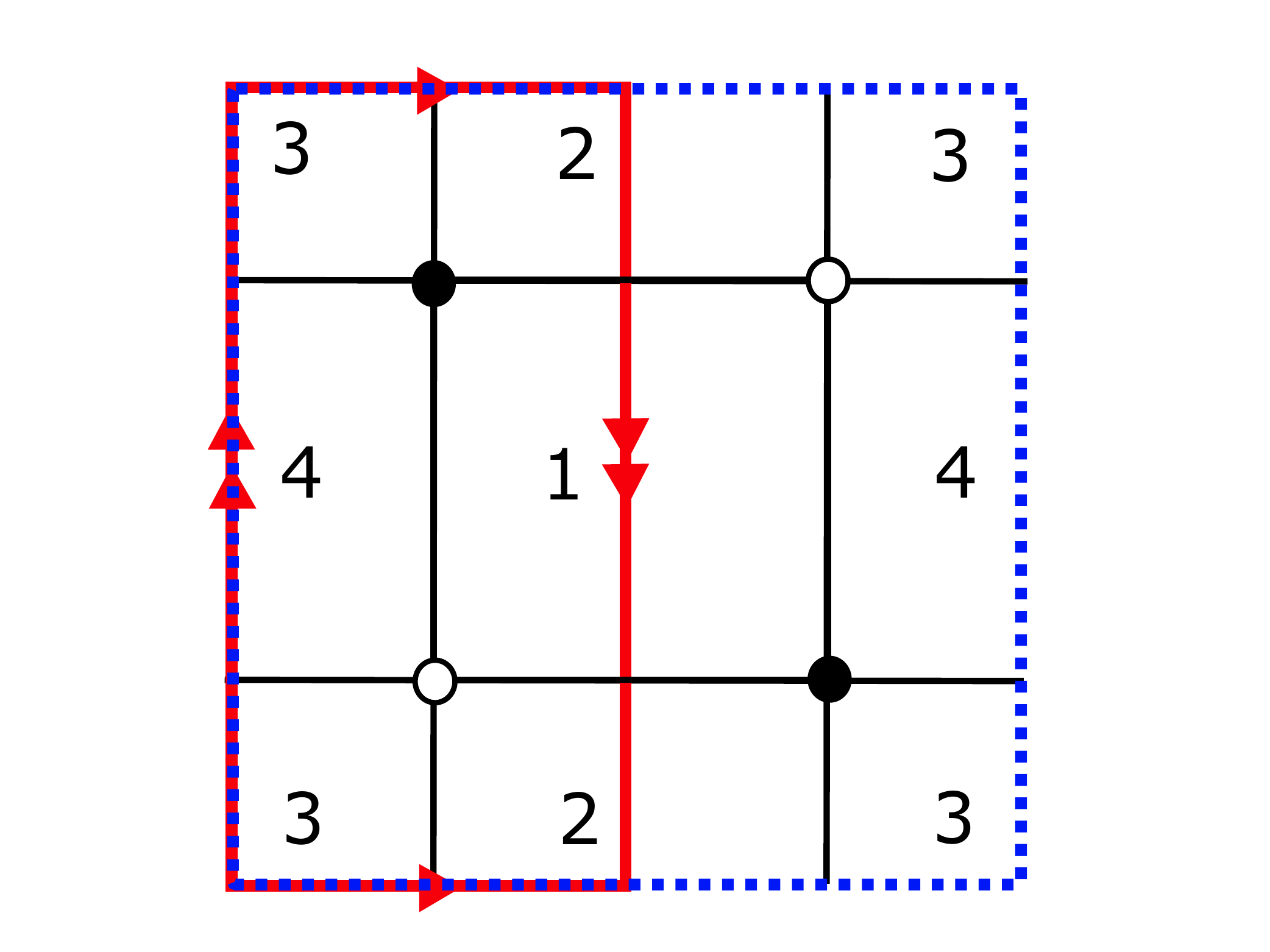}
	\caption{Dimer diagram for the Hirzebruch surface $F_0$. The unit cell is depicted in blue and we show in red the Klein bottle obtained from the orientifold projection.}\label{Fig:F0}
\end{figure}

\paragraph{Zeroth Hirzebruch surface $F_0$.}We show the dimer model and the glide orientifold of $F_0$ in \figref{Fig:F0}.
After projection the gauge group becomes $SU(N_1) \times SU(N_2)$, while the matter content is given by
\begin{equation}
	\begin{array}{ccccccc}
		X & = & (\fund_1 , \antifund_2) \, ,  &\quad  & Y & = & (\fund_1 , \antifund_2)\, , \\
		U_{S,A}  & = & \antisymm_1 , \antiasymm_1\, ,&  \quad & Z_{S,A} & = & \symm_2 , \asymm_2\, , 
	\end{array}
\end{equation}
In this case the ACC impose non-trivial constraints on the gauge group ranks, in particular they must be the same, $N_1=N_2$. The superpotential reads
\begin{align}
	W= X U_S Y^T Z_A - X^TZ_S Y U_A \, . 
\end{align}
The Chan-Paton computation and a proof that the 3-form is odd under the orientifold action are found in \Cref{Sec:F0}.

\paragraph{Deformation fractional branes.} We have seen that the $\mathcal{C}/ \mathbb{Z}_2$ glide reflection admits fractional brane since the ranks of the two gauge groups may be chosen freely. It is in fact a deformation brane \cite{Franco:2005zu,Butti:2006hc} of the parent theory that survives the orientifold projection, in the precise sense described in \cite{Argurio:2020dko}. A natural question is whether such fractional branes may trigger a non-trivial RG-flow giving rise to a cascade of Seiberg dualities \cite{Seiberg:1994pq,Klebanov:2000hb}. We study this process in \Cref{Sec:Cascade} and verify that the cascade steps are: $SU(N+M)_1 \times SU(N)_2 \rightarrow SU(N-M)_1 \times SU(N)_2$, with the same matter content and superpotential, as we flow towards the IR. For $N$ being a multiple of $M$, the deep IR of this gauge theory is expected to reproduce the same features as for a deformed conifold. Notably, on the baryonic branch one finds the vacuum of SYM, displaying confinement and chiral symmetry breaking.

We will see later that it is a fact that the orbifolds of the conifold $\mathcal{C}/ \mathbb{Z}_m \times \mathbb{Z}_n$ compatible with the glide projection preserve some of their deformation branes. The compatibility of fractional branes of the parent theory with the glide reflection is discussed in \Cref{Sec:Toric}.

\subsection{General properties}\label{Sec:KBproperties}

As we have seen, and since the glide reflection leaves no fixed loci, we don't expect any self-identified face (i.e. $SO$ or $USp$ gauge group) to show up in the dimer projection. This restricts the number of gauge groups of the parent theory to be even. A further consequence of not having fixed loci is that there are no self-identified bifundamentals, therefore, tensor matter, if present, always comes in antisymmetric-symmetric pairs, cancelling the contributions to the chiral anomaly. This is precisely what happens in the $\mathbb{C}^2/\mathbb{Z}_2 $ orbifold, where two edges, charged under two identified groups, are identified, leading to a reducible two index tensor, which splits into the sum of a symmetric and an antisymmetric one. We now see how these facts translates in the absence of non-homogenous terms in the anomaly cancellation conditions, allowing always a solution to the latter, and how such projected theories are actually SCFTs.  

Borrowing the notation of \cite{Argurio:2020dko}, we know that the ACC matrix of the projected theory is deduced from that of the parent theory. Denote the latter as,
\begin{align}
A = \left(\phantom{\begin{matrix} \\ B_{11} \\ \\ \hline \\ \\ B_{11} \\ \hline \\ \\ B_{11} \end{matrix}}
\right.\hspace{-1.5em}
\begin{array}{ccc|ccc|ccc}
& & & & & & & & \\
& B_{11} & & & B_{12} & & & B_{13} \\
& & & & & & & & \\
\hline
& & & & & & & & \\
& B_{21} & & & B_{22} & & & B_{23} \\
& & & & & & & & \\
\hline
& & & & & & & & \\
& B_{31} & & & B_{32} & & & B_{33} \\
\undermat{j}{& \text{\quad \quad} &} & \undermat{\large j+k}{& \text{\quad \quad} &} & \undermat{\large b}{& \text{\quad \quad} &} \\
\end{array}
\hspace{-1.5em}
\left.\phantom{\begin{matrix} \\ B_{11} \\ \\ \hline \\ \\ B_{11} \\ \hline \\ \\ B_{11} \end{matrix}}\right)\hspace{-1em}
\begin{tabular}{l}
$\left.\lefteqn{\phantom{\begin{matrix} \\ B_{11} \\ \hline \end{matrix}}}\right\}i$\\ \\
$\left.\lefteqn{\phantom{\begin{matrix} \\ B_{11} \\ \hline \end{matrix}}} \right\}i+k$\\ \\
$\left.\lefteqn{\phantom{\begin{matrix} \\ B_{11} \\ \hline \end{matrix}}} \right\}a$
\end{tabular} \, ,
\end{align}

\vspace{.85 cm}
where indices $i,j=1,\dots,k$ label the gauge groups surviving the orientifold projection and the corresponding entries represent the anomaly contribution of the field between faces $i$ and $j$. Indices $i+k$ and $j+k$ represent gauge groups that are identified with $i$ and $j$ under the orientifold action, respectively. The $a,b$ indices label the self identified gauge groups. Finally, the ACC system takes the form
\begin{align}
A\cdot N = 0 \, ,
\end{align}
where $N$ is a vector whose entries, $N_{(j | j+k | a)}$ are the ranks of the corresponding gauge group.

From what we said earlier, we know that $B_{\star 3}=B_{3 \star}=B_{33}=0$, since there are no self-identified gauge groups. Furthermore, we have no net contributions from tensors to the ACC, meaning that there are no non-homogenous terms in the projected theory ACC. From \cite{Argurio:2020dko}, we know that the projected ACC can be written as
\begin{align}
	\overline{A} \cdot N = \left(
	\begin{array}{ c    }
		B_{11} + B_{12} 
	\end{array}
	\right)\, \cdot N = 0 \, .
\end{align}
It is then easy to see that the all-equal-rank solution in the parent theory is still a solution. Indeed, a general solution for the orientifolded theory has a trivial part, corresponding to a stack of regular branes in the parent theory, and a non-trivial part, corresponding to ``symmetric" fractional branes of the parent theory.

\begin{figure}[h!]
	\centering
	\begin{subfigure}[t]{0.5\textwidth }
		\begin{center} 
			\includegraphics[width=\textwidth]{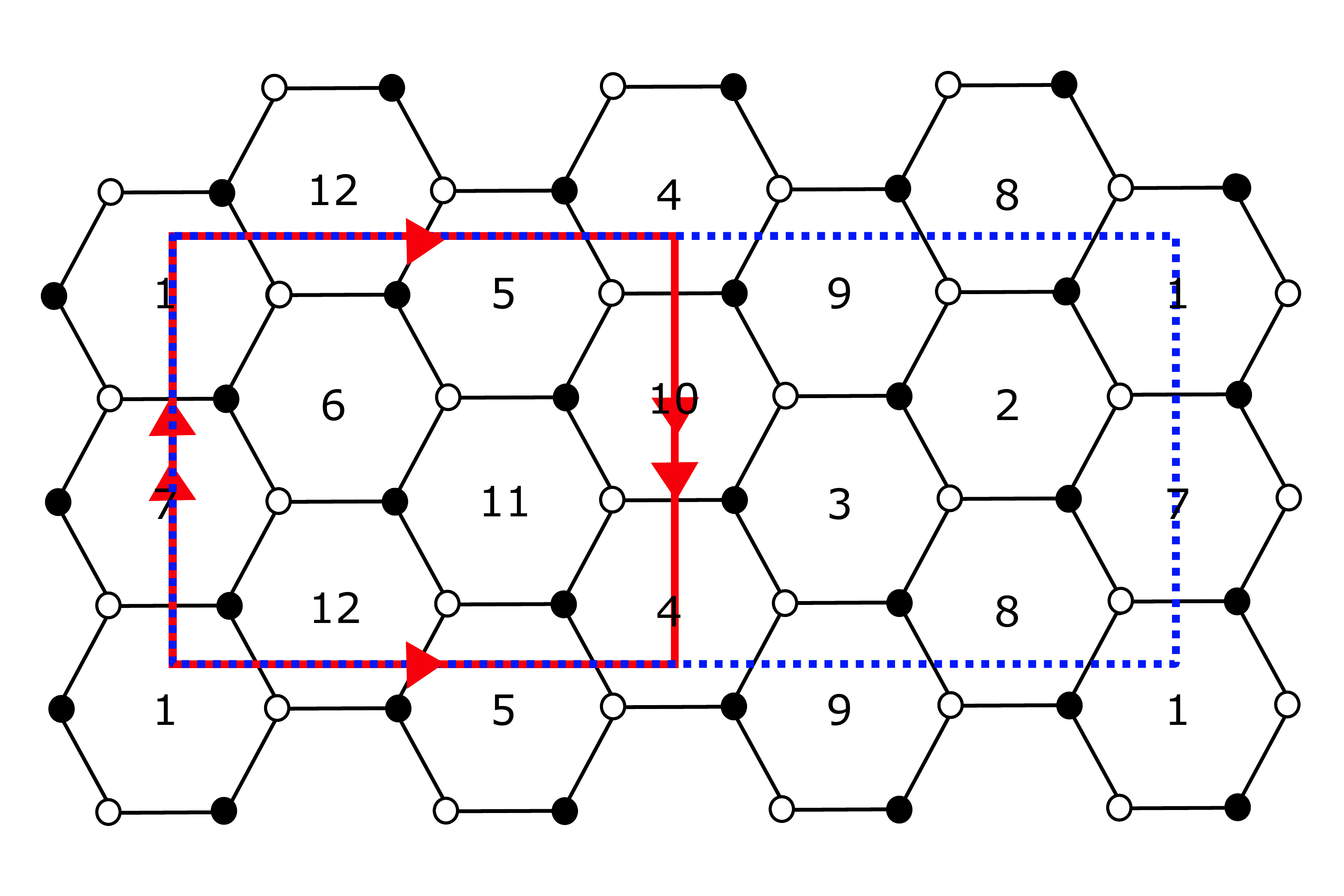}
			\caption{}
			\label{Fig:Z12KleinBottle}
		\end{center}
	\end{subfigure} \hspace{15mm}
	\begin{subfigure}[t]{0.3\textwidth }
		\begin{center} 
			\includegraphics[width=\textwidth]{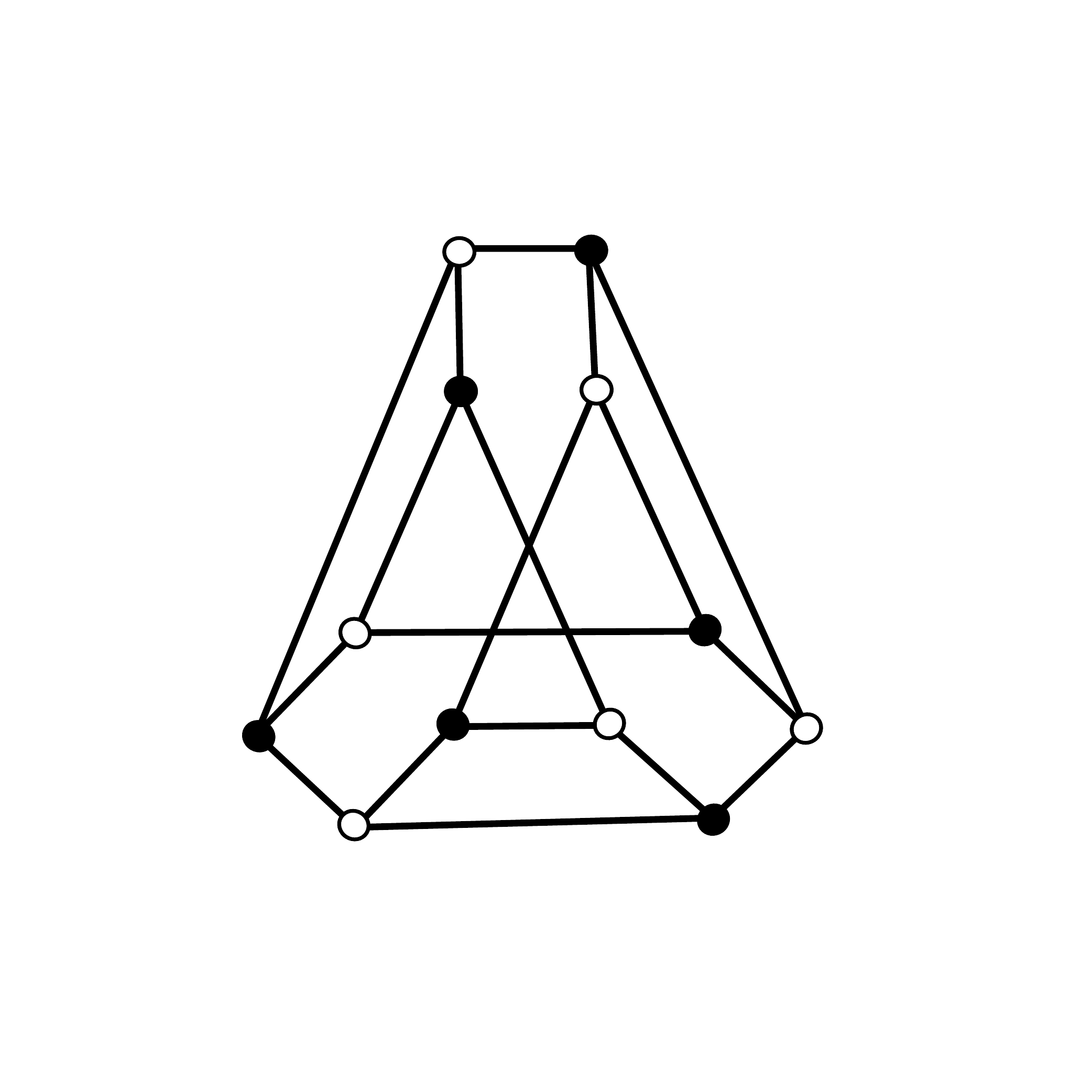}
			\caption{}
			\label{Fig:Franklin}
		\end{center}
	\end{subfigure}
	\caption{(a) Dimer diagram for the orbifold $\mathbb{C}^3/\mathbb{Z}_{12}$ with action $(1,5,6)$. The unit cell and the Klein bottle are depicted in blue and red respectively. (b) The Franklin graph.}\label{Fig:Z12} 
\end{figure}

Fixed loci orientifolds have the remarkable property of producing, in general, non-conformal theories. However, this is not true for glide orientifolds. The theory they describe is an SCFT when the ranks of the gauge groups are all the same. This fact can be seen as follows, consider the $\beta$-function of the parent theory with $N$ probes D-branes,
\begin{align}
	\beta_{SU(N)_i} = 3N-\sum_{i=1}^n \frac{N}{2} (1-\gamma_i)=0  \, , 
\end{align}
where $\gamma_i$ are the anomalous dimensions of the matter fields\footnote{We consider Adj fields as couple of anti-fundamentals fields charged under the same gauge group.}. From this we can read the $\beta$-function of the projected theory whose general form is
\begin{eqnarray}
	\label{betaor1}
	\beta_{SU(N)_i} = 3N-\sum_{i=1}^n \frac{N+b_i}{2} (1-\gamma_i) \, ,
\end{eqnarray}
where the coefficients $b_i$ vanish for fundamental fields and are $\pm2$ for, respectively, symmetric or antisymmetric fields. If we assume that the anomalous dimensions of the fields are the same up to $1/N$ corrections and, since all tensors come in pairs of opposite parity, we see that the $\beta$-function of the gauge groups of the projected theory vanishes as long as all ranks are equal. This dovetails the fact that a Klein Bottle has zero Euler characteristic and, as explained in \cite{Franco:2006gc}, such surfaces may embed a dimer model describing an SCFT\footnote{Other kind of surfaces obtained from orientifolds with fixed loci were found to accommodate SCFTs in \cite{Imai:2001cq,Antinucci:2020yki}.}. Franco and Vegh pointed out that the Franklin graph would be a good candidate to be embedded in a Klein Bottle and host a SCFT not embedded in a torus. Indeed, it can be readily found via a glide reflection of $\mathbb{C}^3/\mathbb{Z}_{12}$, see \figref{Fig:Z12}. This not only confirms their intuition, but it is, to the best or our knowledge, the first instance of such a construction within string theory. 
	
		\section{T-duals of the Glide Orientifold}\label{Sec:Oplane}
	
	\subsection{Type IIA picture and the brane tiling}\label{Sec:Oplane1}
	
	Fixed loci in the dimer have been related to actual orientifold planes in the physical realization of the dimer \cite{Franco:2005rj,Imamura:2008fd}. In fact, one may consider the D3-branes probing a singularity with an orientifold and track the position of the orientifold in the ambient space to the fixed loci in the dimer through T-dualities. An immediate puzzle arises in the case of glide orientifolds, since there are no fixed loci on the torus, i.e. in the brane tiling. In this section we look at $\mathbb{C}^2/\mathbb{Z}_2$ and argue that these orientifolds, which have 8-dimensional fixed loci in the D3 picture (they are O7-planes), don't have a fixed locus in the tiling in the precise sense of \cite{Dabholkar:1996pc,Witten:1997bs}. In the latter reference, the shift action is deduced to be T-dual to a pair of opposite charge O-planes on a circle.
	
	Let us again consider $N$ D3-branes at the tip of a singular toric CY$_3$. As reviewed in the introduction, the dimer presented in \figref{Fig:Z2} is physically realized as a web of D5 and NS5-branes. It is obtained by T-duality along two of the three toric cycles of the toric variety. In particular along those corresponding to mesonic symmetries in the field theory, rather than R-symmetry. Focusing on the case at hand, $\mathbb{C}^2/\mathbb{Z}_2 \times \mathbb{C}$, one may take local coordinates such that $x_7,x_9$ correspond to the two toric cycles that are to be T-dualized. The D-brane configuration is then as in \Cref{Table:D3Singularity} which, after two T-duality should become that of \Cref{Table:BraneTiling}. Note that we have avoided including an orientifold plane in the T-dual, as the dimer shift seems to suggest.
	
	After T-duality one finds D5-branes wrapping the dual cycles with local coordinates $x_7^{\prime}, x_9^{\prime}$. These are in turn identified as the coordinates of the torus $\mathbb{T}^2$ where the 5-brane web lives.
	\begin{table}[!htbp]
		\centering
		\begin{tabular}{ccccccccccc}
			\toprule
			{} & 0 & 1 & 2 & 3 & 4 & 5 & 6 & 7 & 8 & 9 \\
			\midrule
			$\mathbb{C}^2/ \mathbb{Z}_2$ & {} & {} & {} & {} & $\times$ & $\times$ & $\times$ & $\times$ & {} & {} \\
			D3 & $\times$ & $\times$ & $\times$ & $\times$ & {} & {} & {} & {} & {} & {}  \\
			O7 & $\times$ & $\times$ & $\times$ & $\times$ & $\times$ & $\times$ & {} & {} & $\times$ & $\times$  \\
			\bottomrule
		\end{tabular}
		\caption{D3-branes sitting at the tip of $\mathbb{C}^2 / \mathbb{Z}_2$ in the presence of O7-planes.}\label{Table:D3Singularity}
	\end{table}
	\begin{table}[!htbp]
		\centering
		\begin{tabular}{ccccccccccc}
			\toprule
			{} & 0 & 1 & 2 & 3 & 4 & 5 & 6 & 7$^\prime$ & 8 & 9$^\prime$ \\
			\midrule
			NS5 & $\times$ & $\times$ & $\times$ & $\times$ & {} & {} & \multicolumn{4}{c}{$-$ $-$ \ $\Sigma$ \ $-$ $-$} \\
			D5 & $\times$ & $\times$ & $\times$ & $\times$ & {} & {} & {} & $\times$ & {} & $\times$  \\
			\bottomrule
		\end{tabular}
		\caption{The brane tiling. $\Sigma$ is the holomorphic curve in the $67^\prime 89^\prime$-space wrapped by the NS5-brane.}\label{Table:BraneTiling}
	\end{table}
	
	To study the location of the O-plane in the singular geometry, let us introduce the coordinates $z_1, z_2$ and $z_3$ of flat space $\mathbb{C}^3$. We define the coordinates of the variety transverse to the D3-branes, $\mathbb{C}^2 / \mathbb{Z}_2 \times \mathbb{C}$, by constructing invariants under the orbifold action:
	\begin{equation}
		x=z_1^2 \, , \quad y=z_2^2, \quad w=z_1z_2, \quad \text{and}\quad  z=z_3\, ,
	\end{equation}
	with the following relation holding,
	\begin{equation}
		x y=
		w^2 \, .\label{Eq:Relation}
	\end{equation}
As explained in \Cref{Sec:C2Z2}, the orientifold action on the dimer implies that it acts on $z_1, z_2, z_3$ as  $z_1 \leftrightarrow z_2, $. In terms of the orbifold invariant coordinates the orientifold action is then,
	\begin{equation}
		x \leftrightarrow y \, , \quad w \text{ and } z  \text{ fixed}. \label{Eq:Map}
	\end{equation}
	Thus, the orientifold plane extends on the surface defined by $x=y=t$, $t^2=w^2$. 
	From \Cref{{Eq:Relation}} we read two toric $U(1)$ isometries of the orbifold:
	\begin{equation}
		\begin{array}{rcll}
			U(1)_\alpha \, :& \quad x \rightarrow e^{i\alpha} x \, , & \quad y \rightarrow e^{-i\alpha} y \, , & \quad w \rightarrow  w \, , \\
			U(1)_\beta \, :& \quad x \rightarrow e^{i\beta} x \, , & \quad  y \rightarrow e^{i\beta} y \, , & \quad w \rightarrow  e^{i\beta} w \, .
		\end{array}
	\end{equation} 
	We can think about these two isometries as generators of two 1-cycles, $\alpha , \beta$. We can introduce local coordinates parametrizing these cycles, defined whenever they are non-singular, 
	\begin{align}
		\theta_\alpha \equiv \frac{1}{2} \left(\text{Arg}(x) - \text{Arg}(y) \right) \\
		\theta_\beta  \equiv \frac{1}{2} \left(\text{Arg}(x) + \text{Arg}(y) \right)
	\end{align}
	
	We can now identify these two coordinates in terms of the coordinates in \Cref{Table:D3Singularity}: $(\theta_\alpha,\theta_\beta) \sim (x_7,x_5)$. The action  of the orientifold on these two cycles, T-dual to the physical torus, are just $\theta_\alpha \rightarrow - \theta_\alpha$, $\theta_\beta \rightarrow  \theta_\beta$. We thus learn that the orientifold plane spans $x_5$ and is located at $x_7 = 0,\pi$. In fact, there are two orientifold planes of opposite charge such that the total flux cancels with no further sources. One may also argue for the signs being opposite by noting the absence of net RR-charges coming from the O-planes in the dimer picture. This can be seen in the absence of $SO/USp$ groups and the corresponding lack of non-homogeneous terms in the ACC, which can be thought of as Gauss law for compact cycles. Quite remarkably, the T-dual of such a cycle with opposite-charge O-planes, is known to be precisely an orientifold acting as a shift on the T-dual cycle. The absence of fixed loci for this action translates into the absence of O-plane in the dual geometry. This is described in \cite{Witten:1997bs} where T-duality acts as a sort of Fourier transform: the O-planes of opposite charge are related to delta function whose transform are constant and opposite, cancelling each other. This interpretation nicely match the Gauss law analogy we presented earlier.

\begin{figure}
		\centering
		\begin{tikzpicture}[scale=3, >=stealth]
			% Orientifold
			\begin{scope}[canvas is zx plane at y=-0.25]
				\draw[thick, black!50!green, ->] (0,-1)  arc(-90:90:1) node[below right] {$\pi$};
				%\draw[thick, dashed, red, ->] (0,-1) arc(-90:90:1);
			\end{scope}
			
			\begin{scope}[canvas is zx plane at y=0]
				% 1st D4
				\draw (0,0) circle (1);
				% axis 7'
				\draw[thick,->] (0.,-2) arc(-90:90:0.25) node[above right] {$7^\prime$};
			\end{scope}
			
			% 2nd D4
			\begin{scope}[canvas is zx plane at y=0.125]
				\draw (0,0) circle (1);
			\end{scope}
			
			% 3rd D4
			\begin{scope}[canvas is zx plane at y=0.25]
				\draw (1,0) arc(0:90:1) node[right=0.5] {D4's};
				\draw (0,1) arc(90:180:1);
			\end{scope}
			
			% 4th D4
			\begin{scope}[canvas is zx plane at y=-0.125]
				\draw (-1,0) arc(-180:0:1);
			\end{scope}
			
			% 1st NS5
			\begin{scope}[canvas is xy plane at z=-1]
				\draw[black!25!red, thick] (0,-1) -- (0,1) node[above] {NS5};
			\end{scope}
			
			% 2nd NS5
			\begin{scope}[canvas is xy plane at z=1]
				\draw[black!25!red, thick] (0,-0.5) -- (0,1.5) node[above] {NS5};
			\end{scope}
			
			% Axes 8, 9
			\begin{scope}[canvas is xy plane at z=0]
				\draw[thick,->] (-1.85,-0.25) -- (-1.85,0.25) node[above] {8, 9};
			\end{scope}
		\end{tikzpicture}
		\caption{Type IIA picture with the orientifold mapping given by a $\pi$ rotation along $x_7^\prime$  (in green).}\label{Fig:TypeIIA}
	\end{figure}
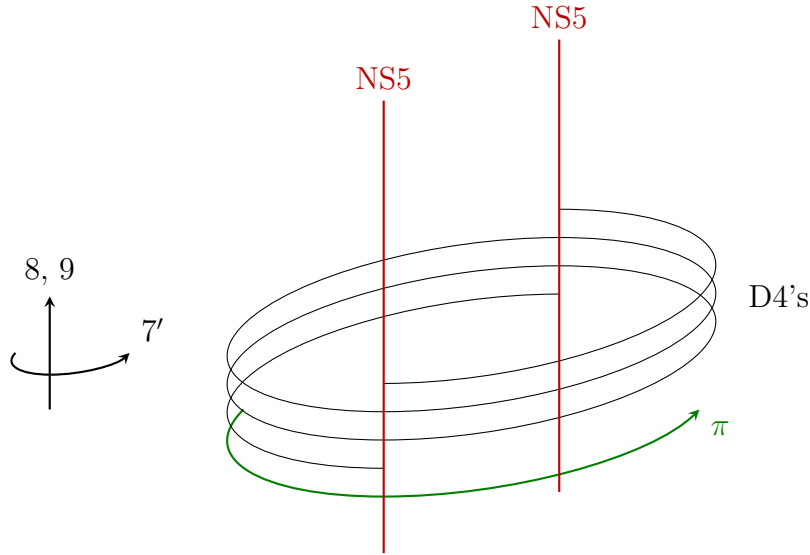

	After one T-duality along $x_7$, the T-dual Type IIA construction is analogous to the ones studied in \cite{Park:1998zh,Park:1999eb,Uranga:1999mb,Ennes:2000fu,Feng:2001rh}. The relevant information is encoded in a cycle $x_7^{\prime}$ where D4-branes are suspended between two NS5-branes. As we explained before, the orientifold action acts now as a shift, rotating halfway the configuration, see \figref{Fig:TypeIIA}. This action is consistent with the mapping of gauge groups and matter fields on the dimer model. 
	
	Finally, if we further T-dualize along the direction spanned by NS5-branes $x_9$, we get to the tiling picture.  After the last T-duality, the orientifold acts on $x_9'$ as a reflection\footnote{This is a standard fact of orientifolds. Upon T-duality along a direction spanned by the O-plane, an O$p$-plane is mapped to an O$(p-1)$-plane, with action $\theta \to - \theta$ on the dual cycle.}. Together with the shift on $x^\prime_7$, these actions reproduce the glide reflection that we see on the tiling.

\subsection{The mirror picture}

\begin{figure}
    \centering
    \begin{subfigure}[t]{0.20\textwidth }
        \begin{center} 
		\includegraphics[width=\textwidth]{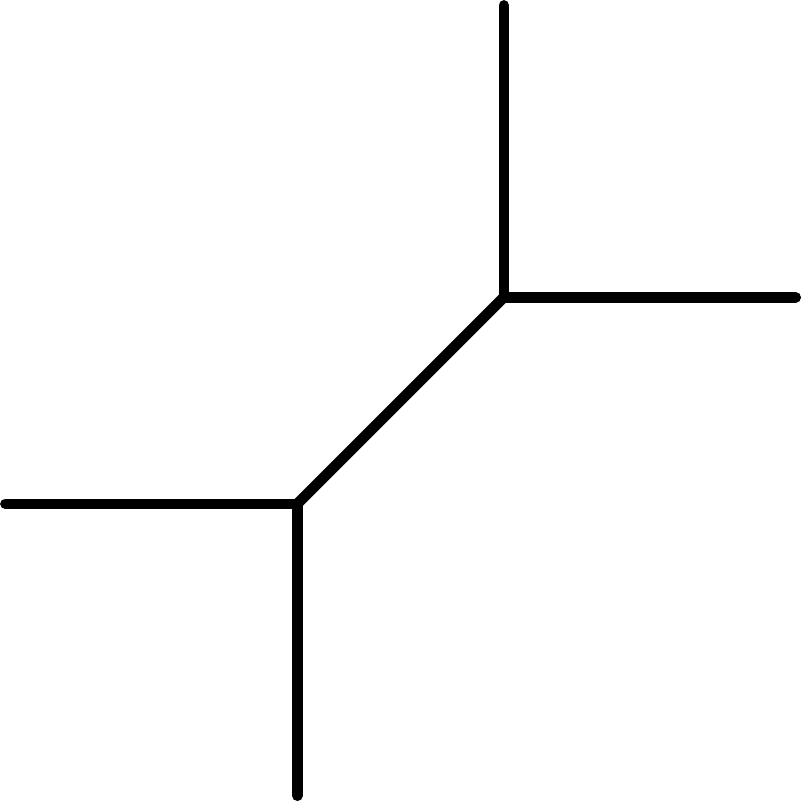}
		\caption{}
		\label{Fig:ConifoldWebDiagram}
		\end{center}
    \end{subfigure} \hspace{6mm}
	\begin{subfigure}[t]{0.20\textwidth }
		\begin{center} 
		\includegraphics[width=\textwidth]{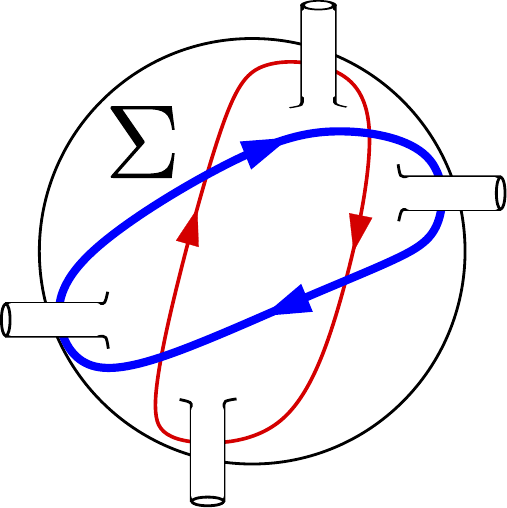}
		\caption{}
		\label{Fig:ConifoldMirrorSurface}
		\end{center}
    \end{subfigure}\hspace{6mm}
    \begin{subfigure}[t]{0.20\textwidth }
        \begin{center} 
		\includegraphics[width=\textwidth]{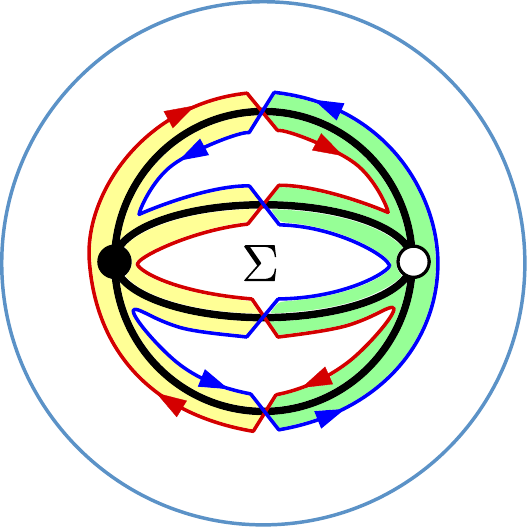}
		\caption{}
		\label{Fig:ConifoldMirrorSurfaceTiling}
		\end{center}
    \end{subfigure} \hspace{6mm}
    \begin{subfigure}[t]{0.20\textwidth }
        \begin{center} 
		\includegraphics[width=\textwidth]{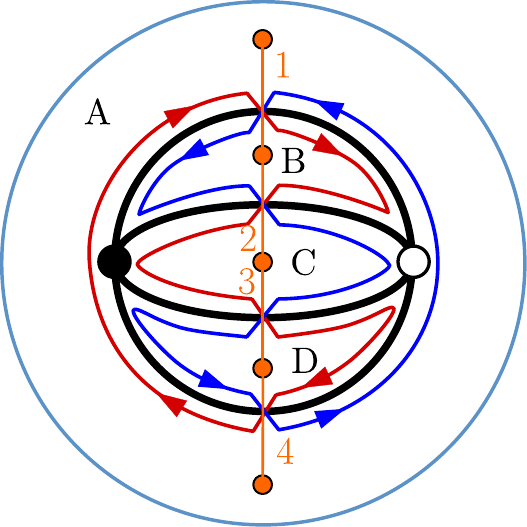}
		\caption{}
		\label{Fig:ConifoldMirrorOplane}
		\end{center}
    \end{subfigure}
    \caption{a) Conifold web diagram. b) shows the actual surface describing the mirror geometry. The Zig-Zag paths in blue and red are 1-cycles where D6-branes may wrap. c) Tiling of the mirror Riemann surface $\Sigma_0$ giving rise to the conifold gauge theory. d) Orientifold of the conifold. Orange segments and dots denote O-planes and punctures where they end, respectively.}\label{Fig:ConifoldMirror} 
\end{figure}

One further T-duality on the remaining toric cycle brings us to the mirror setup of our starting point \cite{Feng:2005gw} (D3-branes at singularities). The mirror geometry is fully specified by a Riemann surface $\Sigma_0$ in which D6-branes, wrapping 1-cycles, give rise to the appropriate field theory\footnote{These 1-cycles are associated to 3-cycles in the full geometry.}. Gauge fields are associated to different 1-cycles, where D6 are wrapped, matter fields to intersections among these cycles and superpotential terms arise from open string worldsheet instantons supported at disks in this Riemann surface. In fact, this Riemann surface can be seen as the ``fattened" version of the web diagram. Furthermore, one can embed the dimer graph on a planar version of it and read immediately both the geometry and the field theory from it. This diagram has been called the shiver. The shiver and the dimer are related by an untwisting procedure \cite{Feng:2005gw}. The physical interpretation is now as follows.
\begin{itemize}
\item Faces in the shiver correspond to ZZPs in the dimer, and represent punctures in $\Sigma_0$ with $(p,q)$ charge given by the winding numbers of the ZZP.
\item ZZP on the shiver correspond to faces on the dimer, and  represent the 1-cycles (Special Lagrangian 3-cycles in the full geometry) where D6-branes wrap. They are hence representing gauge groups.
\item ZZP intersections represent brane intersections where open string massless bifundamental fields are located.
\item Disk on the surface are euclidean disks in the full geometry where open string worldsheet instantons may arise. These generate the superpotential (albeit non-perturbatively!).
\end{itemize}
In \Cref{Fig:ConifoldMirror}  we show the prototypical example, the conifold. 

Already in \cite{Franco:2007ii}, orientifolds were partially understood in this framework. It was noted that the O6-planes should be viewed as stretching between the different punctures. Different pieces of the O-plane would there be assigned different signs. A full understanding of the available sign choices was not achieved and we will not pursue it here. An orientifold of the conifold is shown in \Cref{Fig:ConifoldMirrorOplane}.

\paragraph{A simple example: $\mathbb{C}^2/ \mathbb{Z}_2 \times \mathbb{C}$.} 

\begin{figure}
    \centering
    \begin{subfigure}[t]{0.17\textwidth }
        \begin{center} 
		\includegraphics[width=\textwidth]{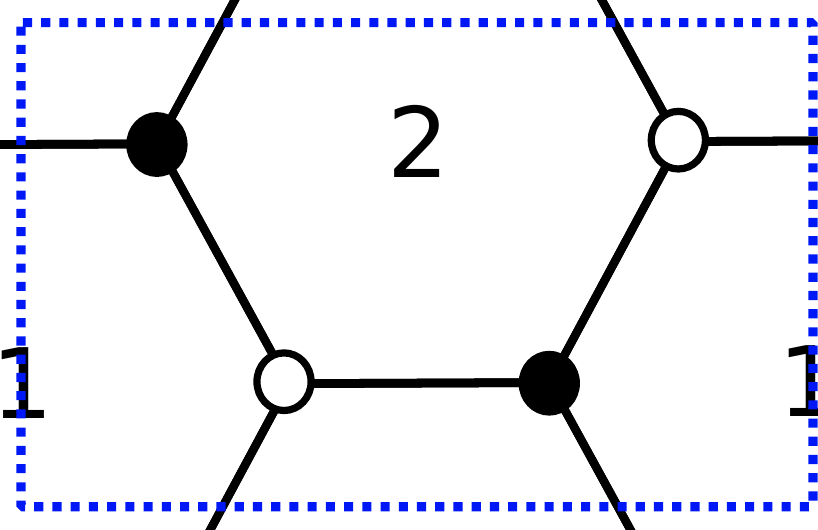}
		\caption{}
		\label{Fig:Z2UnitCell}
		\end{center}
    \end{subfigure} \hspace{3mm}
	\begin{subfigure}[t]{0.17\textwidth }
		\begin{center} 
		\includegraphics[width=\textwidth]{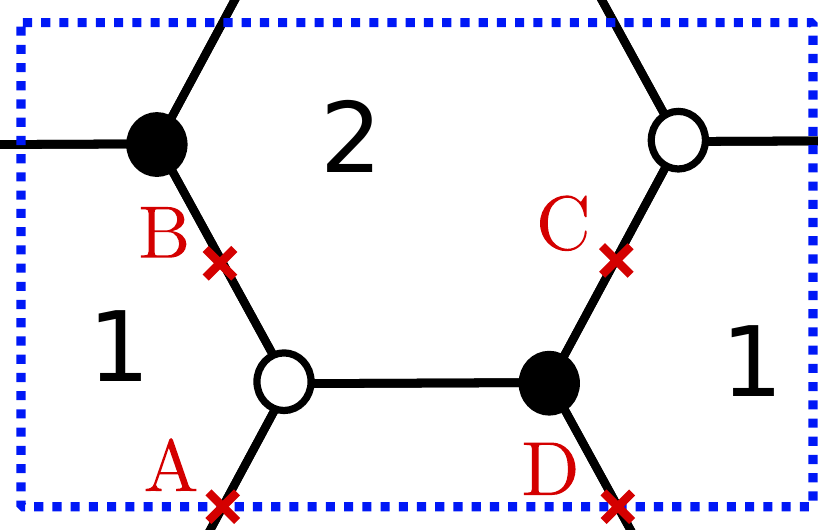}
		\caption{}
		\label{Fig:Z2Point2}
		\end{center}
    \end{subfigure} \hspace{3mm}
	\begin{subfigure}[t]{0.17\textwidth }
		\begin{center} 
		\includegraphics[width=\textwidth]{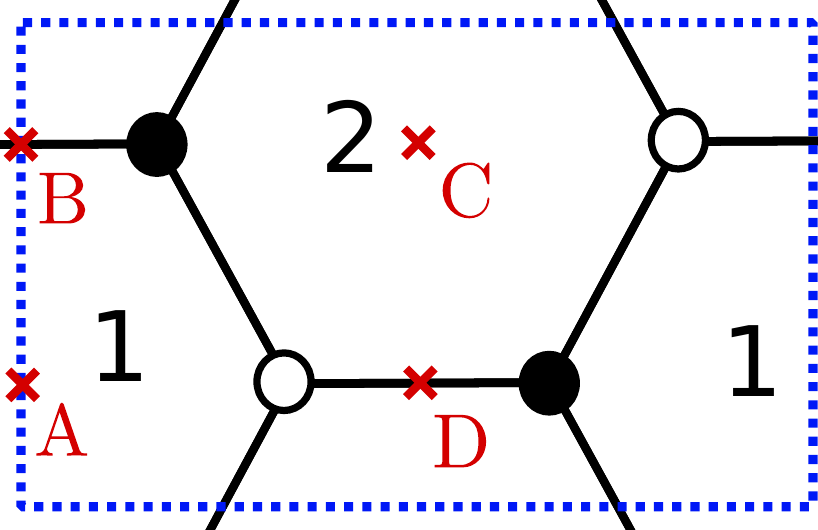}
		\caption{}
		\label{Fig:Z2Point1}
		\end{center}
    \end{subfigure} \hspace{3mm}
	\begin{subfigure}[t]{0.17\textwidth }
		\begin{center} 
		\includegraphics[width=\textwidth]{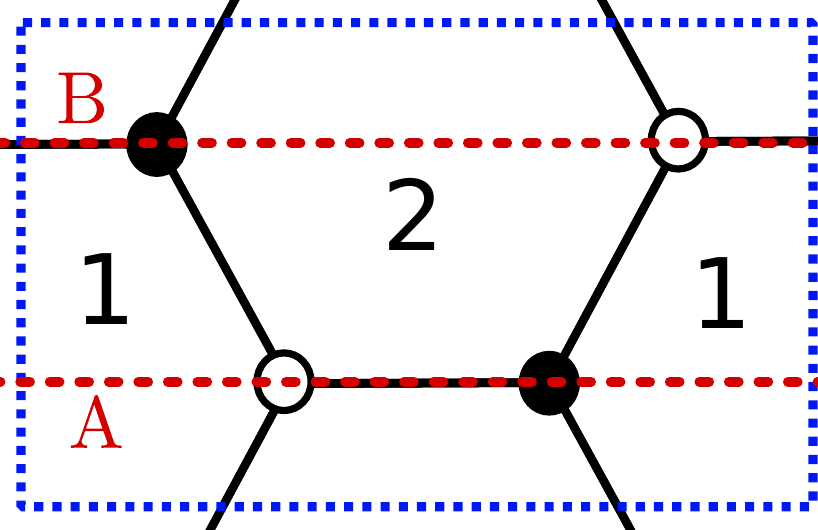}
		\caption{}
		\label{Fig:Z2Line}
		\end{center}
    \end{subfigure}\hspace{3mm}
	\begin{subfigure}[t]{0.17\textwidth }
		\begin{center} 
		\includegraphics[width=\textwidth]{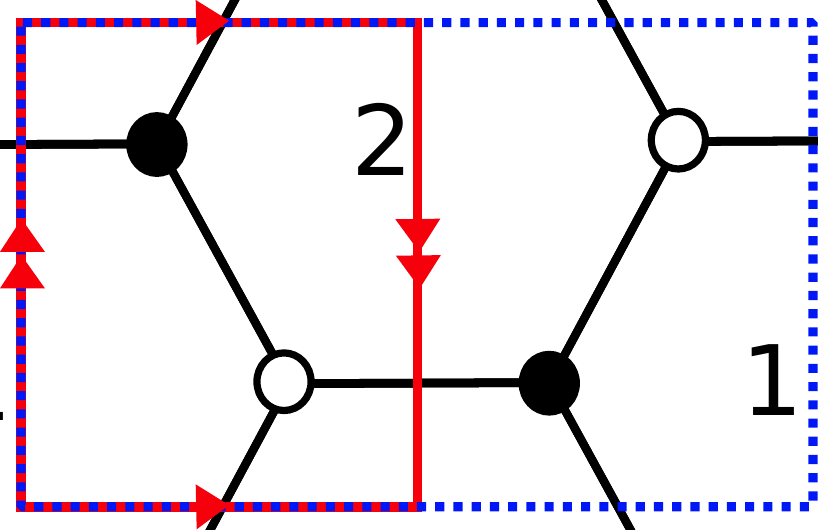}
		\caption{}
		\label{Fig:Z2KleinBottleUnit}
		\end{center}
    \end{subfigure}
    \caption{a) Dimer model unit cell of $\mathbb{C}^2 / \mathbb{Z}_2 \times \mathbb{C}$. Three possible orientifold actions with fixed loci are shown in b), c) and d).  b) and c) correspond to fixed points, while d) is a fixed line. e) is a glide orientifold with no fixed loci.}\label{Fig:C2Z2DimerOrientifolds} 
\end{figure}

In this section we explicitly map all orientifolds in the dimer with those in the T-dual and mirror picture. While we are not able to derive matter field projections on the T-dual nor the mirror, we hope to convey a unifying picture. We will focus on  $\mathbb{C}^2/ \mathbb{Z}_2 \times \mathbb{C}$ for several reasons. It admits all kinds of orientifolds (fixed points, lines and glide reflections), a simple T-dual set-up and its mirror $\Sigma_0$ has genus zero, making it amenable to discussion.  Orientifolds in the dimer can be classified in 4 groups, depending on the fixed loci. There are 2 different fixed point orientifolds, shown in \Cref{Fig:Z2Point2,Fig:Z2Point1} and a fixed line orientifold, in \Cref{Fig:Z2Line}. These were already discussed in \cite{Franco:2007ii}. Different sign assignments, possibly respecting the sign rule, yield the different field contents. Finally, as discussed in \Cref{Sec:Review}, a glide orientifold is realized in the dimer as shown in \Cref{Fig:Z2KleinBottleUnit}.

A complete classification, up to sign permutations and anomaly cancellation, is shown in \Cref{Table:AVeryBigTable}.
\begin{table}
\centering
\begin{tabular}{c|c|c|c|c}
O-Type                        & Figure             & Signs      & Group                           & Tensor Content  \\ 
\hline\hline
\multirow{13}{*}{Fixed Points} & \multirow{5}{*}{\Cref{Fig:Z2Point2}} & $(++++)$ & \multirow{5}{*}{$SU$}            &  $\text{Adj} + 2 \symm + 2 \antisymm $              \\
                              &                    & $(++--)$ &                                 &  $\text{Adj} +  \symm + \antisymm + \asymm + \antiasymm $              \\
                              &                    & $(+--+)$ &                                 &  $\text{Adj} + 2 \asymm + 2 \antisymm $               \\
                              &                    & $(-++-)$ &                                 &  $\text{Adj} + 2 \antiasymm + 2 \symm $               \\
                              &                    & $(----)$ &                                 &  $\text{Adj} + 2 \asymm + 2 \antiasymm $               \\ \cline{2-5}
                              & \multirow{8}{*}{\Cref{Fig:Z2Point1}} & $(++++)$ & $SO \times SO$ &  $ \symm_1 + \symm_2 $                \\
                              &                    & $(++--)$ & $SO \times USp$ &  $ \symm_1 + \asymm_2 $               \\
                              &                    & $(--++)$ & $USp \times SO$ &  $ \asymm_1 + \symm_2 $               \\
                              &                    & $(+-+-)$ & $SO \times SO$ &  $ \asymm_1 + \asymm_2 $               \\
                              &                    & $(-+-+)$ & $USp \times USp$ &  $ \symm_1 + \symm_2 $               \\
                              &                    & $(+--+)$ & $SO \times USp$ &  $ \asymm_1 + \symm_2 $               \\
                              &                    & $(-++-)$ & $USp \times SO$ &  $ \symm_1 + \asymm_2 $               \\
                              &                    & $(----)$ & $USp \times USp$ &  $ \asymm_1 + \asymm_2 $               \\ \hline
\multirow{4}{*}{Fixed Lines}  & \multirow{4}{*}{\Cref{Fig:Z2Line}} & $(++)$   & $SO \times SO$ &  $ \symm_1 + \symm_2 $               \\
                              &                    & $(+-)$   & $SO \times USp$ &  $ \asymm_1 + \symm_2 $               \\
                              &                    & $(-+)$   & $USp \times SO$ &  $ \symm_1 + \asymm_2 $               \\
                              &                    & $(--)$   & $USp \times USp$ &  $ \asymm_1 + \asymm_2 $               \\ \hline
Glide                       & \Cref{Fig:Z2KleinBottleUnit}                  &            & $SU$                           &  $\text{Adj} +  \antisymm + \antiasymm + \symm + \asymm $               
\end{tabular}\caption{Different orientifold projections on the dimer.}\label{Table:AVeryBigTable}
\end{table}

\begin{figure}
    \centering
    \begin{subfigure}[t]{0.17\textwidth }
        \begin{center} 
		\includegraphics[width=\textwidth]{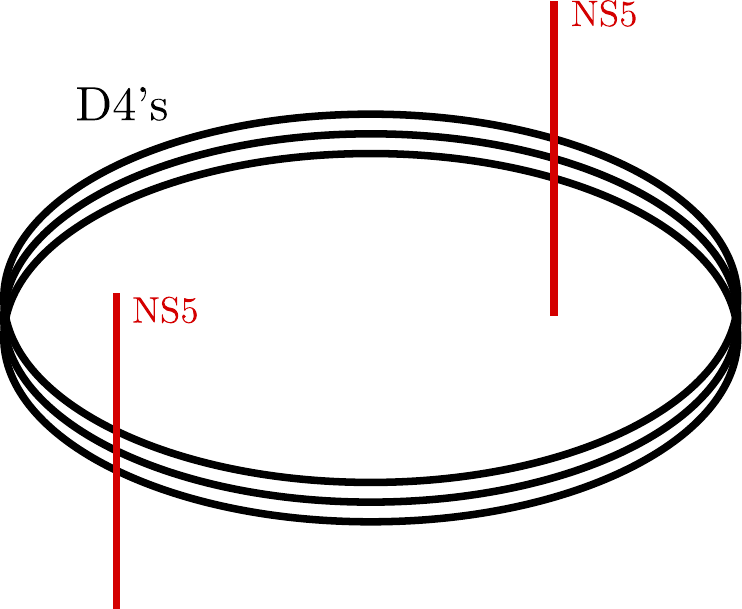}
		\caption{}
		\label{Fig:TDualZ2}
		\end{center}
    \end{subfigure} \hspace{3mm}
	\begin{subfigure}[t]{0.17\textwidth }
		\begin{center} 
		\includegraphics[width=\textwidth]{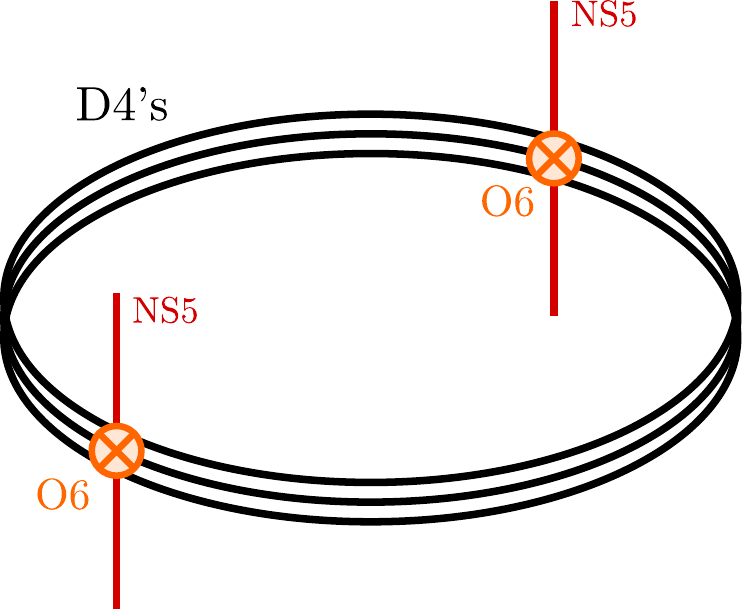}
		\caption{}
		\label{Fig:TDualZ2Point1}
		\end{center}
    \end{subfigure} \hspace{3mm}
	\begin{subfigure}[t]{0.17\textwidth }
		\begin{center} 
		\includegraphics[width=\textwidth]{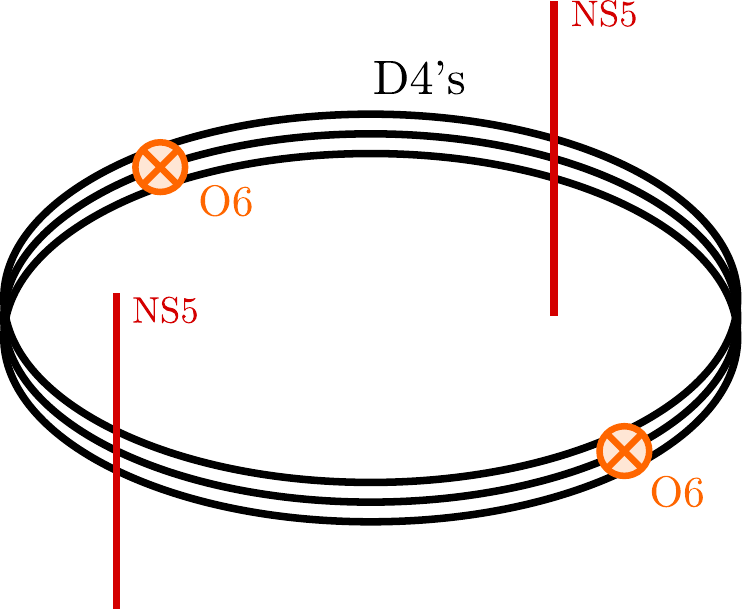}
		\caption{}
		\label{Fig:TDualZ2Point2}
		\end{center}
    \end{subfigure} \hspace{3mm}
	\begin{subfigure}[t]{0.17\textwidth }
		\begin{center} 
		\includegraphics[width=\textwidth]{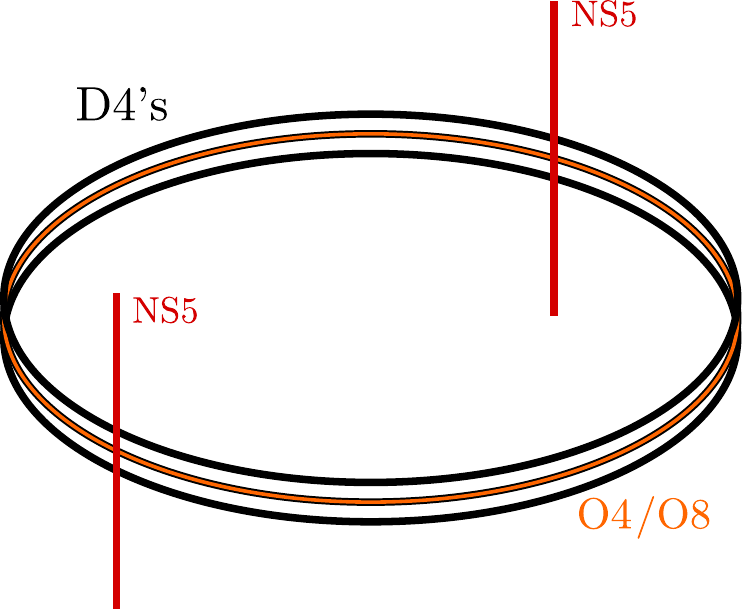}
		\caption{}
		\label{Fig:TDualZ2Oline}
		\end{center}
    \end{subfigure}\hspace{3mm}
	\begin{subfigure}[t]{0.17\textwidth }
		\begin{center} 
		\includegraphics[width=\textwidth]{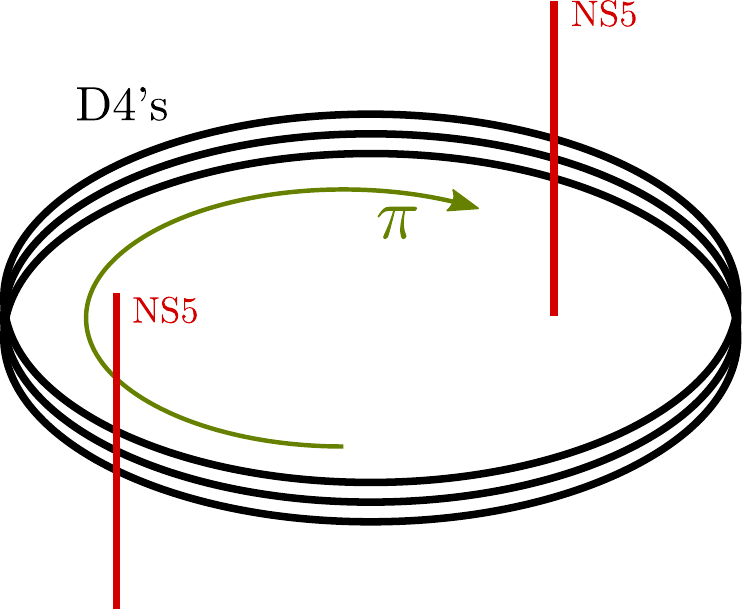}
		\caption{}
		\label{Fig:TDualZ2OShift}
		\end{center}
    \end{subfigure}
    \caption{a) T-dual to $\mathbb{C}^2 / \mathbb{Z}_2 \times \mathbb{C}$. Three possible orientifold actions with fixed loci are shown in b), c) and d).  b) and c) correspond to fixed points, while d) is a fixed line. e) is a shift (glide reflection) with no fixed loci.}\label{Fig:C2Z2TdualOrientifolds} 
\end{figure}

In \Cref{Sec:Oplane1}, we have discussed the setup T-dual to the glide orientifold, shown again in \Cref{Fig:TDualZ2OShift}. The orientifolds in \Cref{Fig:TDualZ2Point1,Fig:TDualZ2Point2,Fig:TDualZ2Oline} have been discussed in the literature \cite{Evans:1997hk,Feng:2001rh,Imai:2001cq}. One can easily identify these three orientifolds in the dimer setup as \Cref{Fig:Z2Point2,Fig:Z2Point1,Fig:Z2Line}, respectively. It is worth mentioning that the fixed lines orientifold with the same sign correspond to the O4-plane in the T-dual, while the case with opposite signs is mapped to an O8-plane. Finally, the glide orientifold, \Cref{Fig:Z2KleinBottleUnit} is identified with \Cref{Fig:TDualZ2OShift}.

\begin{figure}
    \centering
    \begin{subfigure}[t]{0.25\textwidth }
        \begin{center} 
		\includegraphics[width=\textwidth]{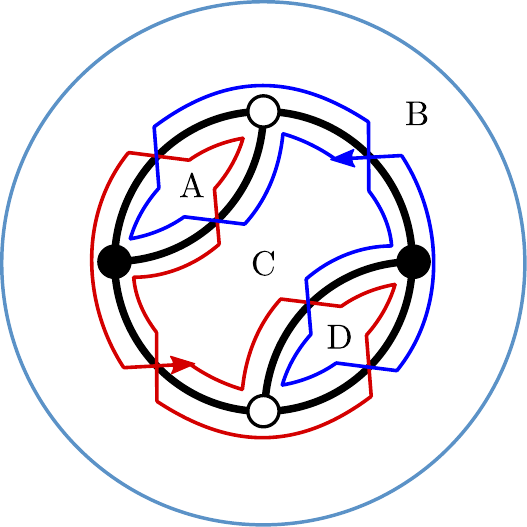}
		\caption{}
		\label{Fig:C2Z2MirrorSurfaceTiling}
		\end{center}
    \end{subfigure} \hspace{3mm}
	\begin{subfigure}[t]{0.25\textwidth }
		\begin{center} 
		\includegraphics[width=\textwidth]{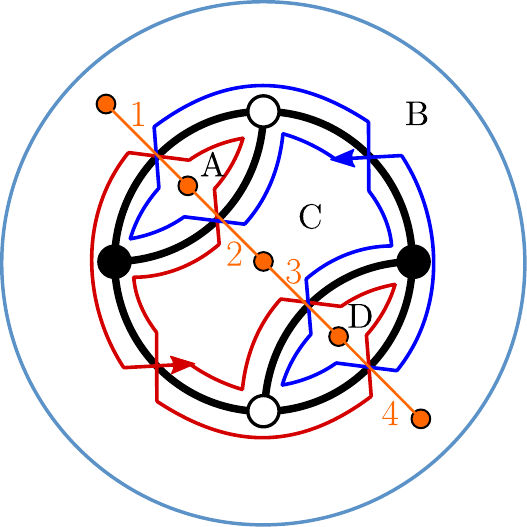}
		\caption{}
		\label{Fig:C2Z2MirrorOrientifold1}
		\end{center}
    \end{subfigure} \hspace{3mm}
	\begin{subfigure}[t]{0.25\textwidth }
		\begin{center} 
		\includegraphics[width=\textwidth]{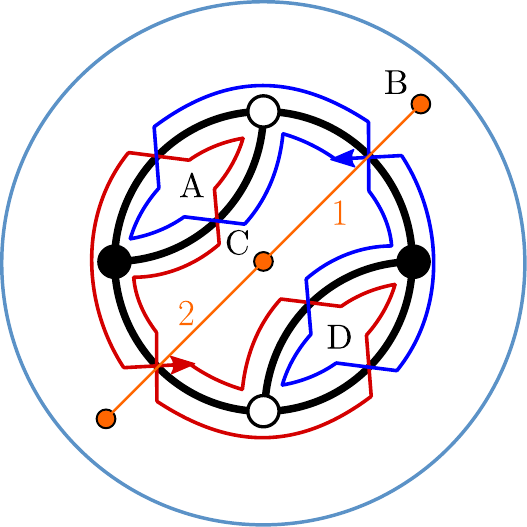}
		\caption{}
		\label{Fig:C2Z2MirrorOrientifold2}
		\end{center}
    \end{subfigure} \hspace{3mm}
	\begin{subfigure}[t]{0.25\textwidth }
		\begin{center} 
		\includegraphics[width=\textwidth]{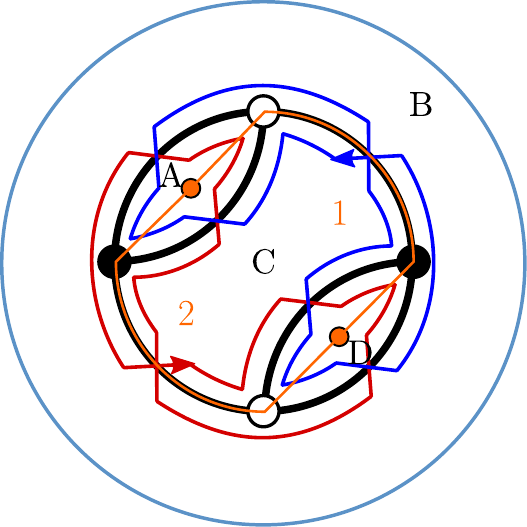}
		\caption{}
		\label{Fig:C2Z2MirrorOrientifold3}
		\end{center}
    \end{subfigure}\hspace{3mm}
	\begin{subfigure}[t]{0.25\textwidth }
		\begin{center} 
		\includegraphics[width=\textwidth]{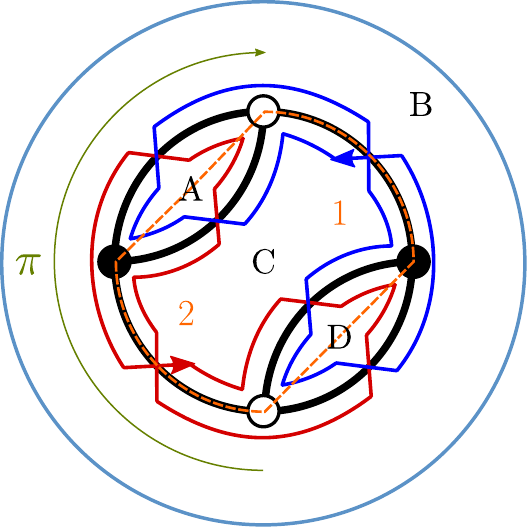}
		\caption{}
		\label{Fig:C2Z2MirrorOrientifoldShift}
		\end{center}
    \end{subfigure}
    \caption{a) Tiling of the mirror Riemann surface $\Sigma_0$ of $\mathbb{C}^2 / \mathbb{Z}_2 \times \mathbb{C}$. Three possible orientifold planes are shown in b), c) and d). The orientifold plane is shown in orange and its different pieces are labelled using orange numbers. In the dimer b) and c) correspond to fixed points, while d) is a fixed line. e) is a shift orientifold with no fixed loci.}\label{Fig:C2Z2MirrorOrientifolds} 
\end{figure}

The shiver is shown in \Cref{Fig:C2Z2MirrorSurfaceTiling}. It is essentially the same as the T-dual with punctures A and D sitting at the NS5 locations. This was to be expected, since the $S^1$ in the T-dual is kept in $\Sigma_0$.  This makes it particularly easy to find the orientifold actions on this surface\footnote{Otherwise one would need to think about the field content on the dimer or the alga map \cite{Feng:2005gw}.}. The four types are shown in \Cref{Fig:C2Z2MirrorOrientifolds} with labels matching those of \Cref{Fig:C2Z2DimerOrientifolds,Fig:C2Z2TdualOrientifolds}. Note, in particular, that the glide reflection, in \Cref{Fig:C2Z2MirrorOrientifoldShift} consists on a $\pi$ rotation and a reflection with respect to the dotted orange circle. The total action thus exchanges punctures B and C, for instance. Unlike the other orientifold actions, there are no fixed loci in this case. While the field content can be deduced from the drawings by assigning a sign to every O-plane piece and assigning the matter projection individually (as in the open string computation), it is not clear how to enforce SUSY in this picture. Without  further knowledge it is not obvious which projections are SUSY-preserving.

\section{Involutions and Zig-Zag Paths}\label{Sec:Toric}

In this section, we first develop, in \Cref{Sec:GlideZZP}, a condition the toric diagram (or equivalently, the ZZP's) of a singularity must satisfy to be compatible with the glide reflection. This enlarges the dictionary between orientifold projections of a given toric singularity and its ZZPs content, as initiated by \cite{Retolaza:2016alb}. Secondly, and with the help of ZZP techniques \cite{Butti:2006hc}, we show in \Cref{Sec:FracZZP} how to detect the presence of fractional branes in the orientifolded singularity. Finally, in \Cref{Sec:ShiftOrient}, we give a general proof that the ``would be" shift orientifold projection is incompatible with the requirement to preserve SUSY. 

\subsection{Glide Orientifold from the Toric Diagram}\label{Sec:GlideZZP}

A glide reflection can be seen as a combination of a shift and a reflection in the dimer model, even if each of them is not a symmetry per se. Starting from what we learned in our examples and using this simple observation, we can understand how this involution acts on the ZZP content of the toric diagram.

\begin{figure}[h!]
	\centering
	\includegraphics[width=0.5\textwidth]{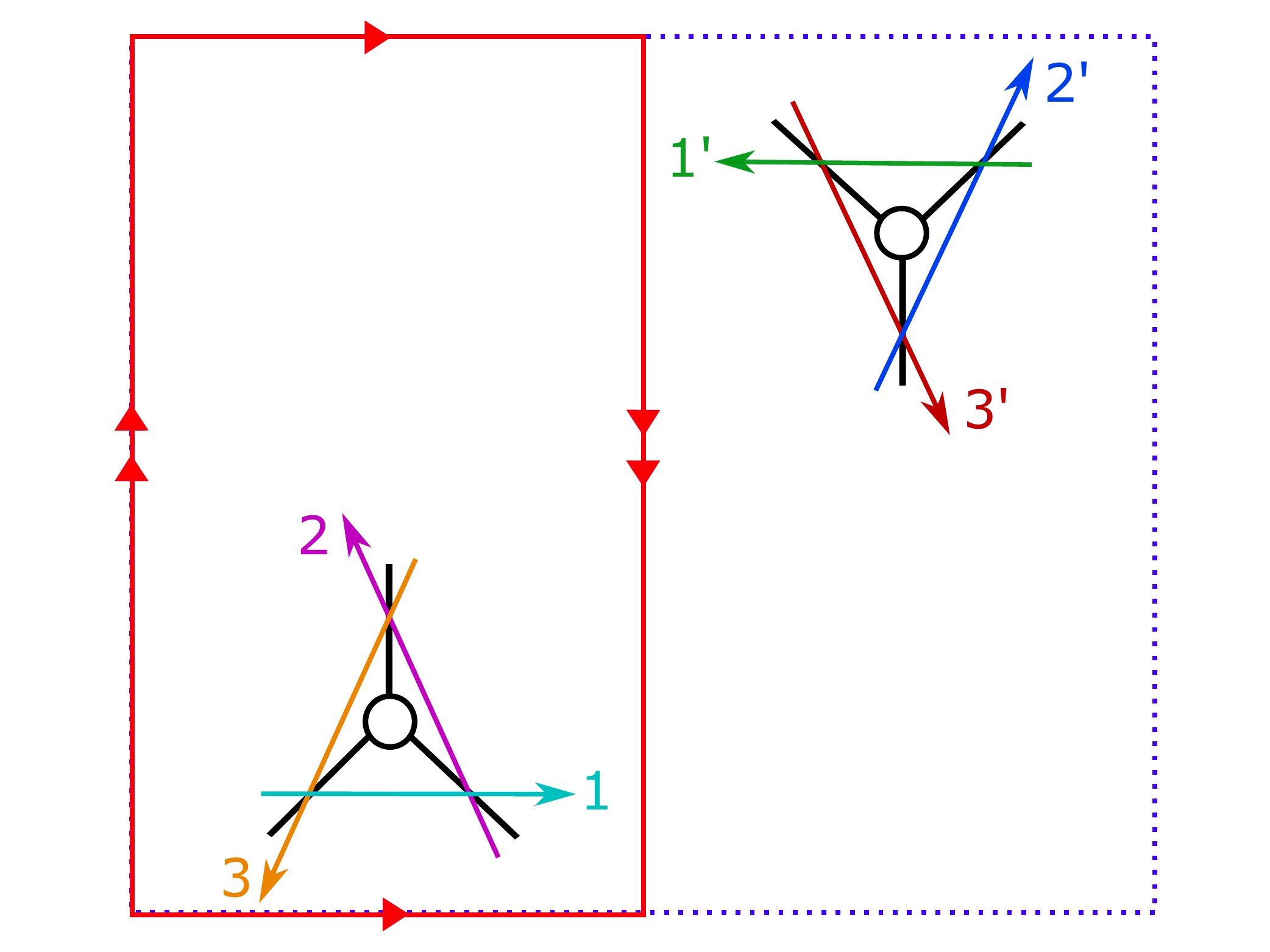}
	\caption{The glide orientifold maps together nodes of the same color. The dashed blue line delineates the unit cell of the parent theory, while the red frame represents the orientifold. The ZZPs $1, 2, 3$ are mapped to $1', 2', 3'$ respectively.}
	\label{Fig:GlideZZP}
\end{figure}

First of all, we notice that the shift and the reflection are performed along the same axis. Consider, for instance, a horizontal shift and axis of reflection as in \figref{Fig:GlideZZP}. The action of the glide reflection reverts the horizontal component of each ZZP. Actually, the glide reflection leaves no fixed ZZPs, since even those perpendicular to the axis are mapped among themselves because of the shift part of the glide reflection.

\begin{figure}[h!]
	\centering
	\includegraphics[width=0.5\textwidth]{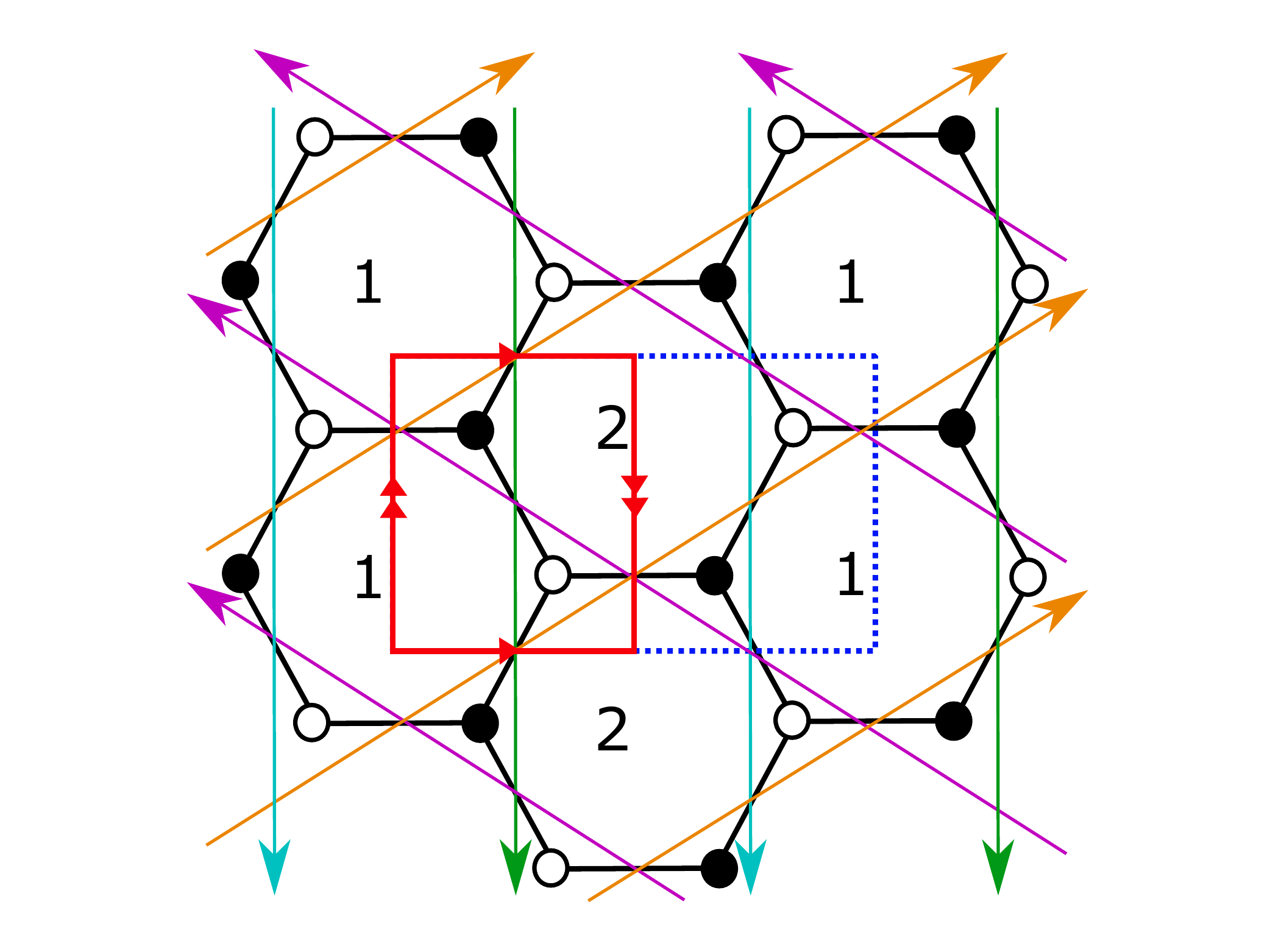}
	\caption{The Klein bottle obtained from the dimer of $\mathbb{C}^2/\mathbb{Z}_2\times \mathbb{C}$ and the corresponding ZZPs.}
	\label{Fig:Z2ZZPKB}
\end{figure}

Putting the two observations together we can say that: if the glide reflection is composed by a \textit{horizontal shift and a reflection axis}, directed as $(1,0)$ in the dimer, ZZPs are mapped as follows: $(p,q)$ is sent to $(-p,q)$ when $p\neq 0$, while all other ZZPs of the form $(0,\pm 1)$ are mapped to one another, preserving the orientation, meaning that they come in even numbers. In our example of \figref{Fig:Z2ZZPKB}, the orange $(1,1)$ and purple $(-1,1)$ ZZPs are interchanged. The same is true for the blue and green ZZPs of the $(0,-1)$ type.

Similarly, in order to construct a Klein bottle with a \textit{vertical shift and reflection axis}, the toric diagram should have ZZPs $(1,0)$ and $(-1,0)$ in even numbers, possibly different, and ZZPs $(p,q)$ with $q\neq 0$ paired with ZZPs $(p, -q)$.

\begin{figure}[h!]
	\centering
	\begin{subfigure}[t]{0.3\textwidth }
		\begin{center} 
			\includegraphics[width=\textwidth]{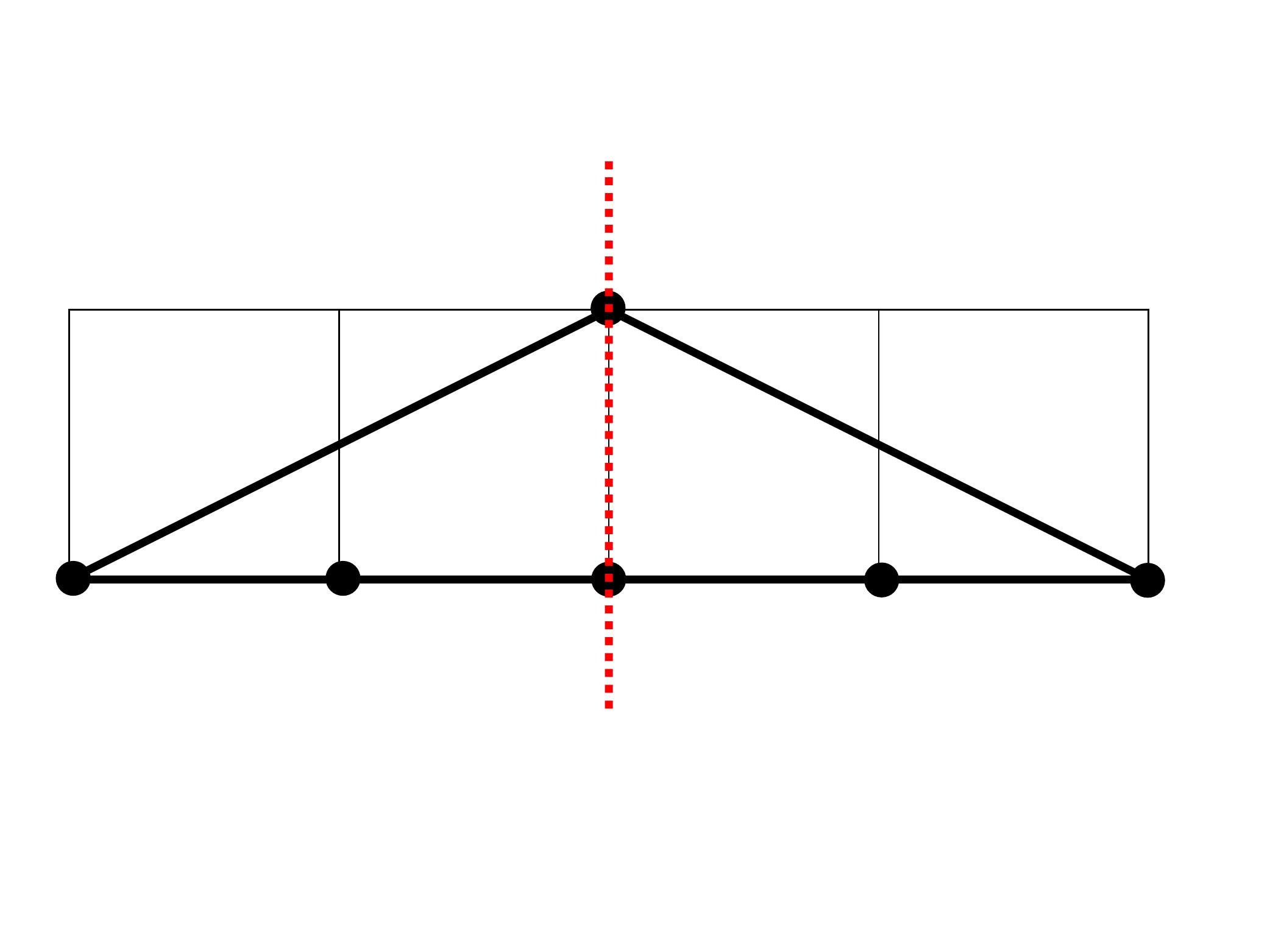}
			\caption{}
			\label{Fig:C2Z4toric}
		\end{center}
	\end{subfigure}
	\begin{subfigure}[t]{0.3\textwidth } 
		\begin{center} 
			\includegraphics[width=\textwidth]{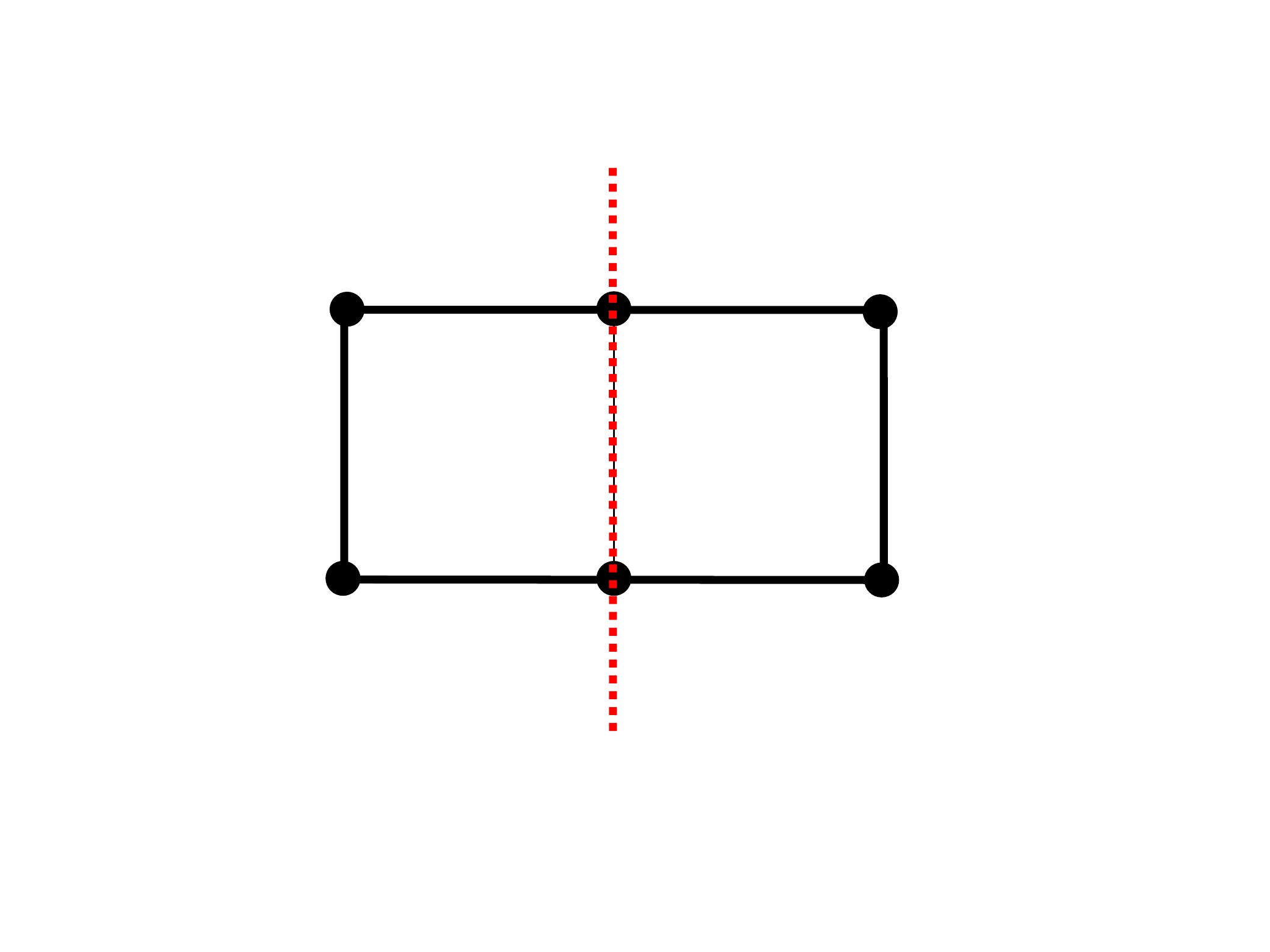}
			\caption{}
			\label{Fig:CZ2toric}
		\end{center}
	\end{subfigure} 
	\begin{subfigure}[t]{0.3\textwidth } 
		\begin{center} 
			\includegraphics[width=\textwidth]{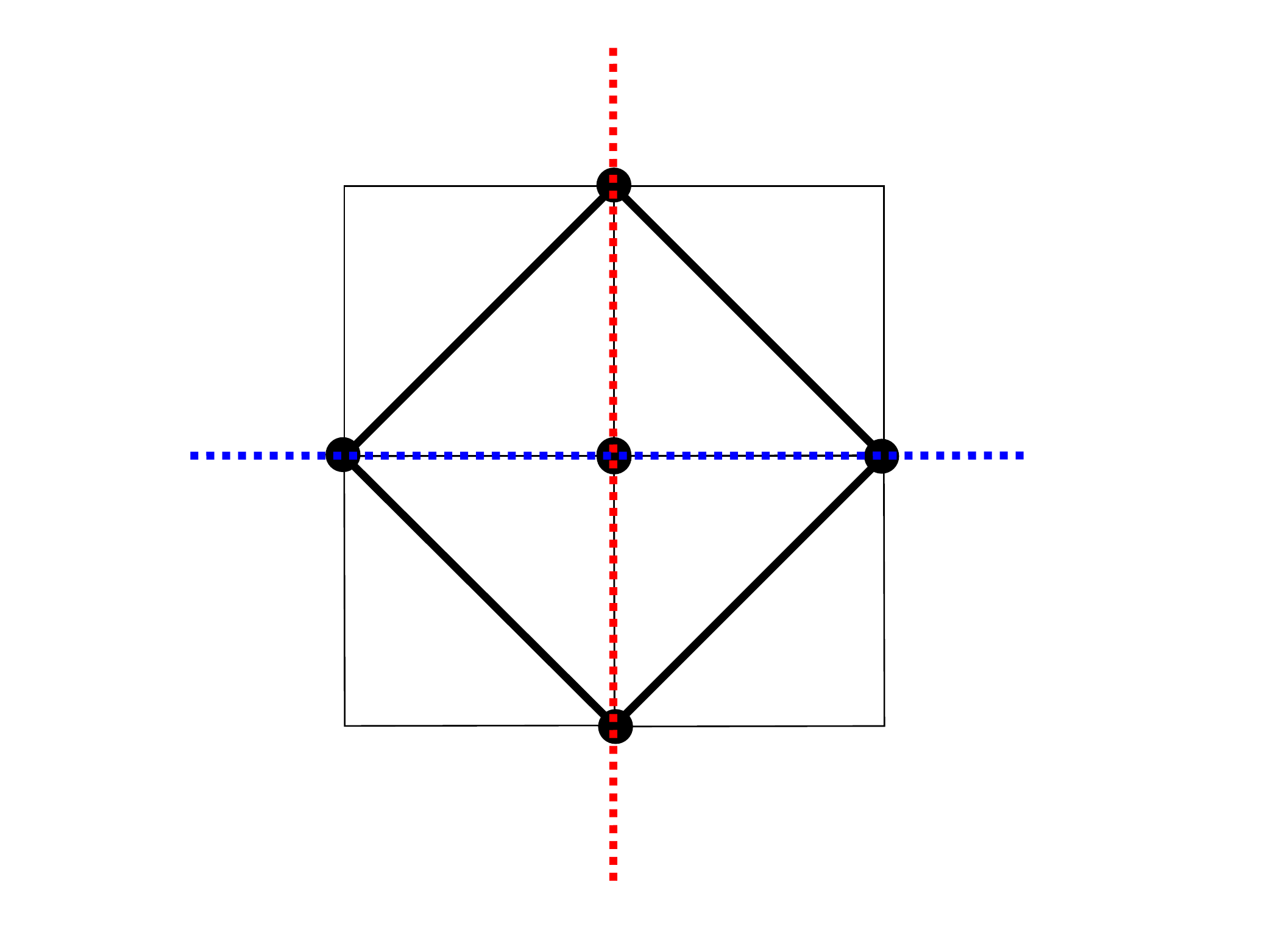}
			\caption{}
			\label{Fig:F0toric}
		\end{center}
	\end{subfigure}
	\caption{Toric diagrams for singularities that satisfy our necessary criterion to admit one (with an axis of reflection in red) or two (with the second axis of reflection in blue) glide projections: (a) orbifold $\mathbb{C}^2/\mathbb{Z}_4$, (b) conifold-like $\mathcal{C}/\mathbb{Z}_2$, and (c) zeroth Hirzebruch surface $F_0$.}
	\label{Fig:toric} 
\end{figure}

These statements can be summarized by saying that the toric diagram should be symmetric with respect to a vertical or horizontal axis. Moreover, each ZZP has to be mapped to another one, imposing that each kind of ZZP parallel to the axis of reflection in the toric plane should come in even numbers. We show in \Cref{Fig:toric} that our examples of \Cref{Sec:Examples} satisfy this criterion.

Lastly, an important remark is that this condition may not be satisfied in some of the $SL(2, \mathbb{Z})$ ``frames'' of the toric diagram, or equivalently, the unit cell in the dimer model may not be symmetric with respect to the glide reflection. Thus, we should state that a generic toric diagram can admit a glide orientifold if it satisfies the conditions above \textit{up to a} $SL(2,\mathbb{Z})$ \textit{action} that can bring its unit cell to a symmetric form with respect to the glide.

\comment{
\AP{Note on the unorbifolding shift ? \figref{Fig:Z2Shift}}
\begin{figure}[h!]
	\centering
	\begin{subfigure}[t]{0.5\textwidth }
		\begin{center} 
			\includegraphics[width=\textwidth]{Z2ZZP.pdf}
			\caption{}
			\label{Fig:Z2ZZP}
		\end{center}
	\end{subfigure} \hspace{15mm}
	\begin{subfigure}[t]{0.4\textwidth }
		\begin{center} 
			\includegraphics[width=\textwidth]{Z2toric.pdf}
			\caption{}
			\label{Fig:Z2toric}
		\end{center}
	\end{subfigure}
	\begin{subfigure}[t]{0.4\textwidth }
		\begin{center} 
			\includegraphics[width=\textwidth]{C3toric.pdf}
			\caption{}
			\label{Fig:C3toric}
		\end{center}
	\end{subfigure}
	\caption{(a) Dimer diagram of the orbifold $\mathbb{C}^2/\mathbb{Z}_{2} \times \mathbb{C}$. The direction of the shift is given in red, and the ZZPs are illustrated in other colours. (b) Toric diagram for the orbifold $\mathbb{C}^2/\mathbb{Z}_{2} \times \mathbb{C}$. (c) Toric diagram for the flat space $\mathbb{C}^3$.}\label{Fig:Z2Shift} 
\end{figure}
}

\subsection{Fractional branes}\label{Sec:FracZZP}

As already mentioned in \Cref{Sec:KBproperties}, these orientifolded theories may admit non-trivial rank assignments, i.e. fractional branes. Their presence can be deduced from the symmetries of the toric diagram and they can be seen as inherited from the ``parent'' theory. Following \cite{Butti:2006hc,Argurio:2020dko}, in what we dub ``Butti's Algorithm'', we can assign a value $v_\Gamma$ to each of the $n$ ZZPs of the toric diagram. These values give rise to anomaly free rank assignments, given that they satisfy the following constraints,
\begin{align}
	\begin{cases}
		\sum_\Gamma v_\Gamma p_\Gamma= 0 \\
		\sum_\Gamma v_\Gamma q_\Gamma = 0 
	\end{cases} \, , \label{Eq:TopConstr}
\end{align}
where the $(p_\Gamma,q_\Gamma)$ are the winding numbers of the ZZP associated to $v_\Gamma$.

Since we know how the glide reflection acts on the ZZPs, we may follow the procedure of \cite{Argurio:2020dko} to see which fractional branes survive the projection. As explained there, only \textit{symmetric} fractional branes survive, in the sense that, given two ZZPs $v_\alpha$ and $v_{\bar{\alpha}}$ mapped to each other under the glide reflection, only rank assignments satisfying the following identification survive,
\begin{equation}
	v_\alpha  =  v_{\bar{\alpha}} \, . \label{Eq:VSym}
\end{equation}
The orientifold projection thus reduces the number of variables $v_\Gamma$ to the subset of $v_\alpha$. Moreover, one can check that \Cref{Eq:TopConstr} leaves only one non-trivial relation: 
\begin{align}
	\sum_\alpha v_\alpha q_\alpha = 0 \, .
\end{align}
Butti's algorithm has a redundancy that allows to perform a global shift on the $v_\alpha$ without affecting the ranks of the gauge groups. Hence, we end up with
\begin{equation}
	\# \text{fractional branes} = n/2 -2
\end{equation}
in the orientifolded theory.

\begin{figure}[h!]
	\centering
	\begin{subfigure}[t]{0.3\textwidth }
		\begin{center} 
			\includegraphics[width=\textwidth]{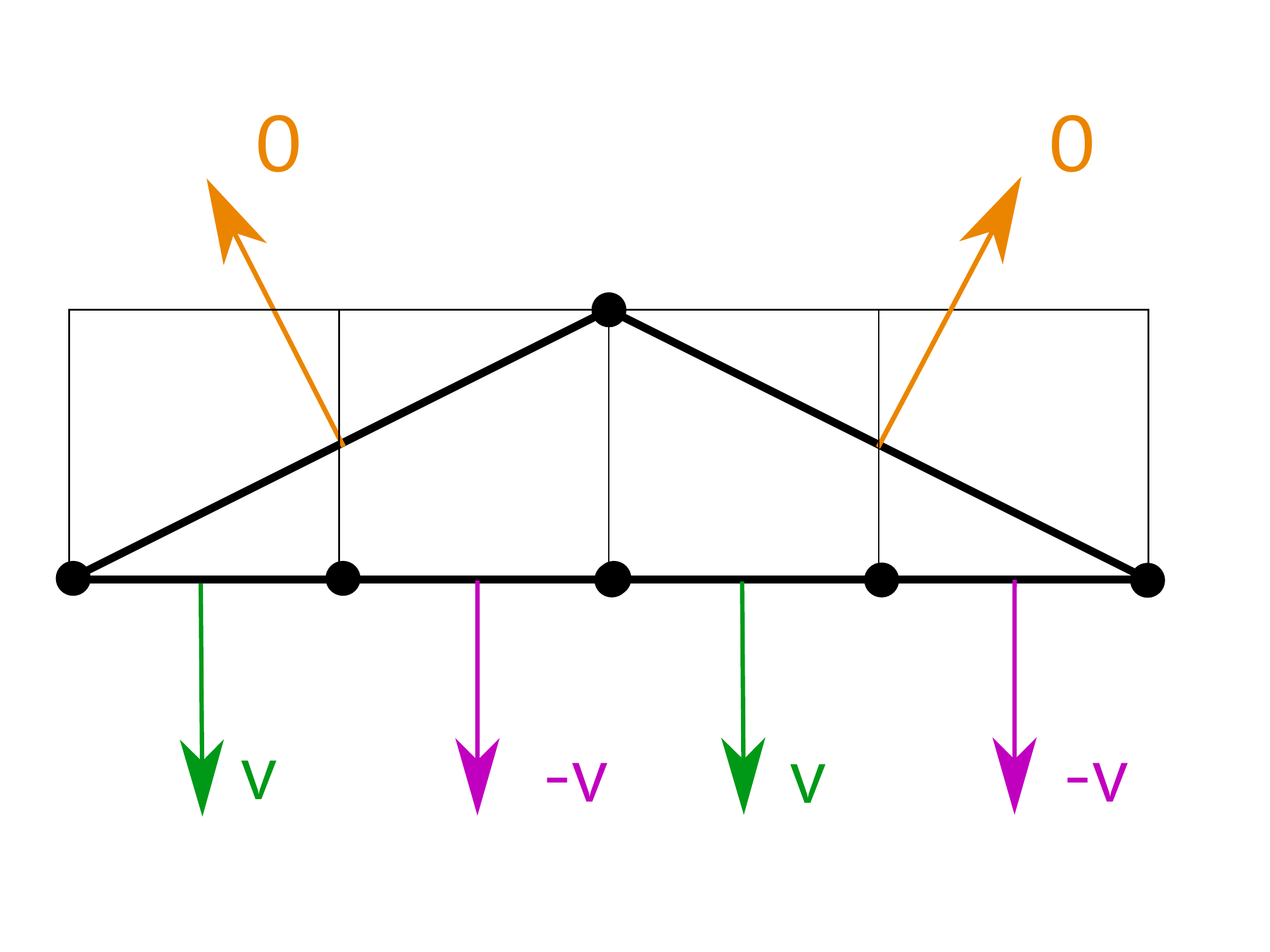}
			\caption{}
			\label{Fig:C2Z4ZZP}
		\end{center}
	\end{subfigure}
	\begin{subfigure}[t]{0.3\textwidth } 
		\begin{center} 
			\includegraphics[width=\textwidth]{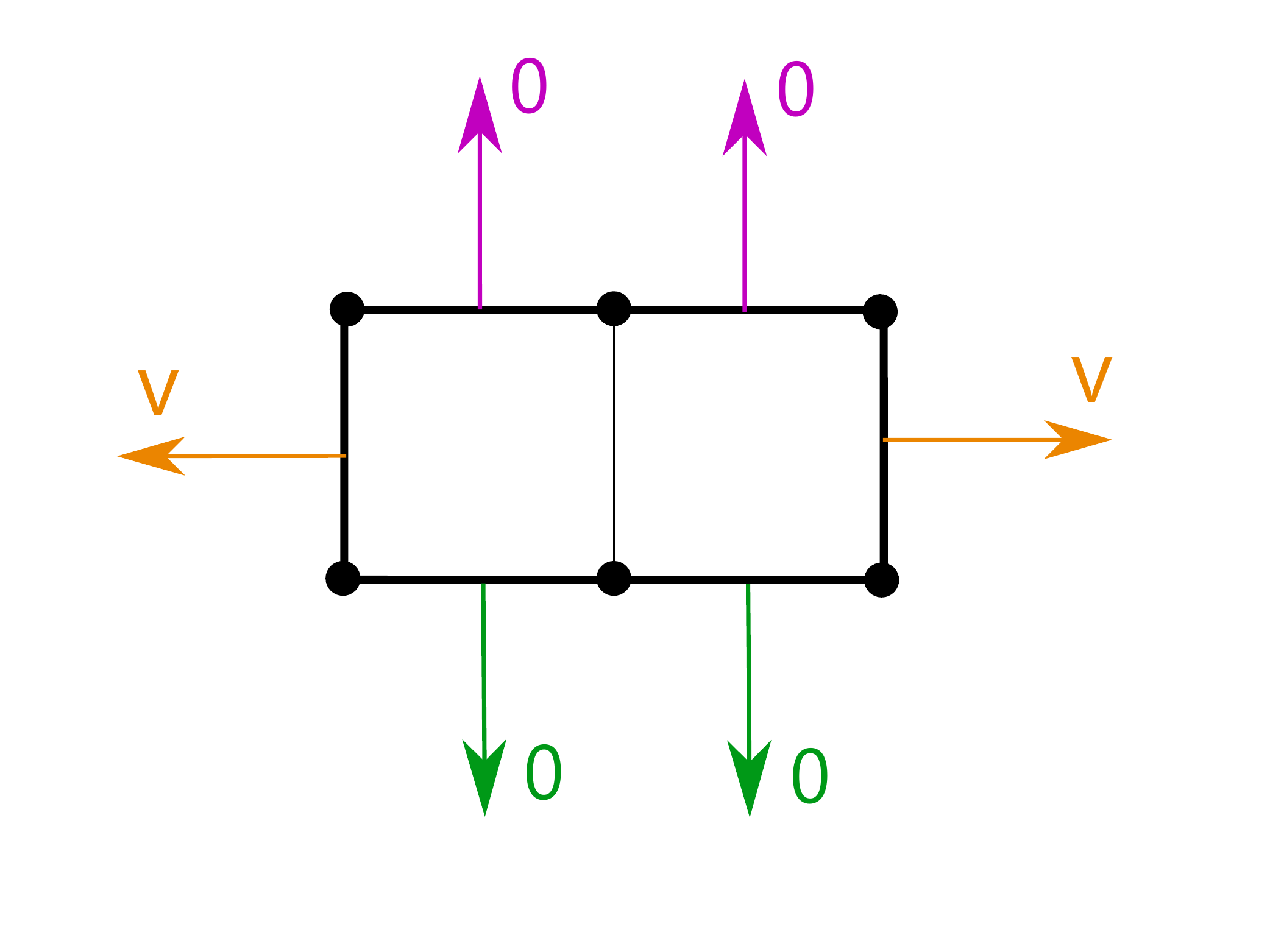}
			\caption{}
			\label{Fig:CZ2ZZP}
		\end{center}
	\end{subfigure} 
	\begin{subfigure}[t]{0.3\textwidth } 
		\begin{center} 
			\includegraphics[width=\textwidth]{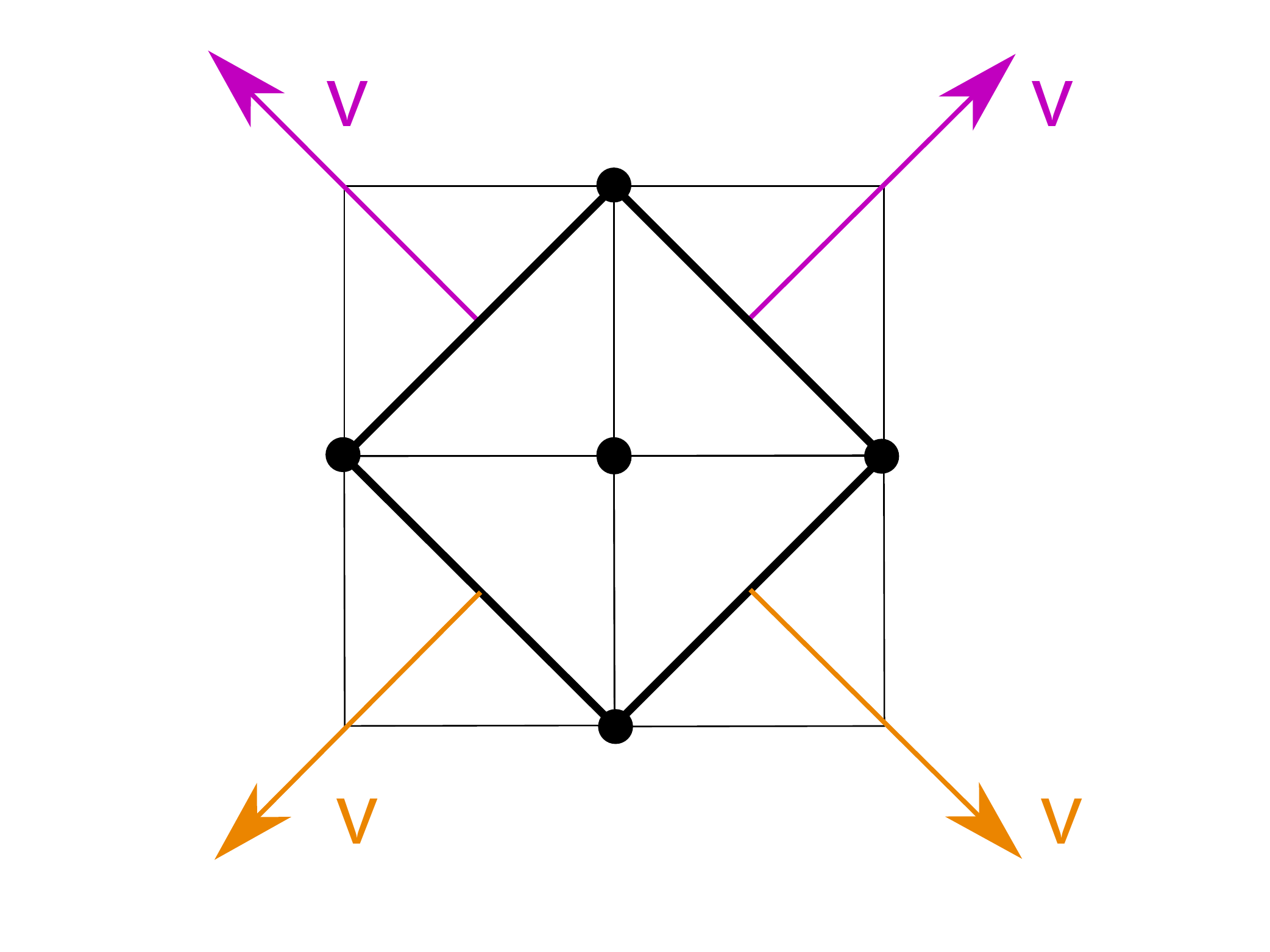}
			\caption{}
			\label{Fig:F0ZZP}
		\end{center}
	\end{subfigure}
	\caption{Symmetric fractional branes in the parent theory lead to fractional branes in its glide orientifolded version. Couples of ZZPs paired by the glide action are drawn in the same color. (a) $\mathcal{N}=2$ fractional brane in $\mathbb{C}^2/\mathbb{Z}_4$, (b) deformation fractional brane in $\mathcal{C}/\mathbb{Z}_2$, and (c) the zeroth Hirzebruch surface $F_0$ admits only the regular brane as a symmetric fractional brane.}
	\label{Fig:ZZPsfractional} 
\end{figure}

Butti's algorithm also tells how to construct different kind of fractional branes in the parent theory by specifying a set of $v_\Gamma$. We now apply this method to theories with a glide reflection orientifold to see when may  $\mathcal{N}=2$ and deformation fractional branes arise.
\begin{itemize}
	\item $\mathcal{N}=2$ fractional branes: The parent theory admits such fractional branes whenever the toric diagram hosts $k>1$ ZZPs with the same winding numbers, say $(p_\mu , q_\mu)$. They are turned on whenever only some of these $v_\mu$, among the whole set of ZZPs $\lbrace v_\Gamma \rbrace$, are non-vanishing. Following \Cref{Eq:TopConstr}, one has
	\begin{equation}
		\sum_{i=1}^k v_{\mu_i} = 0 \, , \quad \text{ and } \,  v_{\nu} = 0 \,\text{ if }\, (p_\nu , q_\nu)  \neq  (p_\mu , q_\mu)\, . \label{Eq:N2}
	\end{equation}
	This condition is compatible with \Cref{Eq:VSym} only if the $k$ ZZPs are sent to ZZPs with the same winding numbers by the glide reflection, restricting to $(0,1)$ or $(0,-1)$ when $(p,q)$ is mapped to $(-p,q)$. Moreover, $k$ should be a multiple of 4, since for each couple of ZZPs with a symmetric assignment $v$, we need a second couple with assignment $-v$ in order to satisfy the sum in \Cref{Eq:N2}. In the examples of \Cref{Sec:Examples}, we found that the singularity $\mathbb{C}^2 / \mathbb{Z}_4$ satisfies this criterion, see \figref{Fig:C2Z4ZZP}.
	\item Deformation fractional branes: The parent theory will have a deformation fractional brane if there is a subset of $m$ ZZPs in equilibrium $\lbrace v_\sigma \rbrace \subset \lbrace v_\Gamma \rbrace$:
	\begin{equation}
		\sum_{i=1}^m (p_{\sigma_i}, q_{\sigma_i}) = 0 \, .
	\end{equation}
	The deformation brane is turned on whenever all $v_\sigma$ have the same non-zero value and all other $v_\tau \notin \lbrace v_\sigma \rbrace$ are vanishing. A glide reflection orientifold theory will have a deformation brane if there is a subset of $m$ ZZPs in equilibrium where each ZZP is accompanied by its image under the glide action, and where $m$ is smaller than $n$.  In the examples of \Cref{Sec:Examples}, we found that $\mathcal{C}/ \mathbb{Z}_2$ satisfies this criterion while the zeroth Hirzebruch surface $F_0$ does not, see \figref{Fig:CZ2ZZP} and \figref{Fig:F0ZZP}.
\end{itemize}

\subsection{Shift Orientifolds}\label{Sec:ShiftOrient}

So far we have only considered orientifolds acting as glide reflections on the dimer. Now we address those acting as a simple shift. We have not discussed these orientifolds earlier because they always break supersymmetry, as we show in the following. In particular, we will see that the holomorphic 3-form $\Omega_3$ is even under such an orientifold action, contradicting the rule of thumb that it should be odd.

\begin{figure}[h!]
	\centering
	\includegraphics[width=0.5\textwidth]{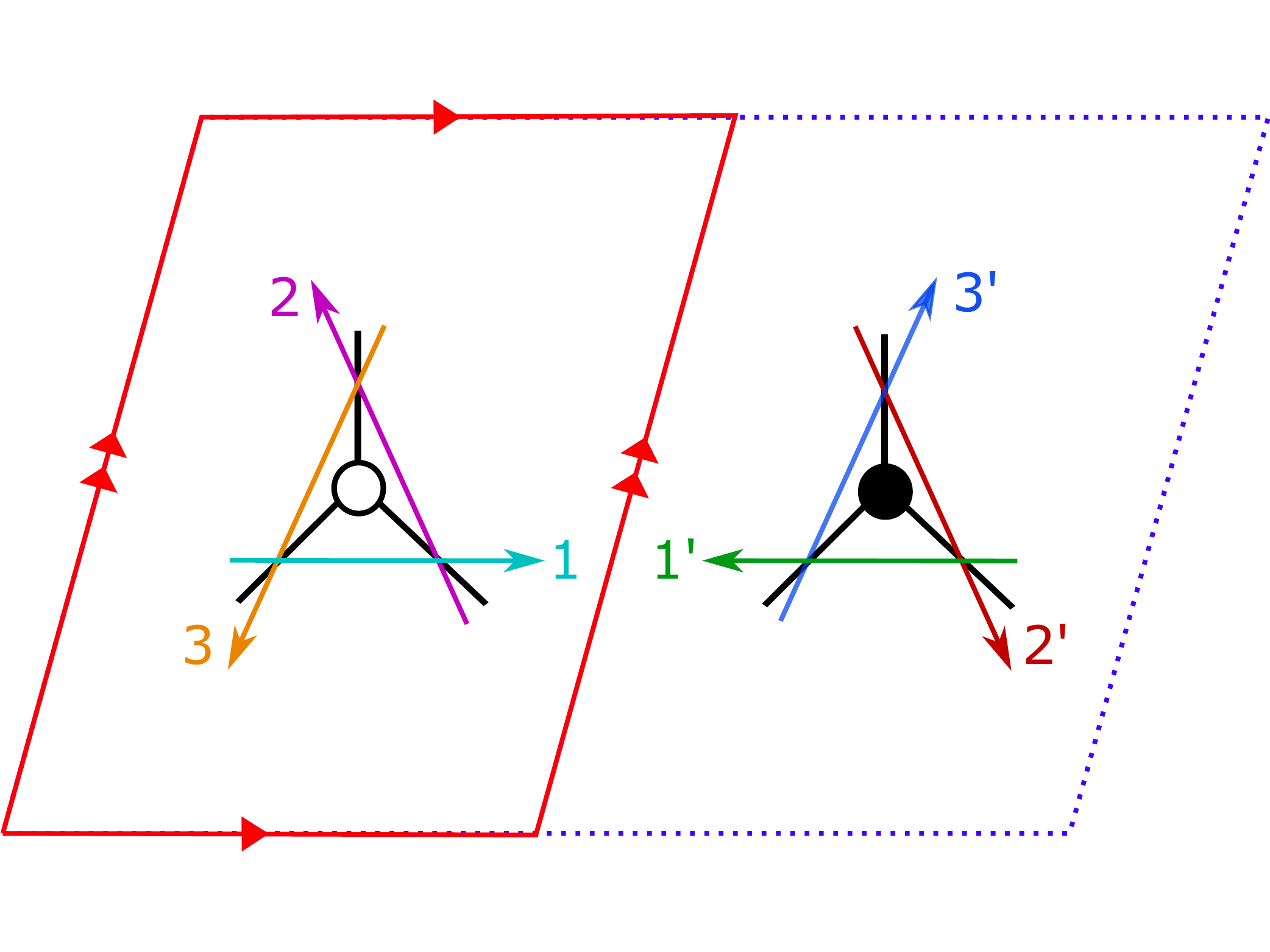}
	\caption{The shift orientifold maps white nodes to black nodes, and vice-versa. The dashed blue line delineates the unit cell of the parent theory, while the red frame represents the orientifold. The ZZPs $1, 2, 3$ are mapped to $1', 2', 3'$ respectively.}
	\label{Fig:ShiftZZP}
\end{figure}

As we observed in \Cref{Torinv}, the shift involution must identify nodes of opposite colors on the dimer, in order to be consistent with the orientifold identification rules. Under such a shift, each ZZP is mapped to a ZZP of opposite winding numbers, $(p,q)\to (-p,-q)$. This can be easily deduced from \Cref{Fig:ShiftZZP}.

From the toric diagram, it is possible to obtain the equations defining associated toric variety probed by the D-branes. To do so we need to compute the integer generators of the dual cone to the toric diagram. This procedure is standard in toric geometry and we refer to \cite{cox2011toric} for all the details. From the lattice vertices on the boundary of the toric diagram $(r_i,s_i)$, we obtain the generators of the cone given by $m_i=(r_i,s_i,1)$. The dual cone is then given by
\begin{align}\label{dualcone}
	S^\vee=\{n \in \mathbb{R}^3 | m_i \cdot n \geq 0 \} \, ,
\end{align}
from which it is easy to see that the vectors $n$ are of the form $(p,q,a)$, where $(p,q)$ are the windings of the ZZPs and $a$ is an integer. Indeed, the generators of the dual cone are nothing but the inward pointing vectors, normal to the faces of the cone generated by the $m_i$. We now need to add the extra generators to span the dual integer cone, $\sigma^\vee=S^\vee \cap \mathbb{Z}^3$. This is achieved by computing linear combinations of the generators with positive rational coefficient and adding all integer vectors we obtain this way. Finally, the equations defining our singularity are given by associating complex coordinates to the generators of the integer dual cone and the relations among them are obtained with the following  identification,
\begin{align}
	n_1+n_2+ \cdots =n_4+n_5+ \cdots \quad \to \quad z_1z_2\cdots=z_4z_5\cdots \, .
\end{align}

For example, let us consider the toric diagram of the conifold, which we place in $\mathbb{Z}^2$ as the square with vertices $(0,0),(0,1),(1,1),(0,1)$. The associated cone is
\begin{equation}
	\sigma = \left\langle \left(\begin{array}{c} 0 \\ 0 \\ 1 \end{array} \right),\left(\begin{array}{c} 0 \\ 1 \\ 1 \end{array} \right),\left(\begin{array}{c} 1 \\ 1 \\ 1 \end{array} \right),\left(\begin{array}{c} 1 \\ 0 \\ 1 \end{array} \right) \right\rangle \subset \mathbb{R}^3 \ ,
\end{equation}
and its dual is
\begin{equation}
	\sigma^\vee = \left\langle n_1=(1,0,0),n_2=(0,1,0),n_3=(-1,0,1),n_4=(0,-1,1) \right\rangle \subset (\mathbb{R}^3)^* \, ,
\end{equation}
from which it is easy to read the equation defining the singularity:
\begin{align}
	n_1+n_3=n_2+n_4 \quad \to \quad z_1z_3=z_2z_4 \,.
\end{align} 
\begin{figure}[h!]
	\centering
	\includegraphics[width=0.3\textwidth]{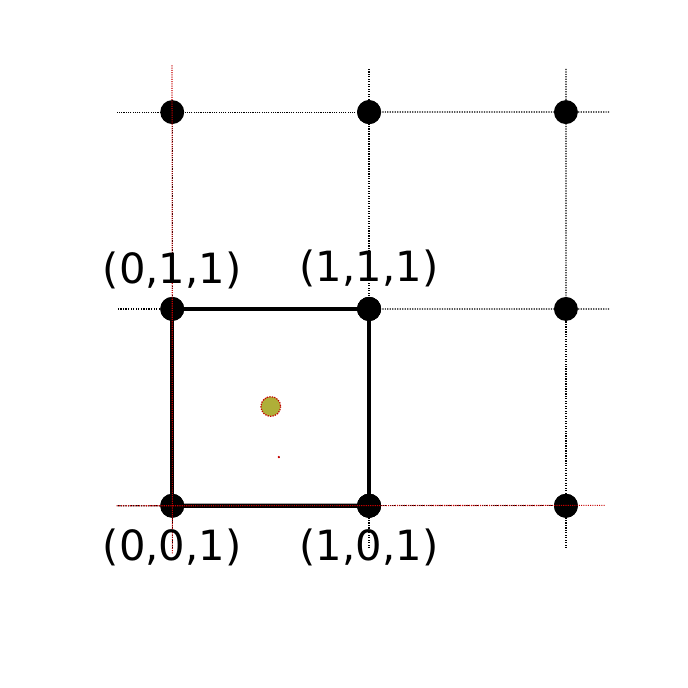}
	\caption{The toric diagram of the conifold.}
\end{figure}

As a second example let us consider the toric diagram of $dP_3$ and the cone it generates:
\begin{equation}
	\sigma = \left\langle \left(\begin{array}{c} 0 \\ -1 \\ 1 \end{array} \right),\left(\begin{array}{c} 1 \\ -1 \\ 1 \end{array} \right),\left(\begin{array}{c} 1 \\ 0 \\ 1 \end{array} \right),\left(\begin{array}{c} 0 \\ 1 \\ 1
	\end{array} \right),\left(\begin{array}{c} -1 \\ 1 \\ 1
	\end{array} \right),\left(\begin{array}{c} -1 \\ 0 \\ 1 \end{array} \right) \right\rangle \subset \mathbb{R}^3 \, .
\end{equation}
It is dual to $\sigma^\vee$ which is the cone:
\begin{align}\label{Eq:dualconedp3}
	\langle n_1=(0,1,1),n_2=(-1,0,1),& n_3=(-1,-1,1),n_4=(0,-1,1),  \nonumber \\
	 & n_5=(1,0,1),n_6=(1,1,1),n_0=(0,0,1) \rangle\ ,
\end{align}
where we added the vector $n_0=(0,0,1)$ since $n_1+n_4=2 n_0$, meaning that we where missing an integer generator.
\begin{figure}[h!]
	\centering
	\includegraphics[width=0.3\textwidth]{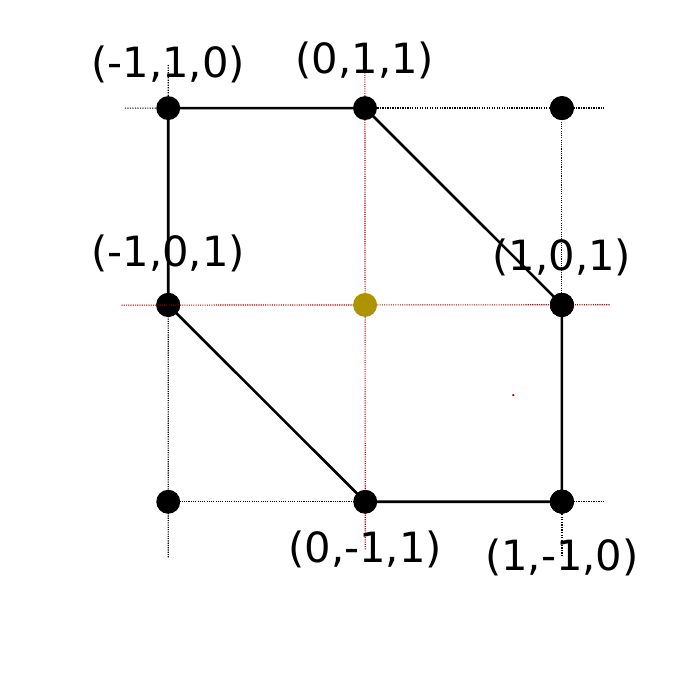}
	\caption{The toric diagram of $dP_3$.}
\end{figure}
The equations of the variety are
\begin{align}\label{Eq:dualconedp3}
	z_1z_4=z_2z_5 & = z_3z_6=z_0^2 \, \nonumber \\
	z_1z_3z_5 & =z_2z_4z_6 \, .
\end{align}

We can use the fact that under the shift involution each ZZP is mapped to a ZZP of opposite winding, hence the corresponding toric diagram must be symmetric under the reflection about its center of mass. Such center of mass has, in general, half-integer coordinates $(\alpha,\beta)$. Under such a reflection, a generic point in the lattice with coordinates $(r,s)\in\mathbb{Z}^2$ is sent to $(2\alpha-r, 2\beta-s)$. Under this operation, the generators of the cone are mapped according to
\begin{equation}
	m'=\left(\begin{array}{ccc}
		-1 & 0 & 2\alpha \\
		0 & -1 & 2\beta \\
		0 & 0 & 1
	\end{array}
	\right) \cdot m
\end{equation}
which maps a generator $m=(a,b,1)$ to $m'=(2\alpha-r, 2\beta-s,1)$. The dual cone $S^\vee$ is in turn invariant under the (right) action of that matrix, which acts as
\begin{align}\label{Eq:symmetry2}
	&n' =\left(\begin{array}{ccc}
		-1 & 0 & 0 \\
		0 & -1 & 0 \\
		2\alpha & 2\beta & 1
	\end{array}\right) \cdot n \, ,
\end{align}
or simply,
\begin{align}\label{Eq:symmetry}
	n = (p,q,a) \quad \to \quad n^\prime = (-p,-q,2\alpha p+  2\beta q + a) \, .
\end{align}

From these observations, we deduce the following properties:
\begin{enumerate}
	
	\item All generators of the dual cone, obtained via \Cref{dualcone},  $n_i=(p_i,q_i,a)$, come paired with another generator $n_{i}^\prime=(-p_i,-q_i,a')$, for some integer $a$ while $a'$ is obtained via \Cref{Eq:symmetry}.
	
	\item Given a generator $n_i=(p_i,q_i,a)$ and its shift image $n_{i}^\prime=(-p_i,-q_i, a')$, we see that a new integer generator that we were missing can be added $n_0=(0,0,1)$, since
	\begin{equation}\label{Eq:Shift1}
	n_i+n_{i}^\prime=(a+a') n_0 \, .
	\end{equation}
	This generator is invariant under the shift. 
	
	\item All other extra generators come in pairs. Given an extra generator $n_l$ such that
	\begin{equation}\label{Eq:Shift2}
	n_i+\dots+n_{j}^\prime + \cdots =b \, n_l \, ,
	\end{equation}
	with $b$ integer, by a symmetry argument, we also need to add $n_{l}^\prime$, since we have\footnote{The transformation law in \Cref{Eq:symmetry} acts linearly on \Cref{Eq:Shift2} such that we obtain \Cref{Eq:Shift3}.}
	\begin{equation}\label{Eq:Shift3}
	n_{i}^\prime+\dots+n_{j} + \cdots =b \, n_{l}^\prime \, .
	\end{equation}	
\end{enumerate}

We now rearrange the generators into two sets: the set of $n_i$ with $i=1,...,k$ and the set $n_{i+k}=n_i^\prime$ of their images under the shift. Moreover we have $n_0$ which is the invariant generator. To each generator $n_i$ we associate a complex coordinate $z_i$. We have $2k+1$ of them, related by $2k-2$ relations, that define the toric 3-fold. We divide these relations in two kinds. The $k$ first kind relations are of the form 
\begin{equation}\label{Eq:Shift5}
z_i z_{i+k}-z_0^{a+a^\prime}=0 \, ,
\end{equation}
and come from \Cref{Eq:Shift1}. We use it to relate every image $z_{i+k}$ to its partner $z_{i}$ and to $z_0$. The second kind relations relate all remaining $z_i$ and $z_0$ together. For example, they may look like
\begin{align}\label{Eq:Shift4}
	z_i z_j z_h^b-z_l z_m z_0^c=0  \, ,
\end{align}
for some integers $b$ and $c$.

Under the shift, relations of the first kind are invariant, those of the second kind are not. However, we can build more symmetric expressions for the latter. As we did when going from \Cref{Eq:Shift2} to \Cref{Eq:Shift3}, \Cref{Eq:Shift4} becomes, under the shift,
\begin{equation}
z_{i+k} z_{j+k} z_{h+k}^b-z_{l+k} z_{m+k} z_0^c=0  \, .
\end{equation}
We can now multiply \Cref{Eq:Shift4} by a term $(z_{i+k} z_{j+k} z_{h+k}^b)$ and use the last equation to find
\begin{equation}
	(z_i z_j z_h^b)(z_{l+k} z_{m+k}) z_0^c - (z_l z_m )(z_{i+k} z_{j+k} z_{h+k}^b)z_0^c = 0 \, ,
\end{equation}
which is now symmetric up to a sign under the shift. We dub these relations the symmetrized second kind relations.

To describe our Calabi-Yau 3-fold, we start with $2k+1$ variables. From the equations of the first kind we can express all the $z_{i+k}$ in terms of the $z_i$ and $z_0$, fixing $k$ variables. Then we can use the symmetrized second kind relations to fix $k-2$ equations, leaving us with only 3 independent variables. Now, the non-vanishing holomorphic 3-form $\Omega_3$ is obtained as the Poincar\'e residue along the CY$_3$ of the meromorphic $(2k+1)$-form in the ambient space $\mathbb{C}^{2k+1}$:
\begin{equation}\label{Eq:ambientform}
\Omega_3	 =  \text{Res}\, \frac{\text{d}z_1\wedge...\wedge \text{d}z_k\wedge...\wedge \text{d}z_{2k}\wedge \text{d}z_0}{\displaystyle \left(\prod_{i=1}^{k} P_i  \prod_{i=1}^{k-2} Q_i \right)} \, ,
\end{equation}
where the $P_i$ are equations of the first kind, while $Q_i$ are of the symmetrized second one. 

Under the action of the shift, the numerator of the 3-form is multiplied by $(-1)^{k}$, since the shift acts on the coordinates exchanging them in pairs. From the denominator we get a factor $(-1)^{k-2}$ coming from the symmetrized second kind equations, cancelling the factor at the numerator and leaving the 3-form invariant. This means that such orientifold projection does not preserve the same supersymmetry as the D3-branes. 

Let us finish this section working out an explicit example. In the case of $dP_3$, one has $k=3$, and the holomorphic 3-form is the residue of the meromorphic 7-form.
\begin{equation}
\Omega_{3}	 =  \text{Res}\, 	\frac{\text{d}z_1\wedge...\wedge \text{d}z_{6}\wedge \text{d}z_0}{(z_1z_4-z_0^2)(z_2z_5-z_0^2)(z_3z_6-z_0^2)(z_1z_3 z_5-z_2 z_4 z_6)}\, .
\end{equation}
Under the involution, the numerator is multiplied by $(-1)^3$. The first three relations are invariants while for the fourth one takes a minus sign. In the end, $\Omega_3$ is even under the symmetry, and hence there cannot be any supersymmetric shift orientifold of $dP_3$.

\section{Conclusion}
In this paper we have studied orientifolds on D3-branes at toric CY$_3$ singularities using dimer models. We established a classification in terms of smooth involutions of the dimer torus, which allowed us to find the last supersymmetric possibility, the glide reflection orientifold. This possibility may also be reached by directly performing the orientifold projection on the open string spectrum. A last possibility existed, a shift orientifold, but it breaks all supersymmetries, as explicitly argued by studying its action on the holomorphic 3-form. Note that these two cases, not considered before, leave no fixed loci. This exhausts the possible orientifolds acting smoothly on the dimer torus.  

Given a toric gauge theory and its associated dimer, one may find the projected theory with the same dictionary as orientifolds with fixed loci. The resulting theories have properties strikingly similar to non-orientifolded theories. 
\begin{itemize}
\item Unlike orientifold theories with fixed loci, glide reflection orientifolds are guaranteed to satisfy the anomaly cancellation conditions for some rank assignment. In fact, these theories are non-chiral. This fact is non-trivial, see \cite{Argurio:2020dko}, and granted by the absence of fixed loci in the glide orientifold that would give raise to tensor matter that could spoil the ACC. From the geometric point of view this boils down to the absence of net RR fluxes sourced by these orientifolds, as there are no fixed loci that can be interpreted as an O-plane. T-duality sheds further light, since the glide orientifold turns to a pair of oppositely charged O-planes on a circle, in the sense of \cite{Dabholkar:1996pc,Witten:1997bs}.
\item Again contrary to intuition, these theories are conformal, as shown by explicit computation of the one loop $\beta$-function, that vanishes identically.
\item Some of these theories admit $\mathcal{N}=2$ or deformation fractional branes. The latter trigger a cascade of dualities \`{a} la Klebanov-Strassler, with a constant step that allows for a UV completion purely in terms of field theory. This is unlike the orientifolds with fixed loci in the literature \cite{Argurio:2017upa} and opens up the possibility of a simple supergravity dual.
\item The glide reflection orientifold may be understood in the T-dual and mirror picture, at least for $\mathbb{C}^2/ \mathbb{Z}_2$, providing a unifying picture.
\end{itemize}

This paper closes the analysis of orientifolds of brane tilings, or at least those acting as smooth involutions on the torus. However, one may consider other kinds of involutions. For example, involutions not respecting the color mapping presented in \Cref{Torinv} or non smooth involutions, can lead to new projections of the tiling, different from the usual orientifold. One may also look for quotients of higher order, in the spirit of what has been done with S-folds \cite{Garcia-Etxebarria:2015wns,Aharony:2016kai}, and their connection with dimer models. These directions are yet to be explored.

Orientifolds have found extensive use in phenomenological applications by allowing for non-perturbatively generated superpotentials or opening the door to SUSY breaking, for instance. We hope our results may shed light in these and related issues.

%========================================
\section*{Acknowledgements}
%========================================

We are very thankful to Riccardo Argurio, Matteo Bertolini and Sebasti\'an Franco for enlightening suggestions and remarks on a preliminary version of the paper. We also thank Ander Retolaza, Angel Uranga for insightful discussions. E.G.-V. and A.P. acknowledge support by IISN-Belgium (convention 4.4503.15) and by the F.R.S.-FNRS under the “Excellence of Science” EOS be.h project n. 30820817. E.G.-V. was also partially supported by the ERC Advanced Grant “High-Spin-Grav”. A.P. is a FRIA grantee of the F.R.S.-FNRS (Belgium). S.M. by the MIUR PRIN Contract 2015 MP2CX4 ``Non-perturbative Aspects Of Gauge Theories And Strings” and by INFN Iniziativa Specifica ST$\&$FI.

\clearpage

\appendix

\section{Worldsheet analysis for the Klein bottle projection of $\mathbb{C}^2/ \mathbb{Z}_4$}\label{Sec:Z4appendix}

In this appendix, we present the worldsheet computations for the K-bottling of $\mathbb{C}^2 / \mathbb{Z}_4$ presented in \Cref{Sec:HigherOrbifolds}.

The open sector of strings on the orbifold before the orientifold projection is obtained as follows:
\begin{equation}
	\begin{array}{cc}
		A_\mu = \left(\begin{array}{cccc}
			A_{1\mu} & 0 & 0 & 0 \\
			0 & A_{2\mu} & 0 & 0 \\
			0 & 0 & A_{3\mu} & 0 \\
			0 & 0 & 0 & A_{4\mu}
		\end{array}\right) \, , \quad & 
		\Phi_1 = \left(\begin{array}{cccc}
			0 & X_{12} & 0 & 0 \\
			0 & 0 & X_{23} & 0 \\
			0 & 0 & 0 & X_{34} \\
			X_{41} & 0 & 0 & 0 
		\end{array}\right) \, , \vspace{0.25cm}\\
		\Phi_2 = \left(\begin{array}{cccc}
			0 & 0 & 0 & Y_{14} \\
			Y_{21} & 0 & 0 & 0 \\
			0 & Y_{32} & 0 & 0 \\
			0 & 0 & Y_{43} & 0 
		\end{array}\right)\, , \quad & \Phi_3 = \left(\begin{array}{cccc}
			Z_{11} & 0 & 0 & 0 \\
			0 & Z_{22} & 0 & 0 \\
			0 & 0 & Z_{33} & 0 \\
			0 & 0 & 0 & Z_{44}
		\end{array}\right) \, .
	\end{array}
\end{equation}

The appropriate orientifold projection, defined as in Equations \eqref{Eq:Z2orientVector} and \eqref{Eq:Z2orientMatter}, is given by 
\begin{equation}
	\gamma_\Omega = \left(\begin{array}{cccc}
		0 & 0 & \mathbb{1}_N & 0 \\
		0 & 0 & 0 & \mathbb{1}_N \\
		\mathbb{1}_N & 0 & 0 & 0 \\
		0 & \mathbb{1}_N & 0 & 0
	\end{array}\right) \, , \quad \text{and} \quad
	R = \left(\begin{array}{ccc}
		0 & 1 & 0\\
		1 & 0 & 0 \\
		0 & 0 & 1
	\end{array}\right) \, .
\end{equation}
It gives the following identification of gauge bosons
\begin{equation}
	A_{1 \, \mu} =  - A_{3 \, \mu}^T \quad \text{and} \quad A_{2 \, \mu} =  - A_{4 \, \mu}^T \, , \\
\end{equation}
the resulting gauge group is $SU(N)_1 \times SU(N)_2$. The matter content follows from
\begin{equation}
	\begin{array}{ccccccl}
		X_{12} &=& Y_{43}^T& \equiv &\mathcal{X}_{12}& \in & (\antifund_1 , \fund_2) \, ,\\
		X_{23} &=& Y_{14}^T& \equiv &\mathcal{X}_{21}& \in & (\antifund_2 , \antifund_1) \, ,\\
		Y_{21} &=& X_{34}^T & \equiv &\mathcal{Y}_{21} & \in & (\antifund_2 , \fund_1) \, ,\\
		Y_{32} &=& X_{41}^T & \equiv &\mathcal{Y}_{12} & \in & (\fund_1 , \fund_2) \, ,\\
		Z_{11} &=& Z_{33}^T & \equiv &\mathcal{Z}_{11} & \in & \mathrm{Adj}_1 \, , \\
		Z_{22} &=& Z_{44}^T & \equiv &\mathcal{Z}_{22} & \in & \mathrm{Adj}_2 \, .
	\end{array} \label{Eq:Z4Orient}
\end{equation}
One can check that the superpotential is the one advertised in Equation \eqref{Eq:Z4supo}.

\section{Computations for the orbifolds of the conifold} \label{Sec:Appendix2}

\subsection{Orbifold of the conifold $\mathcal{C}/\mathbb{Z}_2$}\label{Sec:ConifoldZ2Appendix}
Let be a non-chiral orbifold of the conifold, $\mathcal{C}/ (\mathbb{Z}_l \times \mathbb{Z}_m)$. The general action is given by
\begin{align}
	&\gamma_g V_{1,2} \gamma_g^{-1} = V_{1,2} \nonumber \\
	&\gamma_g A_1 \gamma_g^{-1}=e^{2 \pi i /l}A_1 \, , \quad \gamma_g A_2 \gamma_g^{-1}=A_2   \\
	&\gamma_g B_1 \gamma_g^{-1}=e^{-2 \pi i /l}B_1 \, , \quad \gamma_g B_2 \gamma_g^{-1}=B_2 \, ,  \nonumber 
\end{align}
and
\begin{align}
	&\gamma_g V_{1,2} \gamma_g^{-1} = V_{1,2} \nonumber \\
	&\gamma_g A_1 \gamma_g^{-1}=e^{2 \pi i /m}A_1 \,, \quad \gamma_g A_2 \gamma_g^{-1}=A_2   \\
	&\gamma_g B_1 \gamma_g^{-1}=B_1 \,, \quad \gamma_g B_2 \gamma_g^{-1}=e^{-2 \pi i /m}B_2 \, ,  \nonumber 
\end{align}
where $V_{1,2}$ are the two adjoint vectors related to the gauge groups. In the case of our first example,  $\mathcal{C}/ \mathbb{Z}_2 $, the action gives the following fields 
\begin{equation}
	\begin{array}{ccc}
		V_1 = \left(\begin{array}{cc}
			V_{1} & 0 \\
			0 & V_{3}
		\end{array}\right) \, ,  & \quad
		V_2 = \left(\begin{array}{cc}
			V_{2} & 0 \\
			0 & V_{4}
		\end{array}\right) \, ,  & \quad
		A_1 = \left(\begin{array}{cc}
			0 & A_{14} \\
			A_{32} & 0
		\end{array}\right) \, ,  
		\\
		A_2 =\left(\begin{array}{cc}
			A_{12} & 0 \\
			0 & A_{34}
		\end{array}\right)\, , \quad &  B_1 = \left(\begin{array}{cc}
			B_{21} & 0 \\
			0 & B_{43}
		\end{array}\right) \, , \quad & B_2 = \left(\begin{array}{cc}
			0 & B_{23} \\
			B_{41} & 0
		\end{array}\right)\, ,
	\end{array}
\end{equation}
with a superpotential given by
\begin{align} 
	W = A_1B_1A_2B_2 -A_1B_2A_2B_1 \, .
\end{align}

We consider the following orientifold projection in order to reproduce the glide projection.
\begin{align}
	&V_{1,2} = - \gamma_\Omega V_{1,2}^T \gamma_\Omega^{-1} \, ,  \nonumber\\ 
	&A_{1,2} = \gamma_\Omega B_{1,2}^T \gamma_\Omega^{-1} \, , 
\end{align}
with 
\begin{equation}
	\gamma_\Omega=\begin{pmatrix} 0 && \mathbb{1}_N \\ \mathbb{1}_N  && 0 \end{pmatrix} \, .
\end{equation}

The action on mesons, $x=(A_1B_1)^2$ $y=(A_2B_2)^2,$ $z=A_1B_2$ and $w=A_2B_1$, is
\begin{align}
	&x \leftrightarrow y \, , \hspace{1 cm} z \rightarrow z \, , \hspace{1 cm} w \rightarrow w \, , \nonumber \\
	&\Omega_3=\frac{\text{d}x \wedge \text{d}y \wedge \text{d}z}{2wz^2} \to \Omega_3'=\frac{\text{d}y \wedge \text{d}x \wedge \text{d}z}{2wz^2}=-\Omega_3 \,
\end{align}
which means that the action preserves supersymmetry on the branes.

The gauge group is $SU(N_1)_1 \times SU(N_2)_2$ and matter content given by
\begin{equation}
	\begin{array}{cccc}
		A_{14}=B_{23}^T \equiv & A & = & (\antifund_1 , \antifund_2) \, ,  \\
		B_{41}=A_{32}^T \equiv & B  & = & (\fund_1 , \fund_2)\, , \\
		A_{12}=B_{43}^T \equiv & C  & = & (\antifund_1 , \fund_2)\, , \\
		B_{21}=A_{34}^T \equiv & D  & = & (\fund_1 , \antifund_2)\, , 
	\end{array}
\end{equation}
with superpotential 
\begin{align}
	W= ABCD - BA C^T D^T \, . 
\end{align}

\subsection{Zeroth Hirzebruch surface F$_0$}\label{Sec:F0}
In this case we take the following actions on the fields
\begin{align}
	&\gamma_g V_{1,2} \gamma_g^{-1}= V_{1,2} \\ \nonumber
	&\gamma_g A_1 \gamma_g^{-1}=-A_1 \\ \nonumber
	&\gamma_g A_2 \gamma_g^{-1}=-A_2 \\ \nonumber
	&\gamma_g B_1 \gamma_g^{-1}=B_1 \\ \nonumber
	&\gamma_g B_2 \gamma_g^{-1}=B_2 \, , \\ \nonumber 
	&\end{align}
leading to
\begin{equation}
	\begin{array}{ccc}
		V_1 = \left(\begin{array}{cc}
			V_{1} & 0 \\
			0 & V_{3}
		\end{array}\right) \, ,  & \quad
		V_2 = \left(\begin{array}{cc}
			V_{2} & 0 \\
			0 & V_{4}
		\end{array}\right) \, ,  & \quad
		A_1 = \left(\begin{array}{cc}
			0 & A^1_{14} \\
			A^1_{32} & 0
		\end{array}\right) \, , \nonumber
		\\
		A_2 = \left(\begin{array}{cc}
			0 & A^2_{14} \\
			A^2_{32} & 0
		\end{array}\right)\, , 
		&\quad B_1 = \left(\begin{array}{cc}
			B^1_{21} & 0 \\
			0 & B^1_{43}
		\end{array}\right) \, , &\quad B_2 = \left(\begin{array}{cc}
			B^2_{21} & 0 \\
			0 & B^2_{43}
		\end{array}\right)\, .
	\end{array}
\end{equation}

The orientifold action maps $1\to 4$ and $2 \to 3$, it can be summarized as
\begin{align}
	&V_{1,2} = - \gamma_\Omega V_{1,2}^T \gamma_\Omega^{-1} \, , \nonumber \\ 
	&A_{1} = \gamma_\Omega A_{2}^T \gamma_\Omega^{-1} \, , \\ 
	&B_{1} = \gamma_\Omega B_{2}^T \gamma_\Omega^{-1} \, , \nonumber
\end{align}
with
\begin{equation}
	\gamma_\Omega=\begin{pmatrix} \mathbb{1}_N && 0 \\ 0 && \mathbb{1}_N \end{pmatrix} \, .
\end{equation}
The resulting gauge group is $SU(N)_1 \times SU(N)_2$ and the matter content is given by
\begin{equation}
	\begin{array}{ccccc}
		A^1_{14} = A^{2T}_{14} &\equiv&  U_{S,A}  & = & \antisymm_1 , \antiasymm_1\, , \\
		A^2_{32}=A^{2T}_{32} &\equiv& Z_{S,A} & = & \symm_2 , \asymm_2\,  , \\
		B^1_{21}=B^2_{43} &\equiv& X & = & (\fund_1 , \antifund_2) \, , \\
		B^2_{21}=B^1_{43} &\equiv& Y & = & (\fund_1 , \antifund_2)\, , 
	\end{array}
\end{equation}
with the following superpotential,
\begin{align}
	W= X U_S Y^T Z_A - X^TZ_S Y U_A \, . 
\end{align}

In order to compute the action the 3-form, we compute the equations defining the singularity using the geometrical approach described in \Cref{Sec:ShiftOrient}. The singularity is described by the following equations in $\mathbb{C}^9$
\begin{align}
	z_1 z_3 = z_2 z_4 &= z_0^2\, , \nonumber \\
	z_1 z_2 = z_5 ^2 \, , \hspace{1 cm}& z_2 z_3 = z_7 ^2\, ,  \\
	z_1 z_4 = z_6 ^2 \, ,\hspace{1 cm}& z_3 z_4 = z_8 ^2\, .\nonumber
\end{align}

The action on mesons is
\begin{align}
	z_1 \leftrightarrow z_2 \, , \hspace{1 cm} z_3 \leftrightarrow z_4 \, , \hspace{1 cm} \, z_6 \leftrightarrow z_7 \,
\end{align}
while all other coordinates are invariant. The action on the 3-form is
\begin{align}
	\Omega_3=\text{Res} \frac{\text{d}z_1 \wedge \text{d}z_2 \wedge \text{d}z_3\wedge \text{d}z_4 \wedge \text{d}z_5\wedge \text{d}z_6\wedge \text{d}z_7\wedge \text{d}z_8\wedge \text{d}z_0}{\prod_i P_i} \to -\Omega_3 \, ,
\end{align}
since the polynomials are invariant and in the numerator we are exchanging three pairs of coordinates, resulting in an overall minus sign.

\subsection{A cascade in the glide projection of $\mathcal{C}/\mathbb{Z}_2$}\label{Sec:Cascade}

For a generic choice of ranks, $SU(N+ M)_1\times SU(N)_2$, one finds that the gauge theory has a non-trivial RG-flow and $SU(N+M)_1$ goes more rapidly to strong coupling as we approach the infrared regime of the theory:
\begin{equation}
	\beta_1 = 3M \, , \quad \beta_2 = -3M \, .
\end{equation}
The mesons of the first gauge group are
\begin{equation}
	M_1 = BA \, \quad M_2 = BC \, , \quad M_3 = C^T D^T \quad \text{and} \quad M_4 = DA \, , \label{Eq:Mesons}
\end{equation}
and one thus finds that this gauge theory is Seiberg dual \cite{Seiberg:1994pq} to $SU(N-M)_1\times SU(N)_2$ with a matter content given by the mesons $M_1, M_2, M_3$ and $M_4$ in addition to the following list of bifundamental fields:
\begin{equation}
	a = (\fund_1 , \fund_2) \, , \quad b = (\antifund_1 , \antifund_2) \, , \quad c=(\fund_1 , \antifund_2) \quad \text{and} \quad d = (\antifund_1 , \fund_2) \, .
\end{equation}
The superpotential is given by
\begin{eqnarray}
	W &=& M_2M_4 - M_1 M_3 + M_1 a b + M_2 cb + M_3 d^T c^T + M_4 ad \nonumber \\
	&=&  abd^Tc^T - badc
\end{eqnarray}
where the mesons have been integrated out using F-term relations.

The new gauge theory $SU(N-M)_1\times SU(N)_2$ ends up with the same matter content and superpotential (up to an overall sign) as the initial $SU(N+M)_1\times SU(N)_2$. This can be seen easily with the following mapping:
\begin{equation}
	A \rightarrow b  \, , \quad B \rightarrow a \, , \quad C \rightarrow d \quad \text{and} \quad D \rightarrow c \, .
\end{equation}
The $M$ deformation branes thus trigger a cascade of Seiberg dualities \`a la Klebanov-Strassler \cite{Klebanov:2000hb}. In particular, for $N$ being an integer multiple of $M$, we expect the cascade flow down to $SU(2M)\times SU(M)$ where the physics should the same as for the deformed conifold. Indeed, we can schematically define baryonic operators $\bar{\mathcal{B}} = \left[A C^T\right]^M$, ${\mathcal{B}} = \left[BD^T\right]^M$, and a $2M \times 2M$ squared matrix $\mathcal{M}$ in terms of the mesonic operators of \Cref{Eq:Mesons} that should obey a relation of the form
\begin{equation}
	\mathrm{det}\mathcal{M} - \bar{\mathcal{B}}\mathcal{B} = \Lambda^{4M}_{2M}  \, ,
\end{equation}
where $\Lambda_{2M}$ is the strong coupling scale of $SU(2M)$. Going on the baryonic branch $\bar{\mathcal{B}} = \mathcal{B} = i \Lambda^{2M}_{2M}$, one finds that the mesons decouple, leaving a SYM $SU(M)$ dynamics displaying confinement and chiral symmetry breaking. 

%========================================%
\bibliographystyle{JHEP}
\bibliography{ref}

%========================================%

\end{document}